\documentclass[aps,twocolumn,showpacs]{revtex4} %eqsecnum,

\usepackage{bm}
\input epsf
\newcommand\lsim{\mathrel{\rlap{\lower4pt\hbox{\hskip1pt$\sim$}}
        \raise1pt\hbox{$<$}}}
\newcommand\gsim{\mathrel{\rlap{\lower4pt\hbox{\hskip1pt$\sim$}}
        \raise1pt\hbox{$>$}}}

\newcommand{\sfig}[2]{
\centerline{ \epsfxsize = #2 \epsfbox{#1} }
                }
\newcommand{\Sfig}[2]{
        \begin{figure}[tb]
        \sfig{#1.eps}{0.95\columnwidth}
        \caption{{\small #2}}
        \label{fig:#1}
        \end{figure}
}

\begin{document}

%%%\twocolumn[\hsize\textwidth\columnwidth\hsize\csname
%%%@twocolumnfalse\endcsname

%\draft
%\preprint{HEP/123-qed}
\title{Correlated Fluctuations in Luminosity Distance and the Importance of Peculiar Motion
in Supernova Surveys}

\author{Lam Hui$^{1,2}$ and Patrick B. Greene$^{1,3}$
}

%%%\address{
\affiliation{
$^1$ Institute for Strings, Cosmology and Astro-particle Physics (ISCAP)\\
$^2$ Department of Physics, Columbia University, New York, NY 10027\\
$^3$ Department of Physics and Astronomy, University of Texas at San Antonio, TX 78249\\
{lhui@astro.columbia.edu, Patrick.Greene@utsa.edu}
}

\date{\today}
%%%\maketitle
\begin{abstract}
Large scale structure introduces two different kinds of errors in the luminosity 
distance estimates from standardizable candles such as supernovae Ia (SNe) -- a 
Poissonian scatter for each SN and a coherent component due to correlated 
fluctuations between different SNe. Increasing the number of SNe helps reduce 
the first type of error but not the second. The coherent component has been 
largely ignored in forecasts of dark energy parameter estimation from upcoming SN 
surveys. For instance it is commonly thought, based on Poissonian considerations, 
that peculiar motion is unimportant, even for a low redshift SN survey such as the 
Nearby Supernova Factory (SNfactory; $z = 0.03 - 0.08$), which provides a useful 
anchor for future high redshift surveys by determining the SN zero-point. We show 
that ignoring coherent peculiar motion leads to an underestimate of the zero-point 
error by about a factor of $2$, despite the fact that SNfactory covers almost half 
of the sky. More generally, there are four types of fluctuations: peculiar motion,
gravitational lensing, gravitational redshift and what is akin to the integrated 
Sachs-Wolfe effect. Peculiar motion and lensing dominates at low and high redshifts 
respectively. Taking into account all significant luminosity distance fluctuations 
due to large scale structure leads to a degradation of up to $60 \%$ in the 
determination of the dark energy equation of state from upcoming high redshift SN 
surveys, when used in conjunction with a low redshift anchor such as the SNfactory.
The most relevant fluctuations are the coherent ones due to peculiar motion and the 
Poissonian ones due to lensing, with peculiar motion playing the dominant role.
We also discuss to what extent the noise here can be viewed as a useful 
signal, and whether corrections can be made to reduce the degradation.
%\vskip 0.3truecm
%98.80.-k; 98.80.Es; 98.80.Jk; 95.30.Sf
\end{abstract}  


\maketitle

%%%\vskip 0.5truecm
%%%]

%\narrowtext

\section{Introduction}
\label{intro}

The problem of the cosmological constant, or more generally dark energy, is one of the deepest
problems in cosmology today. While there are by now multiple lines of evidence for the existence
of dark energy \cite{evidence}, the evidence from type Ia supernovae (SNe) was historically 
what convinced a large fraction of the cosmology community that this enigmatic form of energy 
should be taken seriously \cite{Riess:1998,Perlmutter:1999}.
Upcoming and ongoing SN surveys \cite{SNsurveys}, with vastly improved statistics, 
promise to constrain the equation of state of dark energy
to unprecedented precision, thus shedding light on the issue of whether
the apparent acceleration of the universe is caused by the cosmological constant,
a dynamical scalar field or departure from Einstein gravity \cite{gravity}.

There has been much recent work on projections for the determination of dark energy properties
from these SN surveys. By and large, they focus on the following aspects of the error budget:
intrinsic statistical error, systematic error and gravitational lensing induced scatter 
(e.g. \cite{josh0,hutererturner,josh,linder,kim,hutererkim,hl} and references therein).
The intrinsic statistical error refers to the intrinsic spread in SN luminosity even after
suitable standardizing corrections have been applied. It is typical to assume that
the intrinsic spread in magnitude has a (root-mean-squared; rms) size of 
$\sigma^{\rm intr.} = 0.1 - 0.15$ for each SN
\cite{sigintr}.
This kind of intrinsic statistical error can be beaten down by having a large number of SNe.
There are several sources of systematic error, such as Malmquist bias, luminosity evolution,
imperfect corrections for dust extinction, and so on. They are not necessarily diminished by having a large number of SNe, although
a large sample often helps in identifying and characterizing them.
Lastly, gravitational lensing by intervening structures introduces fluctuations in the observed
flux of SNe. So far, the focus has been on how gravitational lensing introduces a Poissonian scatter
rather analogous to the intrinsic spread. This kind of error can likewise be reduced by having a large
sample of SNe \cite{josh0,hl}.

The existing discussion can be improved in two ways. First of all, gravitational lensing
by large scale structure introduces not only a Poissonian scatter to the individual SN flux, but also
correlated flux fluctuations between different SNe. 
One can view the correlated fluctuations as a consequence of the large scale coherence of 
the intervening structures. Second, large scale structure introduces
fluctuations beyond that captured by gravitational lensing, and like lensing, these fluctuations have
a Poissonian component as well as a correlated or coherent component. 
It is worth noting that an expression for all the first order fluctuations in the luminosity distance 
-- first order in metric and energy-momentum perturbations -- has been worked out for 
quite some time e.g. \cite{sasaki,pyne1,pyne2} (with minor corrections; see below).
The full implications for current and future SN surveys, however, have not been explored, with an important
exception (pointed out to us by Dragan Huterer) -- Sugiura, Sugiyama \& Sasaki \cite{sss99} 
computed the anisotropies (the dipole and beyond) in luminosity distance and investigated the
implications for measurements of the decceleration parameter $q_0$.

As we will see, to first order, there are four sources of luminosity distance (or magnitude) fluctuations:
gravitational lensing, gravitational redshift, peculiar motion and an effect akin to the integrated Sachs-Wolfe (ISW) effect.
We will see that for most practical purposes, it is sufficient to consider gravitational lensing and peculiar motion.
They become important at high ($z \gsim 1$) and low ($z \lsim 0.1$) redshifts respectively.

What is particularly interesting, and perhaps surprising, is that peculiar motion plays a significant role
in the degradation of dark energy errors. There is a widespread perception
that the effects of peculiar motion are negligible as long as the median redshift
is greater than $0.05$ or so. Let us take as an example the
Nearby Supernova Factory (SNfactory; other low redshift surveys include the CfA Supernova Program,
Carnegie Supernova Project and LOTOSS, see \cite{SNsurveys}), 
whose redshift range $z = 0.03 - 0.08$ was chosen 
in the hope of making the effects of peculiar motion negligible.
Such a perception seems at first sight quite reasonable: typical peculiar velocities are of the order of
$300$ km/s, and so the ratio of peculiar flow to Hubble flow at $z = 0.055$ is about
$300 / (3 \times 10^5 \times 0.055) \sim 0.02$. Translating this into fluctuations in magnitude 
(details are given 
in \S \ref{dLf}), we have $\delta m \sim 2.17 \times 0.02
\sim 0.04$, which is quite a bit smaller than the intrinsic spread in SN magnitude ($0.1 - 0.15$),
apparently suggesting we can ignore peculiar motion (recall that different sources of errors
are to be added in quadrature).

What such an argument misses is that coherent peculiar flows introduce correlations in
magnitude fluctuations between different SNe. While it is true that peculiar motion introduces
a negligible Poissonian scatter compared to the intrinsic scatter, the correlated component cannot
be ignored as it turns out. One can intuitively understand it as follows. As the number of SNe ($N$) becomes
large, the intrinsic statistical error is beaten down to be quite small in the usual root-$N$ fashion.
Correlated errors, such as that due to correlated/coherent peculiar flows, are not reduced by $N$ at all, and
so there must be some $N$ beyond which the correlated errors become dominant.
We will see that this is indeed the case for the SNfactory. Since a low redshift survey such as the 
SNfactory plays an important role in
constraining the SN zero-point, dark energy determination from higher redshift surveys (where peculiar motion
is less of an issue) is affected indirectly by these considerations as well.

Coherent large scale flows (i.e. bulk flows) have of course been the subject of research for a long
time (see \cite{strauss} for a review).
Particularly relevant to our investigation are discussions of the peculiar velocity monopole, or
what is sometimes referred to as the local Hubble bubble, which incidentally made
use of SNe Ia \cite{shi,shiturner,idit}.
We will see later that for a survey like the SNfactory, which covers roughly half of the sky,
fluctuations in the lower velocity multipoles (the monopole, dipole, etc) 
contribute significantly to the dark energy error budget.

The rest of the paper is organized as follows. 
In \S \ref{prelim}, we set the stage by describing the parameters of interest and
how the fluctuations in magnitude are related to the parameter errors. Some details on 
how to go from the Fisher matrix to actual errorbars of various types are given in Appendix \ref{app:fisher}. 
We describe in \S \ref{flux} how the average magnitude in a given redshift bin fluctuates, and
how these fluctuations can be divided into a Poissonian component and a correlated/coherent component, in effect
defining the magnitude covariance matrix.
To keep the discussion simple, the derivation is relegated to
Appendix \ref{app:poisson}.
In \S \ref{frw}, we derive an expression for the first order luminosity distance fluctuations (\S \ref{dLf}),
and work out explicitly their implications for the magnitude covariance matrix in terms of the
mass power spectrum (\S \ref{dLf2}). 
To keep the discussion simple in the main body, details of these two steps are relegated to
Appendices \ref{app:dL} and \ref{app:poisson2}.
Appendix \ref{app:dL} might be interesting to the more theoretically inclined: 
the explicit expression given here for the luminosity distance fluctuation corrects a minor error
in earlier expressions in the literature.
Appendix \ref{app:poisson2} contains expressions for the velocity window function for an arbitrary
survey geometry that might be of interest to observers who are interested in making predictions for
their own surveys.
In \S \ref{forecasts}, we finally put everything together to make error forecasts. 
It proves illuminating to first focus separately on the contributions to errors
from peculiar motion and lensing (\S \ref{vel} and \ref{lens}), 
and in \S \ref{everything} we make forecasts for a number of ongoing/planned/proposed SN surveys.
The key results are summarized in Fig. \ref{fig:testhistoB}, \ref{fig:testhistoBnoprior}, 
\ref{fig:contour.1.combo} and Table \ref{table}.
We conclude in \ref{discuss} with a brief summary of major results and
a discussion of several issues that naturally arise, some of which are worth exploring further:
\begin{itemize}
\item whether peculiar motion degrades the {\it current} constraints on dark energy (the answer is: not
significantly);
\item how the exact survey geometry impacts dark energy errors;
\item whether internal motion that could add to the peculiar velocity is important (the answer is no);
\item the issue of systematic errors, and how they might change our conclusions;
\item whether realistic redshift measurements are accurate enough for us to have to worry about peculiar velocities 
(the answer is yes);
\item whether corrections can be made for peculiar motion and lensing to reduce the dark energy errors
(the answer is: probably difficult for (the Poissonian part of) lensing, but maybe yes for peculiar motion);
\item whether the noise that we refer to here from peculiar motion and lensing can in fact be turned into
a useful signal (the answer, for lensing, is that the signal is not competitive with the lensing of galaxies).
\end{itemize}

A comment on our terminology: we refer to the fluctuations of interest in this paper as large scale structure induced.
The term large scale structure should be viewed as synonymous with departure from homogeneity. 
Some of the fluctuations discussed here, such as the Poissonian lensing scatter, are in fact dominated
by structures on relatively small scales (galactic scales or smaller).
Also, even though much of the discussion in this paper is phrased in terms of SNe as standard candles,
most of our expressions of course apply equally well to any other distance indicators.

While this paper was in preparation, two preprints \cite{cooray} appeared in the electronic archive that
partially overlap with ours, specifically concerning lensing covariance as noise and signal. 
See also the preprint by \cite{dodelson} on SN lensing as a potentially useful signal. 
There is also a preprint by \cite{bdg} that discusses fluctuations of the luminosity distance in general
as a useful signal (see also \cite{schucker}).

\section{Preliminaries}
\label{prelim}

The relation between luminosity distance $d_L$ and apparent magnitude $m$ is:
\begin{equation}
\label{dLm}
m = 5 {\,\rm log}_{10} d_L + M = 5 {\,\rm ln\,} d_L / {\,\rm ln\,} 10 + M \sim 2.17 {\,\rm ln\,} d_L + M
\end{equation}
where $M$ is the magnitude zero-point. 
Note that it is customary to define the absolute magnitude as differing from our $M$
by some additive constant, whose precise value depends on the unit used for $d_L$.
If we rescale $d_L$ by some multiplicative factor, such as when we switch units or 
when we alter the Hubble constant today $H_0$ (e.g. if we express 
$d_L$ in Mpc/h), the change can be absorbed into
the definition of $M$. Ultimately, in determining cosmological parameters from $m(z)$,
the apparent magnitude as a function of redshift $z$, we 
would marginalize over the zero-point $M$, which means marginalizing over the absolute
magnitude and $H_0$ at the same time. 

What are the cosmological parameters of interest? Here, we are interested in $\Omega_{\rm de}$,
$w_{\rm pivot}$ and $w_a$, which are respectively the dark energy density today (normalized by critical density), 
the dark energy equation of state and its evolution. We use the parametrization proposed by
\cite{polarski,linder0}:
\begin{eqnarray}
\label{linderparm}
w(a) = w_{\rm pivot} + w_a (a_{\rm pivot} - a)
\end{eqnarray}
where $w(a)$ is the equation of state of dark energy at scale factor $a$, 
$w_{\rm pivot}$ is the equation of state at $a = a_{\rm pivot}$, and
$w_a$ is the negative slope $-dw/da$. The scale factor $a_{\rm pivot}$ is chosen such that the errorbars on 
$w_{\rm pivot}$ and $w_a$ are uncorrelated. The precise $a_{\rm pivot}$ therefore varies from experiment to experiment,
but is generally close to but slightly less than unity.
Throughout this paper, we assume a flat universe, so the matter density $\Omega_m$ is not an independent parameter.

To summarize, given a SN experiment yielding $m$ as a function of $z$, 
we can fit for four parameters: $w_{\rm pivot}$, $w_a$, $\Omega_{\rm de}$ and $M$. Let us
label them by $p_\alpha$ with $\alpha$ ranging from $1$ to $4$. 
We will mostly marginalize over $M$ since it is not of cosmological interests. 
Sometimes we marginalize over $\Omega_{\rm de}$ as well, usually with a prior. 
Exactly what prior, if any, is used will be stated explicited in each worked example below.

The Fisher matrix is defined by \cite{scottsbook}
\begin{equation}
F_{\alpha\beta} = \sum_{ij} {\partial m_i \over \partial p_\alpha} {\tilde C}^{-1}_{ij} {\partial m_j
\over \partial p_\beta}
\label{fisher}
\end{equation}
where we imagine that the SNe have been binned up in redshifts labeled by Latin indices: $m_i$ refers to
the averaged $m$ for SNe that fall within a redshift bin centered at $z = z_i$. 
Here $\alpha$ and $\beta$ range from $1$ to $4$, corresponding to the parameters
$w_{\rm pivot}$, $w_a$, $\Omega_{\rm de}$ and $M$.
The (binned) magnitude covariance matrix is
\begin{eqnarray}
\label{tildeCij}
\tilde C_{ij} \equiv \langle \delta m_i \delta m_j \rangle
\end{eqnarray}
All relevant errorbars related to the four parameters, marginalized or otherwise, with or without prior,
can be deduced from 
the Fisher matrix $F_{\alpha\beta}$. 
For instance, the (rms) errorbar on $w_{\rm pivot}$ marginalized over everything else with no prior
\cite{lie}
is
given by $\sqrt{[F^{-1}]_{11}}$. 
Further details are given in Appendix \ref{app:fisher}, especially on how to choose the pivot scale $a_{\rm pivot}$. 
One particularly useful fact:
as long as $a_{\rm pivot}$ is chosen such that the errors on $w_{\rm pivot}$ and $w_a$ are uncorrelated,
the error on $w_{\rm pivot}$ marginalized over $w_a$ is exactly the same as the error on $w$ setting $dw/da$ to
be some fixed value, say zero \cite{wayne}.
This is true even if there are other parameters present, such as $\Omega_{\rm de}$ and $M$.
To be precise: the error on $w_{\rm pivot}$ marginalized over $w_a$ and other parameters (let us call them $p_3$, $p_4$ ...) is
exactly the same as the error on a constant $w$ marginalized over $p_3$, $p_4$ ... (see Appendix \ref{app:fisher}).
This fact is useful for comparing our results with some of those in the literature which
often assume a constant $w$.

A common alternative parametrization, $w = w_0 + w'(z-z_{\rm pivot})$, would result in an errorbar for $w_0$ that is
similar to $w_{\rm pivot}$, and an errorbar for $w'$ that is typically about 
half of that for $w_a$.

Any error projections for surveys necessarily assume a fiducial model. We will assume
throughout a flat cosmological constant dominated model with matter density $\Omega_m = 0.27$,
dark energy density $\Omega_{\rm de} = 0.73$, $w_{\rm pivot} = -1$ and $w_a = 0$ \cite{wmap3}.

Our next task is to calculate the magnitude covariance matrix $\tilde C_{ij}$.

\section{The Magnitude Covariance Matrix -- Poissonian and Coherent Components}
\label{flux}

Let us recall that we have from observations a vector of numbers: the apparent
magnitude $m_i$ averaged over redshift bin $i$, with 
$z_i$ being the average redshift of that bin \cite{binning}.
As we show in Appendix \ref{app:poisson}, the magnitude covariance matrix is
\begin{eqnarray}
\label{tC}
\tilde C_{ij} \equiv \langle \delta m_i \delta m_j \rangle = 
\delta_{ij} {{({\sigma^{\rm intr.}})^2 + ({\sigma^{\rm Poiss.}_i})^2}\over N_i} + C_{ij}
\end{eqnarray}
where $N_i$ is the number of SNe in the $i$-th redshift bin,
$\sigma^{\rm intr.}$ is the intrinsic dispersion of SN magnitude which we will
take to be either $0.1$ or $0.15$ \cite{sigintr}, and $\sigma^{\rm Poiss.}_i$ is the Poissonian
dispersion induced by large scale structure for each SN (note that it depends on $i$, or
the redshift $z_i$, in general). The intrinsic scatter and the Poissonian large scale structure induced scatter
add in quadrature, and both scale with $1/N_i$ as expected i.e. they are both Poissonian in nature.
The symbol $C_{ij}$ quantifies the contribution to $\tilde C_{ij}$ from
correlated/coherent large scale structure fluctuations, and it does not scale inversely with the number of SNe.
In principle, one should also add to $\tilde C_{ij}$ a term that describes systematic errors.
We will discuss this briefly in \S \ref{discuss}, but will leave it out for most of our discussion.
We refer to $C_{ij}$ simply as the correlation matrix, to be distinguished from
$\tilde C_{ij}$ which we call the magnitude covariance matrix i.e. $C_{ij}$ is the non-Poissonian, or coherent,
part of $\tilde C_{ij}$.

The Poissonian large scale structure induced variance is simple to write down:
\begin{eqnarray}
\label{sigP}
&& ({\sigma^{\rm Poiss.}_i})^2 = 
\left[{5 \over {\,\rm ln\,} 10}\right]^2 \int {dz \over \Delta z_i} \langle \left[ \delta_{d_L} (z)\right]^2 \rangle \\ \nonumber
&& \sim
\left[{5 \over {\,\rm ln\,} 10}\right]^2 \langle \left[ \delta_{d_L} (z_i)\right]^2 \rangle
\end{eqnarray}
which follows from eq. (\ref{dLm}). Here, $\delta_{d_L} (z) \equiv \delta d_L/\bar d_L (z)$ is the fractional fluctuation in luminosity distance
due to departure from homogeneity (i.e. large scale structure) for an observed redshift $z$. 
Note that this is the fractional fluctuation at one point in the sky i.e. for one SN. The
ensemble average $\langle \,\, \rangle$ is
over realizations of the universe. The integration in the first line of eq. (\ref{sigP})
is over the redshift bin centered at $z_i$ with width $\Delta z_i$.
For a sufficiently narrow $\Delta z_i$, 
since $\langle \left[ \delta_{d_L} (z)\right]^2 \rangle$ typically varies slowly with $z$,
the second line is a good approximation.

The correlated/coherent component concerns $\delta_{d_L}$ in different parts of the sky:
\begin{eqnarray}
\label{Cij}
C_{ij} = \left[{5 \over {\,\rm ln\,} 10}\right]^2 \int {dz {d^2 \theta} dz' {d^2 \theta'} \over \Delta z_i A_i \Delta z_j A_j}
\langle \delta_{d_L} (z, \bm{\theta}) \delta_{d_L} (z', \bm{\theta'}) \rangle
%\left[{5 \over {\,\rm ln\,} 10}\right]^2 \int {{d^2 \theta} {d^2 \theta'} \over A_i A_j}
%\langle \delta_{d_L} (z_i, \theta) \delta_{d_L} (z_j, \theta') \rangle
\end{eqnarray}
where we have added an extra argument $\bm{\theta}$ or $\bm{\theta'}$ to $\delta_{d_L}$ to remind ourselves
that the fluctuation depends on the redshift as well as the angular position. The angular integration is done
over the area of the survey, and in general, the total area of the survey could be different at different redshifts, so
we allow the area $A$ to carry an index $i$ or $j$. The redshift integration is done over the respective redshift bins. 
In cases where $\langle \delta_{d_L} (z, \bm{\theta}) \delta_{d_L} (z', \bm{\theta'}) \rangle$ does not vary
rapidly over the respective redshift bins (such as in the lensing contributions to $\delta_{d_L}$), the 
expression above is well approximated by
$[{5 / {\,\rm ln\,} 10}]^2 (A_i A_j)^{-1} \int {{d^2 \theta} {d^2 \theta'}}
\langle \delta_{d_L} (z_i, \bm{\theta}) \delta_{d_L} (z_j, \bm{\theta'}) \rangle$.

It is crucial that the Poissonian variance
scales as $1/N_i$ whereas $C_{ij}$ does not (eq. [\ref{tC}]; this is justified in Appendix \ref{app:poisson}).
Naively one would expect $C_{ij}$ to be quite small especially if the area $A_i$ or $A_j$ is large. 
However, for future surveys where $N_i$ can be quite large, it is entirely possible for the Poissonian
term to be smaller than $C_{ij}$, as we will see.

It is also important to emphasize that $C_{ij}$, even when $i=j$, is non-Poissonian, in the sense that
$i=j$ only tells us that we are looking at the same redshift bin, but $\bm{\theta}$ (or $z$) can still of course differ from 
$\bm{\theta'}$ (or $z'$).
In other words, when we talk about correlated/coherent/non-Poissonian fluctuations, we mean fluctuations
that are correlated between different SNe, at different redshifts or different angular positions or both.

As far as we know, the importance of $C_{ij}$ has been overlooked in existing error forecasts for SN surveys.
Moreover, the only large scale structure contribution to the magnitude covariance matrix 
$\tilde C_{ij}$ (eq. [\ref{tC}]) that has been considered in the context
of SN surveys is that from lensing, through the Poissonian term $(\sigma_i^{\rm Poiss.})^2$
(but see \cite{cooray} for preprints that partially overlap with ours concerning correlated lensing fluctuations).

\section{The Luminosity Distance Fluctuation and its Contribution to the Magnitude Covariance Matrix}
\label{frw}

Our goal in this section is two-fold. First, we write down
an expression for $\delta_{d_L}$, the fractional luminosity distance fluctuation, that
is accurate to first order in perturbations. Second, we work out its
two-point correlation and second moment as they show up in the magnitude covariance matrix
(eq. [\ref{tC}], [\ref{sigP}] and [\ref{Cij}]).
The main results are eq. (\ref{ddL2}) - (\ref{Wvellarge}).
Readers not interested in details can skip ahead to \S \ref{forecasts}.

\subsection{The Luminosity Distance Fluctuation}
\label{dLf}

There exist a large literature on the 
luminosity distance in a weakly perturbed Friedmann-Robertson-Walker universe \cite{pfrw}. 
Sasaki \cite{sasaki} derived the general expressions using the optical scalar equations, and
gave an explicit integration in the case of an Einstein-de-Sitter universe.
More recently, Pyne and Birkinshaw \cite{pyne1,pyne2}, using a different technique, derived explicit expressions
for more general cases, including a non-flat universe as well as a universe with dark energy.
In Appendix \ref{app:dL}, we present a derivation closely following that of \cite{pyne1,pyne2}, but
somewhat simplified. The end result differs slightly from \cite{pyne2}, but only in terms
that are subdominant compared to the gravitational lensing and peculiar motion terms, and so this
difference has no impact on the main conclusions of this paper. The origin of the difference is explained
in Appendix \ref{app:dL}. Relegating the technical details to that Appendix, let us
focus here on a somewhat heuristic but more intuitive 
derivation of the dominant terms, those
arising from peculiar motion and lensing. 

Recall the following standard relation between the observed flux $F$ and the
intrinsic luminosity $L$, valid for a universe with or without inhomogeneities
\cite{peebles}:
\begin{eqnarray}
\label{peebles}
F(z) = {L \over 4\pi (1+z)^4} {\delta\Omega_0 \over \delta A_e}
\end{eqnarray}
where $\delta A_e$ is the proper area of the emitter,
$\delta \Omega_0$ is the solid angle at the observer subtended by light rays from the emitter,
and $z$ is the observed redshift.
The angular diameter and luminosity distances are respectively defined to be
\begin{eqnarray}
\label{dLdA}
d_A = \sqrt{\delta A_e / \delta\Omega_0} \quad , \quad d_L = d_A (1+z)^2
\end{eqnarray}
We emphasize again that these expressions are equally valid in a homogeneous or
an inhomogeneous universe.

In a homogeneous universe, the above equations take the form
\begin{eqnarray}
\label{peeblesHomogeneous}
&& \bar F (\bar z) = {L \over 4\pi (1+ \bar z)^4 [\bar d_A (\bar z)]^2} \\ \nonumber
&& \bar d_A (\bar z) = \chi_e /(1+\bar z) \\ \nonumber
&& \chi_e \equiv \chi (\bar z) = \int_0^{\bar z} dz'/H(z') \\ \nonumber
&& \bar d_L (\bar z) = \bar d_A (\bar z) (1+\bar z)^2
\end{eqnarray}
where the bar on top of quantities reminds us that they are defined in
a homogeneous universe. Here $\chi$ is the usual comoving distance in a homogeneous
universe, which is an integral over redshift of the inverse Hubble parameter $H$, and
$\chi_e$ is the comoving distance to the emitter. 

Suppose we perturb such a universe by introducing peculiar motion, excluding
for the moment other possible sources of fluctuations.
This has two effects. First, $\bar z$ is Doppler shifted to a new value $z$:
\begin{eqnarray}
\label{shiftshift}
1 + z = (1 + \bar z) (1 + {\bf v_e}\cdot {\bf n} - {\bf v_0} \cdot {\bf n})
\end{eqnarray}
where ${\bf v_e}$ and ${\bf v_0}$ are the peculiar velocities of the emitter and the observer
respectively, and ${\bf n}$ is the unit vector from observer to emitter.
This is accurate to first order in peculiar velocities.
Note that the speed of light is set to unity.

The second effect of introducing peculiar motion is to modify the angular diameter
distance. 
Peculiar motion modifies $\delta \Omega_0$ but not $\delta A_e$ (to first order) in eq. (\ref{dLdA}),
leading to
\begin{eqnarray}
\label{dAshift}
d_A (z) = \bar d_A  (\bar z) (1 + {\bf v_0} \cdot {\bf n})
\end{eqnarray}
Peculiar motion causes $\delta \Omega_0 \rightarrow \delta \Omega_0 (1 - 2 {\bf v_0} \cdot {\bf n})$,
which can be derived by performing boosts.

Making use of eq. (\ref{dLdA}), (\ref{shiftshift}) and (\ref{dAshift}), we therefore have \cite{diffway}
\begin{eqnarray}
\label{dLshift}
d_L (z) = \bar d_L (\bar z) (1 + 2{\bf v_e} \cdot {\bf n} - {\bf v_0} \cdot {\bf n})
\end{eqnarray}

%From SNlens/Code/hubble.ps, using hubble.plot
\Sfig{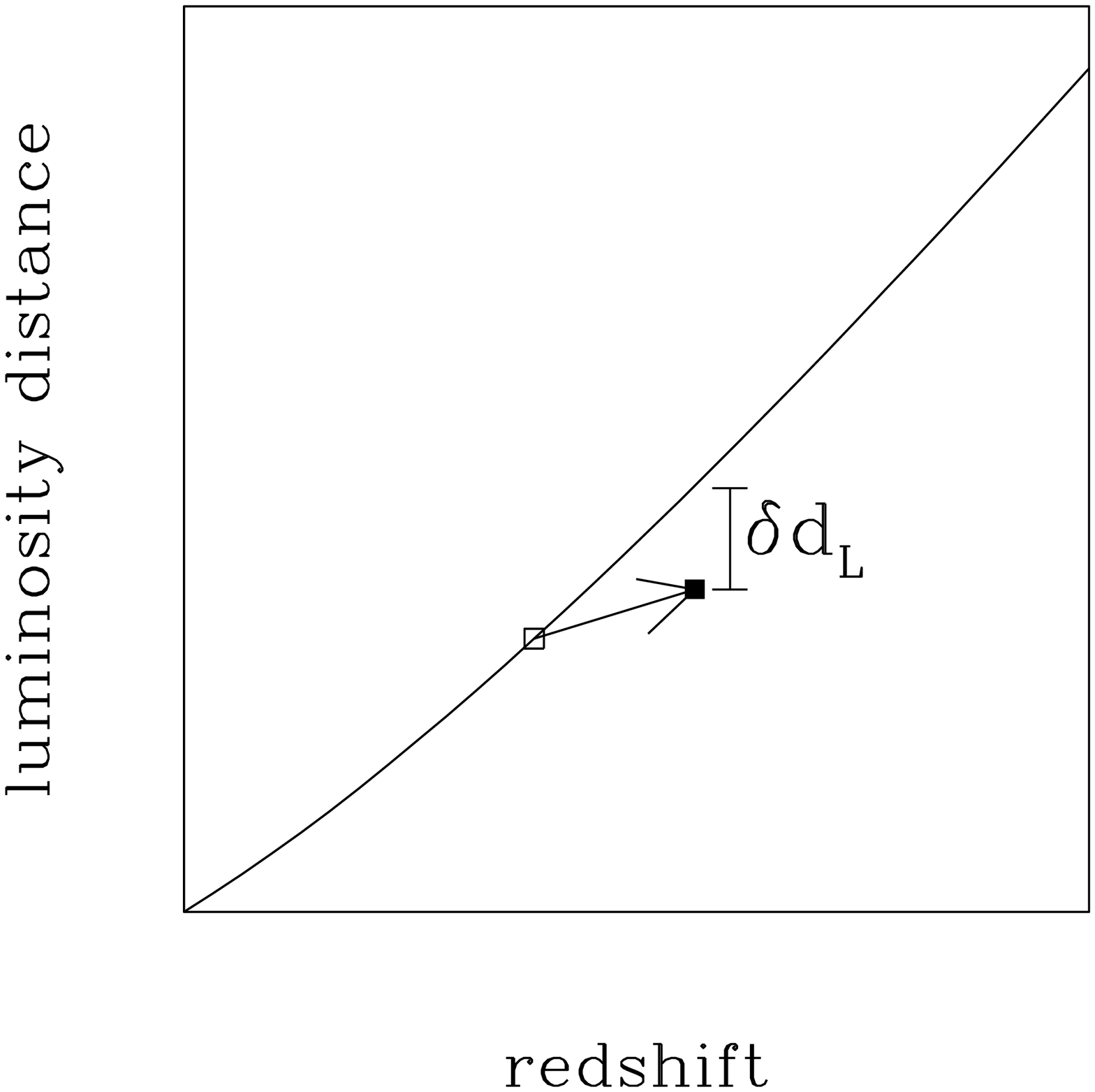}{A schematic Hubble diagram to illustrate the effects of peculiar motion.
The solid line represents the luminosity distance in a homogeneous universe.
Consider a SN represented by the open square in such a universe.
Suppose this SN is given a peculiar velocity moving away from the observer.
This SN would then be displaced to the position of the solid square -- note that both
the redshift and the luminosity distance are changed. The luminosity distance difference
that we are interested in is $\delta d_L (z) = d_L (z) - \bar d_L (z)$, where
$z$ is the actual observed redshift, $d_L (z)$ is the actual observed
luminosity distance, and $\bar d_L (z)$ is the luminosity distance
given by the solid line, at the {\it same} redshift $z$. The
fractional luminosity distance fluctuation that we study is
$\delta_{d_L} = \delta d_L(z)/\bar d_L(z)$.
}

It is important to keep in mind that 
the luminosity distance fluctuation that concerns us is
$\delta_{d_L} (z) = [d_L (z) - \bar d_L (z)]/ \bar d_L (z)$, where
$d_L$ and $\bar d_L$ are at the {\it same} redshift $z$.
Fig. \ref{fig:hubble} illustrates this point.

A Taylor series expansion of $\bar d_L$ tells us
\begin{eqnarray}
\label{dLtaylor}
\bar d_L (z) = \bar d_L (\bar z) \left(1 + [1 + {a_e \over a'_e \chi_e}]
[{\bf v_e} \cdot {\bf n} - {\bf v_0} \cdot {\bf n}] \right)
\end{eqnarray}
Note that the scale factor $a_e$, its derivative with respect to conformal time $a'_e$
and the comoving distance $\chi_e$ can be evaluated at either $\bar z$
or $z$: the changes in the expression above are of second order in the perturbations.

Eq. (\ref{dLshift}) and (\ref{dLtaylor}) together imply \cite{asymmetry} 
\begin{eqnarray}
\label{ddLvel}
\delta_{d_L} (z,{\bf n}) = 
{\bf v_e}\cdot {\bf n} - {a_e \over a'_e \chi_e} ({\bf v_e}\cdot {\bf n} - {\bf v_0} \cdot {\bf n})
\end{eqnarray}
This is accurate to first order in peculiar velocities, ignoring all other possible fluctuations.
We have added the argument ${\bf n}$ to $\delta_{d_L}$ to emphasize the fact that
$\delta_{d_L}$ depends on direction (or angle) in addition to redshift.

Gravitational lensing modifies the observed flux of an object without changing
its redshift. The gravitational lensing magnification $F^{\rm lensed}/F^{\rm unlensed}$
is worked out in many places e.g. \cite{kaiser,scottsbook}:
\begin{eqnarray}
\label{magnification}
{\rm magnification} = 1 + 2 \int_0^{\chi_e} d\chi {(\chi_e - \chi)\chi \over \chi_e} \nabla^2 \phi(\chi)
\end{eqnarray}
where $\phi$ is the gravitational potential fluctuation, and $\nabla^2$ is the Laplacian in comoving space.

The gravitational lensing contribution to $\delta_{d_L}$ is therefore
\begin{eqnarray}
\label{lensonly}
\delta_{d_L} = - \int_0^{\chi_e} d\chi {(\chi_e - \chi)\chi \over \chi_e} \nabla^2 \phi(\chi)
\end{eqnarray}

In summary, the total peculiar motion and lensing contributions to $\delta_{d_L}$ are
\begin{eqnarray}
\label{ddL2}
&& \delta_{d_L} (z, {\bf n}) = {\bf v_e} \cdot {\bf n} 
- {1\over \chi_e} \left[{a\over a'}\right]_e ({\bf v_e} \cdot {\bf n} - {\bf v_0} \cdot {\bf n}) \\ \nonumber
&& \quad - \int_0^{\chi_e} d\chi {(\chi_e - \chi)\chi \over \chi_e} \nabla^2 \phi(\chi)
\end{eqnarray}
To reiterate: 
${\bf v_e}$ and ${\bf v_0}$ are the peculiar velocities of the emitter and observer, and
${\bf n}$ is the line of sight unit vector pointing away from the observer
(${\bf n}$ here plays the role of $\bm{\theta}$ in eq. [\ref{Cij}]); the comoving distance to emitter $\chi_e$, the
scale factor at emission $a_e$ and its derivative with respect to conformal time
$a'_e$ are evaluated at redshift $z$.
One can see from above that for small $\chi_e$ or at a low redshift, the peculiar motion term
proportional to $1/\chi_e$ becomes
important, while at a large redshift, the lensing term (second line) is more important. 
A more rigorous derivation of $\delta_{d_L}$, together with an explanation of
why other first order contributions can be ignored, is given in Appendix \ref{app:dL}.

\subsection{From $\delta_{d_L}$ to the Magnitude Covariance Matrix}
\label{dLf2}

Our next task is to compute the second moment of $\delta_{d_L}$ for eq. (\ref{sigP}), and the
two-point correlation of $\delta_{d_L}$ for eq. (\ref{Cij}).
This is carried out in detail in Appendix \ref{app:poisson2}.
Let us summarize the results here.

%%%%%%%%%%%%
%{\bf Will erase later: notes 05 III supplement p. 1 to 4.}
%%%%%%%%%%%%

From eq. (\ref{sigP}) and eq. (\ref{ddL2}), it can be shown that
\begin{eqnarray}
\label{sigPfull0}
({\sigma_i^{\rm Poiss.}})^2 = ({\sigma_i^{\rm Poiss.,\, lens}})^2 + ({\sigma_i^{\rm Poiss.,\, vel.}})^2
\end{eqnarray}
\begin{eqnarray}
\label{sigPfull}
&& ({\sigma_i^{\rm Poiss.,\, lens}})^2 \equiv 
\left[{5 \over {\,\rm ln \,} 10}\right]^2 \left[{3 H_0^2 \Omega_m \over 2}\right]^2 \times \\ \nonumber
&& \quad \quad 
\int_0^{\chi_i} {d\chi \over a^2} 
\left[{(\chi_i - \chi)\chi \over \chi_i}\right]^2 \int {d^2 k_\perp \over (2\pi)^2} P(k_\perp, a)
\\ \nonumber 
&& ({\sigma_i^{\rm Poiss.,\, vel.}})^2 \equiv
\left[{5 \over {\,\rm ln \,} 10}\right]^2 \times \\ \nonumber
&& \quad \quad \left[1 - {a_i \over a'_i \chi_i}\right]^2 
(D'_i)^2
\int {d^3 k \over (2\pi)^3} {k_z^2 \over k^4} P(k, a=1)
\end{eqnarray}
where $P(k,a)$ and $P(k_\perp,a)$ represent the 
mass power spectrum at scale factor $a$ and 
at wavenumber $k$ and $k=k_\perp$ respectively,
$k_z$ and $k_\perp$ denote the line-of-sight and transverse components of
the wave vector ($k_\perp$ is the norm of the 2D vector ${\bf k_\perp}$), 
$D$ is the linear growth factor and $D'$ is its derivative
with respect to conformal time, and any quantity with the subscript $i$ is evaluated at $z=z_i$. 
The symbol $(\sigma_i^{\rm Poiss.\,,lens.})^2$ stands for
the Poissonian lensing term that is often studied: it gives the variance in convergence, up to a factor of $[5/{\,\rm ln\,}10]^2$.
The term 
$(\sigma_i^{\rm Poiss.\,,vel.})^2$
gives the variance in luminosity distance due to peculiar motion, up to the same factor.
Note that there is no cross velocity-lensing term -- it vanishes because the lensing
projection forces $k_z = 0$. 

Likewise, from eq. (\ref{Cij}) and eq. (\ref{ddL2}), using the plane-parallel approximation
(more discussion below), we find
\begin{eqnarray}
\label{Cijfull}
&& C_{ij} = C_{ij}^{\rm lens} + C_{ij}^{\rm vel.} \\ \nonumber
&& C_{ij}^{\rm lens} \equiv \left[{5 \over {\rm ln} 10} 
{3 H_0^2 \Omega_m \over 2}\right]^2 \int_0^{{\rm min}(\chi_i, \chi_j)}
{d\chi \over a^2} \left[{(\chi_i - \chi)\chi \over \chi_i}\right] \\ \nonumber
&& \quad \quad \left[{(\chi_j - \chi)\chi \over \chi_j}\right] 
\int {d^2 k_\perp \over (2\pi)^2} P(k_\perp, a) W^{\rm lens}_{ij} ({\bf k_\perp}, \chi) \\ \nonumber
&& W^{\rm lens}_{ij} ({\bf k_\perp}, \chi) \equiv 
\int {d^2 \theta d^2 \theta' \over A_i A_j} e^{-i{\bf k_\perp} \cdot \chi ({\bm{\theta}} - {\bm{\theta'}})}
\\ \nonumber 
&& \quad \quad =
{2 J_1(k_\perp \chi \theta_i^{\rm max.}) \over k_\perp \chi \theta_i^{\rm max.}}
{2 J_1(k_\perp \chi \theta_j^{\rm max.}) \over k_\perp \chi \theta_j^{\rm max.}} \\ \nonumber
&& C_{ij}^{\rm vel.} \equiv 
\left[{5 \over {\,\rm ln \,} 10}\right]^2
\left[1 - {a_i \over a'_i\chi_i} \right] 
\left[1 - {a_j \over a'_j \chi_j}\right] D'_i D'_j \\ \nonumber
&& \quad \quad \int {d^3 k \over (2\pi)^3} {k_z^2 \over k^4} P(k, a=1) W^{\rm vel.}_{ij} ({\bf k}) \\ \nonumber
&& W^{\rm vel.}_{ij} ({\bf k}) \equiv \int
{d^3 x d^3 x' \over V_i V_j} e^{-i{\bf k} \cdot ({\bf x} - {\bf x'})} \\ \nonumber 
&&
= {\,\rm \, cos\,}[k_z(\chi_i - \chi_j)] 
\left[ {2\over k_z \Delta \chi_i}{\,\rm sin\,} {k_z \Delta\chi_i \over 2}\right] \\ \nonumber
&& \,\,
\left[ {2\over k_z \Delta \chi_j}{\,\rm sin\,} {k_z \Delta\chi_j \over 2}\right]
{2 J_1(k_\perp \chi_i \theta_i^{\rm max.}) \over k_\perp \chi_i \theta_i^{\rm max.}}
{2 J_1(k_\perp \chi_j \theta_j^{\rm max.}) \over k_\perp \chi_j \theta_j^{\rm max.}}
\end{eqnarray}
where $C_{ij}^{\rm lens}$ and $C_{ij}^{\rm vel.}$ are the lensing and peculiar motion
contributions to the correlation matrix $C_{ij}$.
The window function for the lensing term $W_{ij}^{\rm lens}$ depends on the survey geometry:
the integration of $\bm{\theta}$ and $\bm{\theta'}$ is over the survey area, which is allowed to depend on
the redshift bin $i$ or $j$ for the sake of generality. The expression given for $W_{ij}^{\rm lens}$
in terms of the Bessel function $J_1$ assumes the survey spans a circular patch
on the sky, with
angular radius $\theta_i^{\rm max.}$ or $\theta_j^{\rm max.}$ (i.e. $A_i = \pi (\theta_i^{\rm max.})^2$,
$A_j = \pi (\theta_j^{\rm max.})^2$).
The window function for the peculiar motion term $W_{ij}^{\rm vel.}$ 
depends upon an integration over the comoving volumes of the redshift bins: $V_i$ and $V_j$.
The expression given for $W_{ij}^{\rm vel.}$ in terms of the Bessel function
and trigonometric functions assumes also that the survey spans a circular patch 
in the sky (see Appendix \ref{app:poisson2} for expressions appropriate for more general geometries).
Here $\Delta \chi_i$ and $\Delta \chi_j$ are the widths of the respective redshift bins in comoving distance.
Note also that $W_{ij}^{\rm vel.}$ strictly speaking should have an imaginary part, which
is odd in $k_z$ and so integrates to zero.
For more complicated survey geometries, $W_{ij}^{\rm lens}$ and $W_{ij}^{\rm vel.}$ can always
be worked out from the definitions above.

The expressions in eq. (\ref{Cijfull}) assume small angles or the plane parallel approximation.
For the lensing term, this is acceptable since, as we will see, lensing is generally important
only at high redshifts ($z \gsim 1$) where practical surveys of the future cover a sufficiently small area.
However, for the velocity term, which is generally important at low redshifts where an ongoing survey
such as the SNfactory covers a large area (almost half of the sky), it is useful to have
an expression that does not assume the plane parallel approximation:
\begin{eqnarray}
\label{cijvellarge}
&& C_{ij}^{\rm vel.} \equiv \left[{5 \over {\,\rm ln\,} 10}\right]^2
\left[1 - {a_i \over a'_i\chi_i} \right] 
\left[1 - {a_j \over a'_j \chi_j}\right] D'_i D'_j \\ \nonumber
&& \quad \quad \int {d^3 k \over (2\pi)^3} {1\over k^2} P(k, a=1) W^{\rm vel.}_{ij} ({\bf k}) \\ \nonumber
&& W^{\rm vel.}_{ij} ({\bf k}) \equiv \int
{d^3 x d^3 x' \over V_i V_j} ({\bf \hat k} \cdot {\bf n})
({\bf \hat k} \cdot {\bf n'})
e^{-i{\bf k} \cdot ({\bf x} - {\bf x'})} 
\end{eqnarray}
where ${\bf n}$ and ${\bf n'}$ are unit vectors pointing in the direction
${\bf x}$ and ${\bf x'}$, and ${\bf \hat k}$ is the same for ${\bf k}$.
Since $P(k,a=1)/k^2$ does not depend on direction, one can replace $W^{\rm vel.}_{ij} ({\bf k})$ by
its average over the solid angle of ${\bf k}$:

%%%%%%%%%%%
%{\bf Will erase later: notes 05 III p. 86}
%%%%%%%%%%%

\begin{eqnarray}
\label{Wvellarge}
&& W^{\rm vel.}_{ij} (k) = 
\sum_{\ell=0}^\infty (2\ell + 1)  
\left[ \int_{\chi_i - \Delta \chi_i/2}^{\chi_i + \Delta \chi_i} {d\chi\over \Delta\chi_i} j'_\ell (k\chi)\right]
\\ \nonumber
&& \quad 
\left[ \int_{\chi_j - \Delta \chi_j/2}^{\chi_j + \Delta \chi_j} {d\chi\over \Delta\chi_j} j'_\ell (k\chi)\right]
\left[ \int_0^{\rm \theta^{\rm max.}_i} {d\theta {\,\rm sin}\theta P_\ell ({\,\rm cos}\theta) 
\over 1 - {\,\rm cos}\theta^{\rm max.}_i}
\right]
\\ \nonumber
&& \quad \left[ \int_0^{\rm \theta^{\rm max.}_j} {d\theta' {\,\rm sin}\theta' P_\ell ({\,\rm cos}\theta') 
\over 1 - {\,\rm cos}\theta^{\rm max.}_j}
\right]
\end{eqnarray}
where $j_\ell$ is the spherical Bessel function, and $j'_\ell$ is its derivative with
respect to the argument, and $P_\ell$ is the Legendre polynomial.
Note that the survey area $A_i = 2\pi
(1 - {\,\rm cos}\theta^{\rm max.}_i)$ ($\sim \pi (\theta^{\rm max.}_i)^2$ only for
small $\theta^{\rm max.}_i$). 

Eq. (\ref{Wvellarge}) assumes a contiguous and circularly symmetric survey geometry.
Expressions for a more general geometry are given in Appendix \ref{app:poisson2} (eq. [\ref{WvellargeSNf}] \&
[\ref{WvellargeGeneral}]).

{\it Throughout this paper, all results presented come from using the exact expression
for the velocity term (eq. [\ref{cijvellarge}]) rather than the small angle approximation
for $C_{ij}^{\rm vel.}$
(eq. [\ref{Cijfull}]).}
It is an interesting question how well the small angle expression 
for the velocity term pproximates the exact expression. 
The answer is: surprising well even for surveys that cover
large portions of the sky. This is discussed in the context of a concrete example in \S \ref{vel}.

Note that if $\theta^{\rm max.}_i$ or $\theta^{\rm max}_j$ equals $\pi$ (i.e. an all sky survey), 
only the monopole $\ell = 0$ term in eq. (\ref{Wvellarge}) would be non-zero. More generally, if the SN survey
covers large portions of the sky, only the low multipoles would be significant.

%%%%%%%%%%%%%%%%%%%%
%{\bf Will erase later: for the effect of whether to put the slow functions inside
%the $\chi$ integrals, see Code/guide on discussion of TestCvel2againc, TestCvel2againd, TestCvel2againd2.
%Bottom line is that it makes little difference, probably changes final rms error on $M$ at the level of
%a few percent. Or the degradation by velocity is good to the first decimal place.}
%%%%%%%%%%%%%%%%%%%%

Eq. (\ref{sigPfull0}), (\ref{sigPfull}), (\ref{Cijfull}), (\ref{cijvellarge}) and (\ref{Wvellarge}) 
are the main results of this section. 
Their derivation from eq. (\ref{ddL2}) is given in Appendix \ref{app:poisson2}.
They are to be substituted into eq. (\ref{tC})
to obtain the magnitude covariance matrix.

One subtlety: eq. (\ref{sigPfull0}) - (\ref{cijvellarge})
assume the ${\bf v_0}$ term in eq. (\ref{ddL2}) can be removed, because the observer's peculiar motion is
fairly well-known from the dipole of the microwave background \cite{ned}.
However, when the SN is sufficiently close to the observer, care must be taken to take into account the
correlation between ${\bf v_e}$ and ${\bf v_0}$ in eq. (\ref{ddL2}). This is discussed in Appendix \ref{app:poisson2}.
For realistic SN surveys, we have checked that the correlation makes a negligible difference.
An alternative error estimate is to allow ${\bf v_0}$ to be
a random variable and include its contribution to the variance of the magnitude fluctuations -- doing
so would increase the errorbar.

It is worth emphasizing that the above expressions are derived using linear perturbation theory. 
For instance, the velocity terms in eq. (\ref{sigPfull0}) - (\ref{cijvellarge}) assume
linear perturbation growth. This is acceptable for $C_{ij}^{\rm vel.}$ 
(eq. [\ref{Cijfull}] and [\ref{cijvellarge}]) since
the velocity integrals are dominated by large scale modes for relevant survey areas.
This is more questionable for the Poissonian velocity contribution $(\sigma_i^{\rm Poiss.,\, vel.})^2$
(eq. [\ref{sigPfull0}] and [\ref{sigPfull}]), 
since this depends on the one-point (small scale) velocity dispersion. 
However, as we will see in \S \ref{vel}, the velocity contribution to
the Poissonian term is relatively unimportant compared to the intrinsic contribution, and 
it is not necessary
to get it exactly right. To be precise: we will use the linear power spectrum for the velocity
terms (Poissonian and otherwise) throughout this paper.
We use the transfer function from \cite{ehu}, assuming a scalar spectral index of $n_S = 0.95$,
a Hubbble constant of $h=0.7$, a matter density of $\Omega_{\rm m}=0.27$, 
a baryon density of $\Omega_{\rm b}=0.046$, and a normalization of $\sigma_8 = 0.8$ \cite{wmap3}.
(Our fiducial cosmology is always a flat cosmological constant dominated universe
with $\Omega_{\rm de} = 0.73$; see the end of \S \ref{prelim}.)

For the lensing integrals in eq. (\ref{sigPfull0}) - (\ref{Cijfull}), it is not uncommon
to use the nonlinear mass power spectrum in the integrands \cite{bhuv}, 
and assume the results are a good approximation even on small angular scales (a rigorous justification
can be found in \cite{scott}).
Whether one uses the linear or nonlinear power spectrum does not matter
much for $C_{ij}^{\rm lens.}$ (eq. [\ref{Cijfull}]) which is dominated by large scale modes
anyway for realistic survey areas. For the Poissonian lensing term $(\sigma_i^{\rm Poiss.,\, lens})^2$ 
(eq. [\ref{sigPfull0}] and [\ref{sigPfull}]), the nonlinear mass power spectrum would provide
a more accurate estimate. Moreover,
it is possible that the result is further enhanced (beyond
that predicted by the standard nonlinear mass power spectrum)
by the adiabatic contraction of halos due to the cooling of 
baryons, presence of MACHOs, etc. This will be further discussed in \S \ref{lens}.
Throughout this paper, the lensing integrals are always done using the {\it nonlinear} power spectrum
prescribed by \cite{smith} (with suitable enhancement for the Poissonian term to account
for plausible effects of the baryons; see below).

%%%%%%%%%%%%
%{\bf Will erase later: notes 05 III p. 76 contains a summary of everything.
%Lack of cross-term shown in notes 05 III p. 32.}
%%%%%%%%%%%%

\section{Putting Everything Together -- from Luminosity Distance Fluctuations to
Error Forecasts}
\label{forecasts}

Our goal in this section is to make error forecasts by putting together the results from the last 3 sections, 
encapsulated in eq. (\ref{fisher}) for the Fisher matrix, eq. (\ref{tC}) for the magnitude
covariance matrix , and eq. (\ref{sigPfull0}) - (\ref{Wvellarge}) for explicit expressions for
components of the magnitude covariance matrix in terms of the power spectrum (with a small
modification to eq. [\ref{sigPfull0}]; see below).
It will prove illuminating to first study separately
the velocity and lensing contributions to the covariance matrix, and their effects
on the parameter errors.
This is done in \S \ref{vel} and \ref{lens}. 
In \S \ref{everything}, we make error forecasts for several ongoing/planned/proposed SN surveys,
and discuss the relative importance of the velocity- and lensing-induced luminosity distance
fluctuations.

\subsection{Peculiar Motion}
\label{vel}

%From SNlens/Code/TestCvel1/testcvelBspline.ps
%Original was from SNlens/Code/TestCvel1/testcvelB.ps
\Sfig{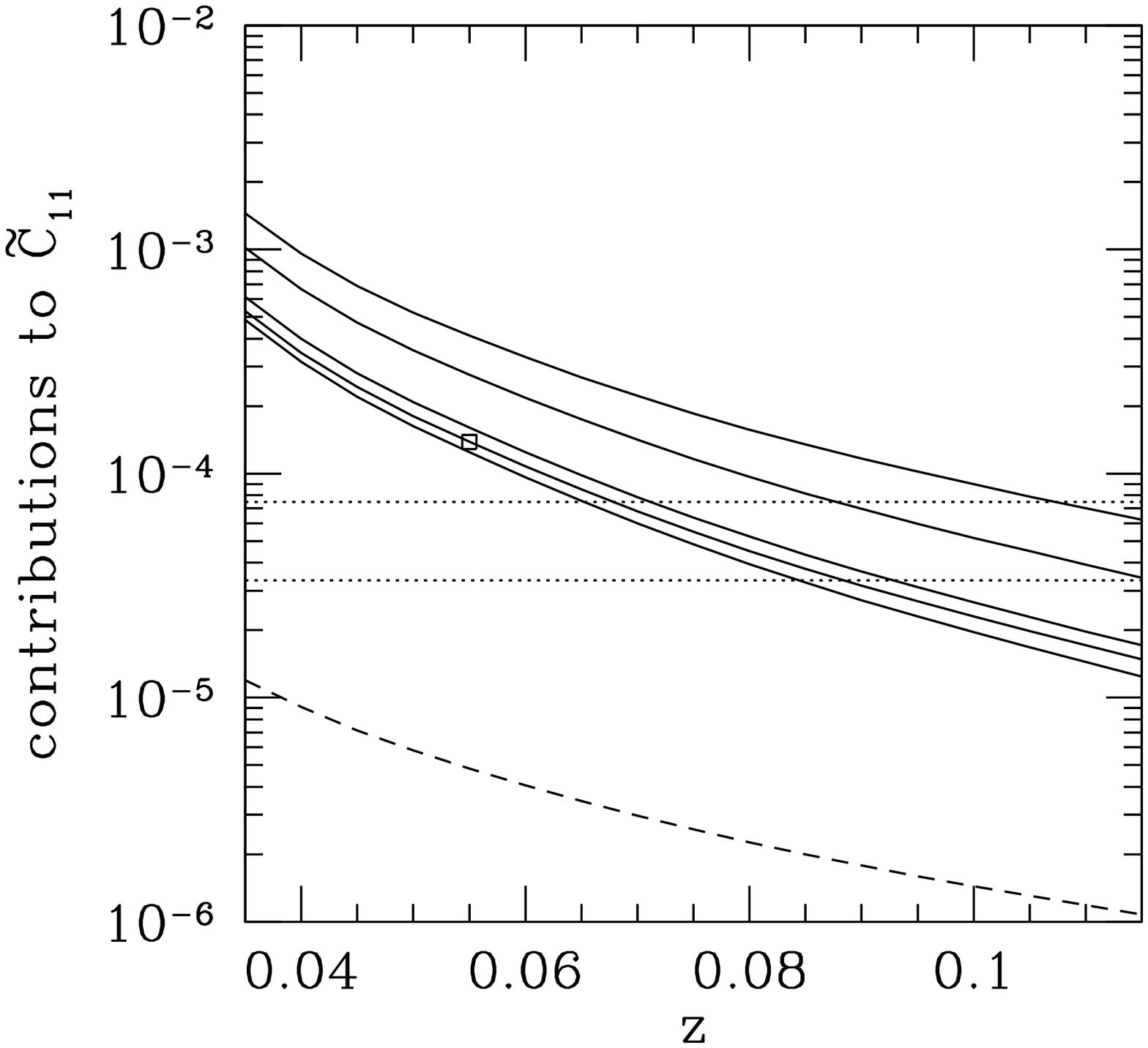}{Various contributions to the magnitude covariance matrix $\tilde C_{11}$ (eq. [\ref{tC}]) for a low
redshift SN survey characterized by one redshift bin with $\Delta z = 0.05$.
The five solid lines show the coherent/correlated velocity term $C_{11}^{\rm vel.}$ (eq. [\ref{Cijfull}])
as a function of the mean redshift $z$, for a survey area of, from top to bottom,
$1000$, $5000$, $20000$, $20000$b and $41000$ (full sky) square degrees. A contiguous circularly symmetric
geometry is assumed, except for the case of $20000$b, which has the $20000$ square degrees split
into two patches, one centered at the north pole and the other south (i.e. a galactic cut of about $\pm 30^0$).
Note that the SNfactory (SNf) 
has a mean $z$ of $0.055$ ($z = 0.03 - 0.08$), and a total area of about $20000$ sq. degrees
(denoted by the open square).
The dashed line shows the Poissonian velocity contribution (eq. [\ref{sigPfull}]) to the
magnitude covariance matrix: $(\sigma_1^{\rm Poiss.,\, vel.})^2$ divided by $N=300$ SNe; 
note that it is independent of 
the survey area, but depends on the mean $z$.
The dotted lines show the contribution from the intrinsic magnitude scatter: 
$(\sigma_1^{\rm intr.})^2 /N$, where $N=300$, and $\sigma_1^{\rm intr.}$ equals $0.1$ for the lower line
and $0.15$ for the upper line. Lensing contributions to $\tilde C_{11}$ are negligible at these redshifts.
}

%%%%%%%%%%%%
%{\bf Will erase later: for how binning changes the results. See discussion in 
%Code/guide on TestM1again vs TestM1again2.}
%%%%%%%%%%%%

Let us consider a survey with only one redshift bin \cite{bin} with a width of $\Delta z = 0.05$.
We are interested in the velocity contributions to the Poissonian and non-Poissonian
terms in the magnitude covariance matrix (eq. [\ref{tC}]; with one redshift bin, 
the covariance matrix is just a number $\tilde C_{11}$).
Note that at the low redshifts we consider in this section, the lensing contributions are negligible.
The coherent/non-Poissonian velocity term $C_{11}^{\rm vel.}$ (eq. [\ref{Cijfull}]) is plotted as solid lines in 
Fig. \ref{fig:testcvelB}. It is shown as a function of the mean redshift of the survey.
In other words, $\Delta z$ is fixed at $0.05$, but the central redshift is allowed to vary, and
$C_{11}^{\rm vel.}$ is shown as a function of that central redshift. The various solid lines
span from top to bottom a survey area of $1000$ to $41000$ square degrees
\cite{area}. A contiguous circuliarly symmetric geometry is assumed, except for the second solid line
from the bottom: it consists of two circular patches one centered at the north and the other south,
with a total area of $20000$ square degrees, in other words, it has a galactic cut of about $\pm 30^0$.

For comparison, we show as a dashed line the Poissonian velocity term
$(\sigma_1^{\rm Poiss.,\, vel.})^2 /N$ (eq. [\ref{sigPfull0}] and [\ref{sigPfull}]), with $N = 300$ SNe.
Note that this term is independent of the survey area.
We also show as dotted lines the intrinsic Poissonian term 
$(\sigma_1^{\rm intr.})^2/N$, for $\sigma_1^{\rm intr.} = 0.1$ (lower line) and
$\sigma_1^{\rm intr.} = 0.15$ (upper line). 

The above numbers are chosen for a reason: the Nearby Supernova Factory (SNfactory) is a survey
of about $300$ SNe that covers roughly half of the sky ($\sim 20000$ square degrees), and
is centered around $z \sim 0.055$ spanning a width of $\Delta z \sim 0.05$.
(Other similar surveys include the Carnegie Supernova Project and LOTOSS, see \cite{SNsurveys}.)
To be precise, henceforth, whenever we discuss the SNfactory, we assume a geometry that coincides
with the next to bottom solid line of Fig. \ref{fig:testcvelB}. (The moral of the bottom three solid
lines is that neither the precise area nor the precise geometry matters much, as long as the survey
covers a significant fraction of the sky {\it and} does not have many holes or edges.)
One can see that for these parameters, the coherent fluctuation term $C_{11}^{\rm vel.}$ is larger than the Poissonian (intrinsic and velocity)
terms, and becomes even more important if the survey were done at a lower redshift.
The Poissonian velocity term (dashed line), on the other hand, 
is always subdominant compared to the intrinsic term \cite{linearVSnl} -- this is probably the
reason for the common
perception that peculiar motion can be ignored as part of the
error budget for SN surveys. {\it This perception is incorrect because it ignores the coherent velocity
fluctuations quantified by $C_{11}^{\rm vel.}$ (solid lines).}

%%%%%%%%%
%{\bf Will erase later: for discussions on how linear and nonlinear power makes a difference, see discussion in
%Code/guide on TestCvel2again vs TestCvel2NL, as well as footnote \cite{linearVSnl}.}
%%%%%%%%%

It is useful to understand qualitatively why the different contributions to $\tilde C_{11}$ depicted in 
Fig. \ref{fig:testcvelB} take the values they do.
The contribution from the Poissonian intrinsic scatter is the simplest: 
$0.1^2/300$ or $0.15^2/300$ giving
$3.3 \times 10^{-5}$ or $7.5 \times 10^{-5}$ (dotted lines).
The contribution from the Poissonian velocity term is also easy to understand.
At low redshifts, the term $\sigma_1^{\rm Poiss.,\, vel.}$ (eq. [\ref{sigPfull}])
is roughly $2.17 \times v/(c z)$ where $v$ is the typical peculiar velocity 
($\sim 300$ km/s) and $c z$ is
the Hubble flow. For instance, at $z = 0.055$, this amounts to 
$\sigma_1^{\rm Poiss.,\, vel.} \sim 0.04$, and therefore
$(\sigma_1^{\rm Poiss.,\, vel.})^2/300 \sim 5 \times 10^{-6}$ (dashed line). 
For the non-Poissonian velocity term $C^{\rm vel.}_{11}$ (eq. [\ref{cijvellarge}]),
let us focus on the case corresponding to the SNfactory, with a 
total area of $20000$ square degrees and a mean redshift of $z=0.055$
(the lowest black line). The large survey area means that the window function 
$W^{\rm vel.}_{11} (k)$ is dominated by the low order multipoles (eq. [\ref{Wvellarge}]).
Let us consider the monopole $\ell = 0$,
which picks out $k \sim 0.005$ h/Mpc corresponding to a mean distance of
$\chi \sim \Delta \chi \sim 200$ Mpc/h. 
The integral over power spectrum (second line of
eq. [\ref{cijvellarge}]) can therefore be approximated by
$4\pi k P(k)/(2\pi)^3$ evaluated at $k \sim 0.005$ h/Mpc, giving
roughly $4 \,({\rm Mpc/h})^2$. The prefactors in the first line of eq. (\ref{cijvellarge})
equal $\sim (2.17)^2 \times (1/0.055)^2 \times (0.5/3000)^2 ({\,\rm h/Mpc})^2$,
where we have made use of the fact that $D'$ is roughly half the inverse Hubble radius 
$\sim 0.5/(3000 {\,\rm  Mpc/h})$ (recall that the speed of light is set to one).
Putting all these together yields $C^{\rm vel.}_{11} \sim 2 \times 10^{-4}$.

%%%%%%%%%%%%%%
%{\bf Will erase later: for a more detailed discussion of the above points. See
%Code/guide on TestCvel2againb.}
%%%%%%%%%%%%%%

A low redshift survey such as the SNfactory provides an important anchor for 
surveys at higher redshifts in that it helps determine the zero-point
$M$ (eq. [\ref{dLm}]). As we will see, 
combining high redshift SN surveys with a low redshift survey such as the SNfactory
often reduces the error on the equation of state of dark energy by a factor of about $2$.
It is therefore important to ask: to what extent does peculiar motion, particularly
coherent peculiar motion, increase the projected error on $M$ from a survey like the SNfactory?

%From SNlens/Code/TestM1/testMspline.ps
%Old version from SNlens/Code/TestM1/testM.ps
\Sfig{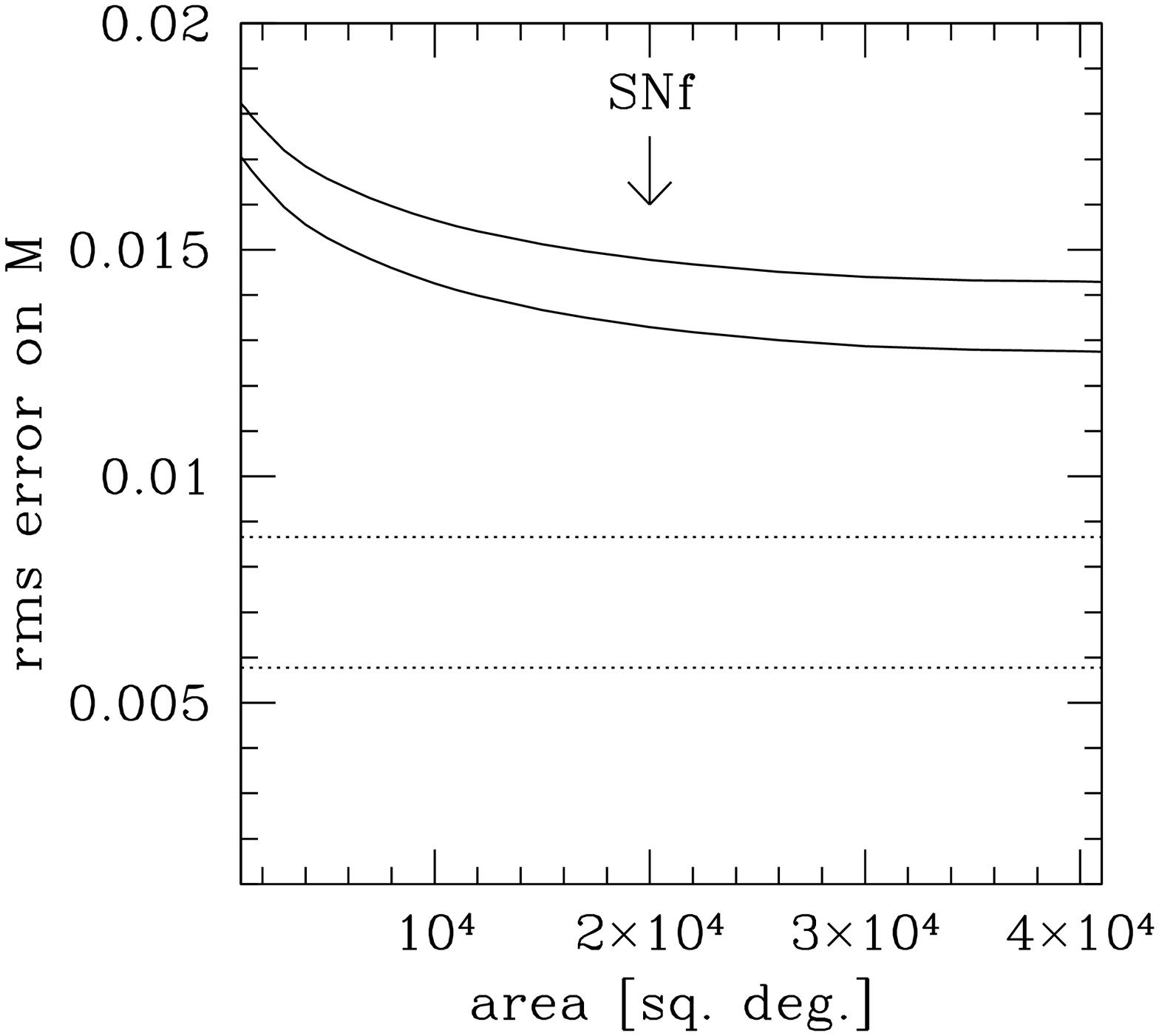}{The zero-point ($M$ in eq. [\ref{dLm}]) rms error
as a function of survey area (keeping all other parameters fixed). 
The survey redshift coverage is fixed: $z = 0.03 - 0.08$,
and the number of SNe is 300. The upper pair of solid lines allow for the effects of
peculiar motion, while the lower pair of dotted lines do not.
Within each pair, the upper line uses an intrinsic scatter of 
$\sigma^{\rm intr.} = 0.15$ and the lower one uses $\sigma^{\rm intr.} = 0.1$.
Note that the SNfactory (SNf) covers half of the sky, which is about $20000$ square degrees.
}

%From SNlens/Code/TestCvel1/testcveldwBspline.ps 
%Originally from SNlens/Code/TestCvel1/testcveldwB.ps
%Other related figures are testcveldwpBspline.ps and testcvelDESdwBspline.ps (see Code/guide for descriptions).
\Sfig{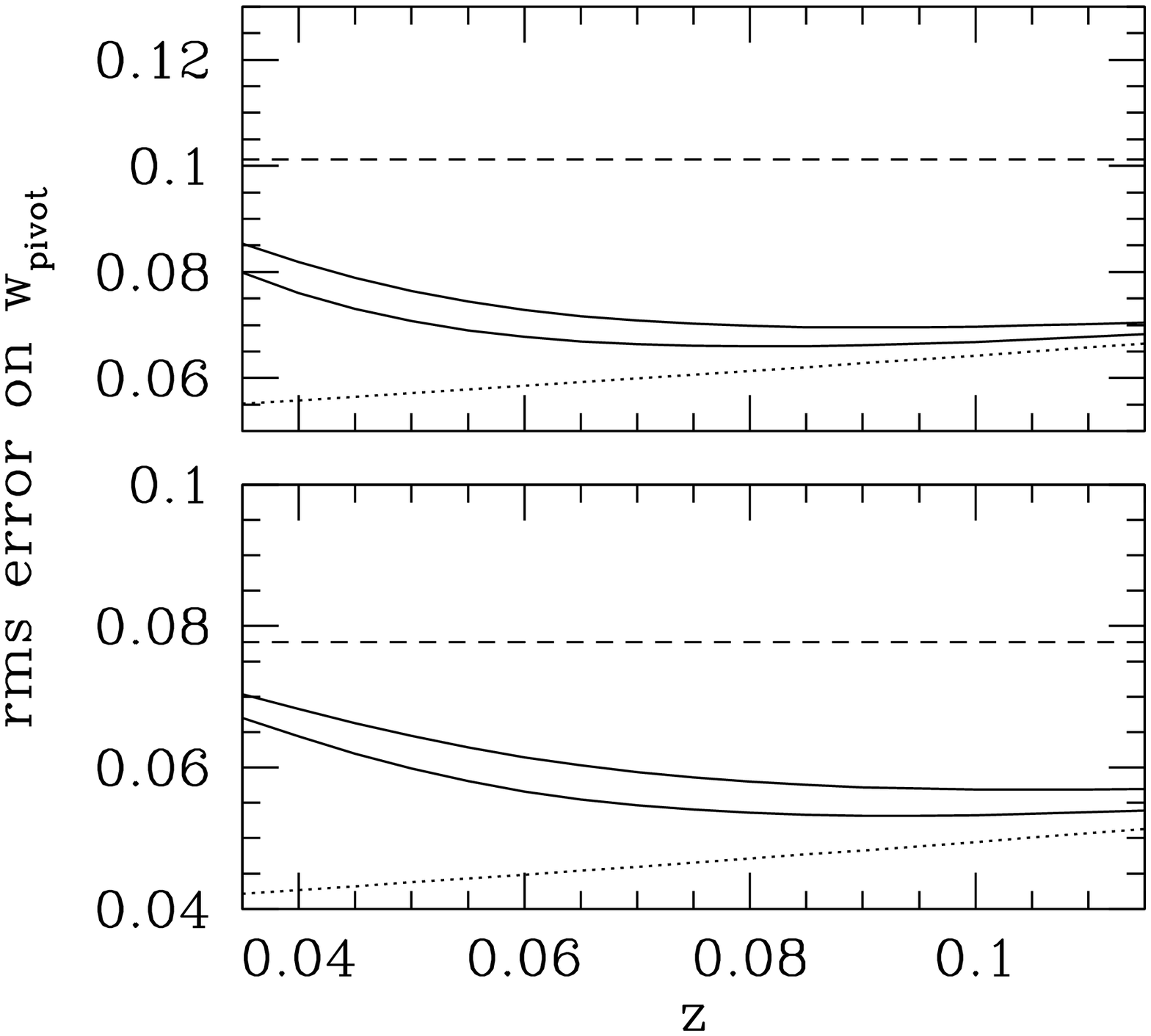}{Marginalized error on $w_{\rm pivot}$ as a function of the mean redshift
of a low $z$ survey of $300$ SNe, with $\Delta z = 0.05$.
The upper panel assumes an intrinsic magnitude dispersion of $\sigma^{\rm intr.} = 0.15$, and the lower panel uses $\sigma^{\rm intr.} = 0.1$. 
Here, it is assumed the low $z$ survey is combined with
a high $z$ SNAP-like survey whose parameters are kept fixed (see Table \ref{tabsurveys}).
The solid lines show the error on $w_{\rm pivot}$ when peculiar motion induced
fluctuations (dominated by the coherent ones) are taken into account: the upper solid line
is for a low $z$ survey area of $1000$ square degrees and the lower one is 
for $20000$ square degrees. The dotted line shows the same when peculiar motion is ignored,
hence the result is independent of survey area. For comparison, the horizontal dashed line
shows the error on $w_{\rm pivot}$ if one uses only the data from the high $z$ SNAP-like survey.
Gravitational lensing is taken into account in all cases above.
A prior of $\delta \Omega_{\rm de} = 0.03$ (flat universe) is assumed.
}

Fig. \ref{fig:testM} provides the answer.
The dotted lines show the errorbar on $M$ (keeping all other parameters fixed \cite{fixed}) from
a survey of 300 SNe that spans $z = 0.03 - 0.08$, {\it ignoring peculiar motion} i.e. only 
the intrinsic magnitude scatter is taken into account: the upper dotted line is for an intrinsic
scatter of $\sigma^{\rm intr.} = 0.15$, and the lower dotted line is for $\sigma^{\rm intr.} = 0.1$.
With only the intrinsic scatter taken into account, the error on $M$ is independent of survey area.
The solid lines show the same, except this time including peculiar motion induced fluctuations.
As before, the upper line of the pair uses $\sigma^{\rm intr.} = 0.15$ and lower line uses $\sigma^{\rm intr.} = 0.1$.
(At these redshifts, other sources of large scale structure fluctuations such as lensing 
are negligible.) For a survey like the SNfactory ($\sim 20000$ square degrees), one can see that
peculiar motion increases 
the error on $M$ by about a factor of $2$, depending on the intrinsic scatter assumed.
This result makes good sense because we can see from Fig. \ref{fig:testcvelB} that including
the coherent velocity contribution $C_{11}^{\rm vel.}$ raises the total magnitude covariance by
a factor of $3 - 4$.
{\it The lesson: peculiar velocity has a significant impact on the determination of the SN zero-point
from a low redshift anchor.}

Note that in all our computations of the coherent/correlated velocity term $C_{ij}^{\rm vel.}$, we
use the exact expression that allows for large angles (eq. [\ref{cijvellarge}]). 
We find that using the plane parallel approximation (eq. [\ref{Cijfull}]) leads to an underestimate
of $C_{ij}^{\rm vel.}$ by only about $10 \%$, even for a survey with high sky coverage
like the SNfactory.

%%%%%%%%%%%%
%{\bf Will erase later: for small vs large angles, see discussion in Code/guide on TestCvel2again vs TestCvel2small.}
%%%%%%%%%%%%

Fig. \ref{fig:testcvelB} might give one the impression that one is 
better off moving the low redshift anchor to a higher $z$
where the peculiar motion induced magnitude fluctuations are smaller,
largely because the ratio of peculiar velocity to Hubble flow is smaller.
However, 
to measure the equation of state $w_{\rm pivot}$ accurately,
it is advantageous to have a long lever arm
in redshift. In other words, a large redshift span (from the low redshift anchor
like the SNfactory to a high redshift SN survey) is preferable -- recall that 
the discovery of cosmic acceleration comes from comparing the
low redshift part ($z \lsim 0.1$) of the Hubble diagram with the high redshift part ($z \sim 1$). 
From this point of view, it is not immediately obvious that moving the low redshift anchor
to a higher $z$ actually helps.
Fig. \ref{fig:testcveldwB} addresses this question.

The solid lines show the marginalized error 
on $w_{\rm pivot}$ 
when a high z SNAP-like survey of 2000 SNe (see Table \ref{tabsurveys} for details)
is combined with a low z survey of 300 SNe,
width $\Delta z = 0.05$ and a mean $z$ as shown on the x-axis (with a geometry like that used
for the SNfactory, see Fig. \ref{fig:testcvelB}).
A prior of rms $\delta \Omega_{\rm de} = 0.03$ is assumed.
The two solid lines are for a low $z$ survey area of $1000$ (upper) and $20000$ (lower) square degrees
respectively. They allow for all sources of magnitude fluctuations we have discussed:
peculiar motion, gravitational lensing and 
intrinsic scatter (the upper panel is for $\sigma^{\rm intr.} = 0.15$ and the lower panel uses
$\sigma^{\rm intr.} = 0.1$).
One can see that the optimal mean redshift for the low $z$ survey is about $0.08$.
However, as long as the mean redshift is above $0.06$ or so, the precise redshift does not
appear to matter much. The upturn of the error at $z \lsim 0.06$ is due to coherent
peculiar motion as discussed before (Fig. \ref{fig:testcvelB}). The flatness of the solid
lines at $z \gsim 0.06$ is due to the rough cancellation of two opposing effects
mentioned above: a higher redshift (for the low redshift anchor) is good for suppressing peculiar motion 
induced fluctuations while a lower redshift is useful
for creating a long lever arm.
The SNfactory has a mean redshift which is sufficiently close to optimal
that it is probably not worth moving it to a higher redshift where observations are more challenging.
%How much difference does it actually make? z=0.055 gives 0.0690121, z=0.075 gives 0.0660891,
%z=0.095 gives 0.0664712. Therefore, fractional change from 0.055 to 0.075 is 4 % or so. Not much indeed.
%This is for sigma_intr=0.15.
%For sigma_intr=0.1: the change is from z=0.0555 gives 5.80415E-02, to z=0.075 gives 5.40491E-02,
%which is about 7 %. Still small compared to 60 % that we are talking about in this paper.

The dotted line shows the error on $w_{\rm pivot}$ for exactly the same set up as above but
with peculiar motion ignored (the result is independent of the survey area
of the low $z$ survey). One can see that by ignoring peculiar motion, 
one might reach the erroneous conclusion that the SNfactory should be moved to a lower redshift. 
We emphasize that it is the coherent peculiar motion that matters here -- the Poissonian
velocity induced fluctuations are simply too small to be of consequence (see Fig. \ref{fig:testcvelB}).
For comparison, we also show in Fig. \ref{fig:testcveldwB} with a horizontal dashed line
the error on $w_{\rm pivot}$ if one uses only the data from the high $z$ survey. 
The difference between the dashed line and the solid/dotted lines illustrates the
benefit of having a low $z$ anchor.

Rather similar conclusions are reached about the optimal redshift of the low $z$ anchor
when we examine the impact on the error of $w_a$, or when we combine the low $z$ survey
with other high $z$ surveys e.g. those in Table \ref{tabsurveys}.

\subsection{Gravitational Lensing}
\label{lens}

%From SNlens/Code/TestCvelLens3/testcvellensB.ps
\Sfig{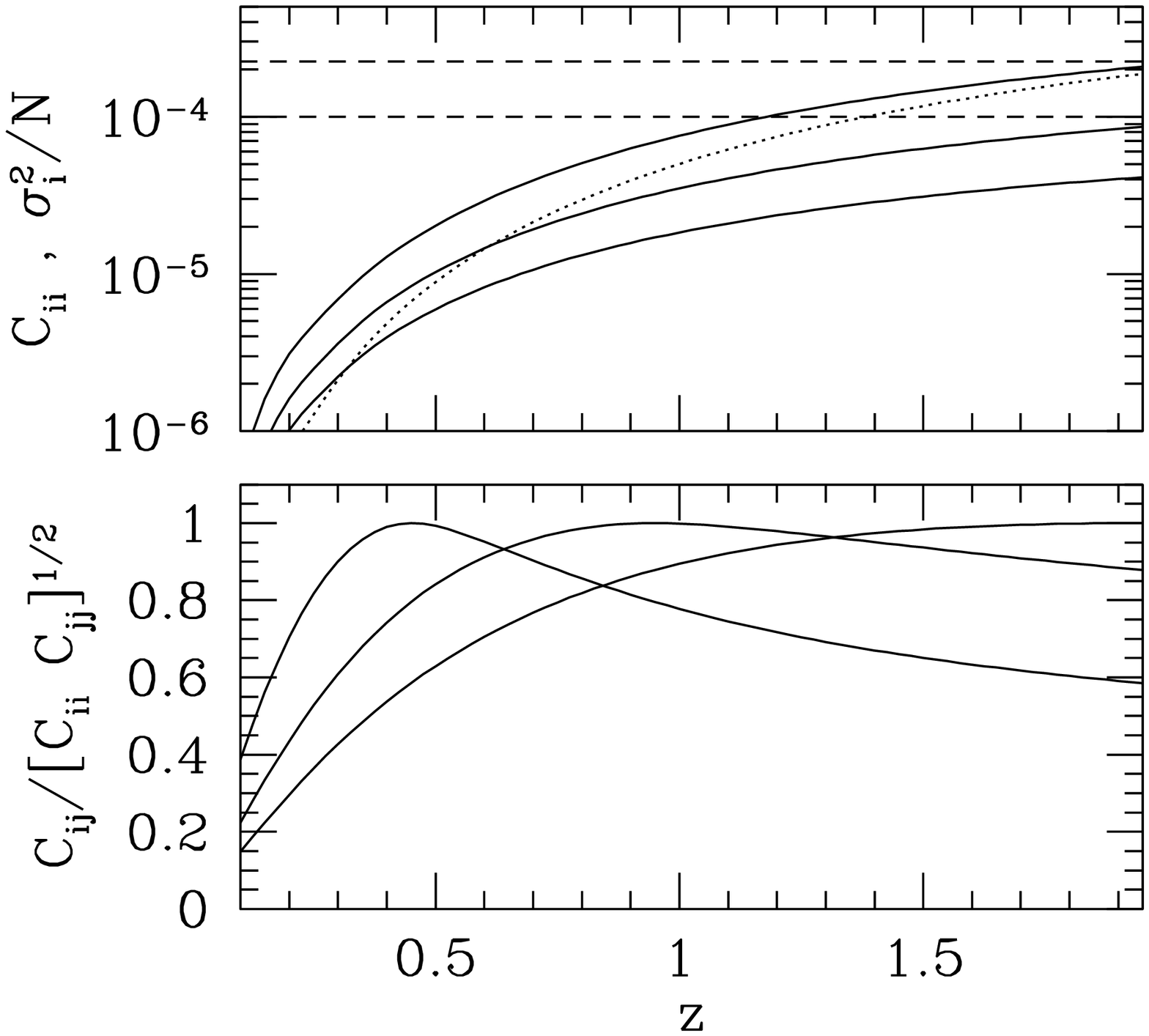}{Various lensing contributions to the magnitude covariance matrix (eq. [\ref{tC}]).
Upper panel: the solid lines show the non-Poissonian lensing term $C^{\rm lens}_{ii}$ 
(eq. [\ref{Cijfull}]) as a function of $z = z_i$ for a survey of $1$, $5$ and $15$ square degrees (top to bottom);
the dotted line shows the Poissonian lensing term $2 \times (\sigma^{\rm lens}_i)^2 / N$ 
(eq. [\ref{tC}] \& [\ref{sigPfull}]), 
for $N = 100$; for reference, the dashed lines show the expected contribution from intrinsic scatter
$(\sigma^{\rm intr.})^2/N$, for $\sigma^{\rm intr.} = 0.15$ (upper line) and $0.1$ (lower line). 
Lower panel: the normalized lensing correlation across redshifts i.e.
$C^{\rm lens}_{ij}/[C^{\rm lens}_{ii} C^{\rm lens}_{jj}]^{1/2}$ as a function of $z = z_i$, for
three different values of $z_j$: $0.045$, $0.95$ and $1.95$.
}

%From SNlens/Code/TestJEDIallI3/testcvellensC.ps
\Sfig{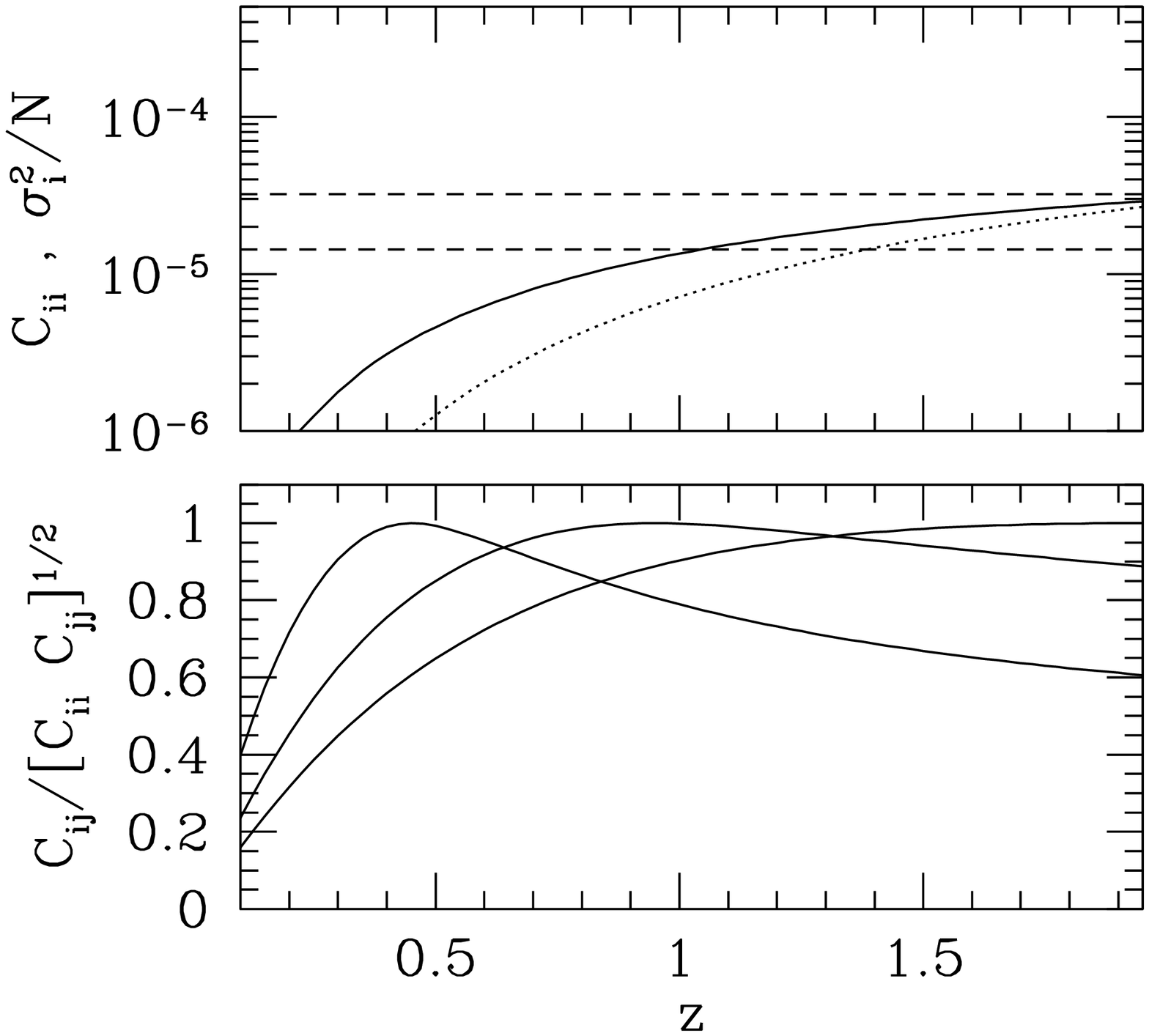}{Similar to Fig. \ref{fig:testcvellensB}, except that the survey area is
$24$ square degrees, and $N = 700$, numbers that are motivated by JEDI
(think of $N$ as roughly the number of SNe per $\Delta z$ of $0.1$).
}

Here, we are interested in the relative importance of the Poissonian and non-Poissonian lensing induced 
fluctuations,
$(\sigma_i^{\rm Poiss.,\, lens})^2$ and $C_{ij}^{\rm lens}$. The latter has been largely overlooked 
in error forecasts for SN surveys. As we will see, unlike the case of peculiar motion, it is
the Poissonian term that is important for lensing. 
Because this term 
is sensitive to small scale fluctuations, as mentioned in \S \ref{dLf2}, we always use the {\it nonlinear}
mass power spectrum to compute all lensing quantities.

To facilitate comparison with previous treatments of the Poissonian
lensing fluctuations, we introduce one modification to our previous expressions: replace eq. (\ref{sigPfull0}) by
\begin{eqnarray}
\label{sigPfull0b}
({\sigma_i^{\rm Poiss.}})^2 = 2 ({\sigma_i^{\rm Poiss.,\, lens}})^2 + ({\sigma_i^{\rm Poiss.,\, vel.}})^2
\end{eqnarray}
The boost factor of $2$ appears to be necessary to roughly reproduce the results of \cite{hl} who used 
a halo approach instead of following a power spectrum approach like we do. (The results of \cite{hl} are
about $2 - 3$ times higher than ours without the boost factor.) In other words,
\cite{hl} effectively used a different power spectrum from ours.
Physically, such a boost could arise from non-gravitational physics (recall that the nonlinear power spectrum
we use \cite{smith} arises purely from gravitational instability). For instance, the adiabatic contraction
of dark matter halos due to the radiative cooling of baryons can enhance the power spectrum on small scales
\cite{martin}. There could even be a significant population of MACHOs which can lens the SNe 
(see e.g. \cite{hl}). 
On the other hand, there is at least some evidence suggesting that Cold Dark Matter models might
overpredict the amount of small scale structure. The precise boost factor is therefore
a bit uncertain. The choice of $2$ is somewhat arbitrary, but it seems prudent to include some
enhancement of power due to well-motivated physical effects \cite{martin}. 
Ultimately, 
high redshift SNe themselves will tell us what the right level of Poissonian lensing
fluctuation is. In any case, the boost leads to 
a more conservative errorbar. Throughout this paper, eq. (\ref{sigPfull0b}) is used in place of
eq. (\ref{sigPfull0}) when making error forecasts.

It is interesting to note that if the small scale power spectrum has a shape similar to that
seen in N-body simulations \cite{smith}, 
the integral for $({\sigma_i^{\rm Poiss.,\, lens}})^2$ (eq. [\ref{sigPfull}])
is dominated by $k \sim 10$ h/Mpc. This coincides with the scale where the adiabatic contraction
of halos due to baryon cooling, in some sense the most plausible non-gravitational effect, 
is expected to become important \cite{martin}. On larger scales, gravity is almost certainly the only
significant shaping force of large scale structure. This is why no boost factor is necessary
for the non-Poissonian lensing term $C_{ij}^{\rm lens}$ (eq. [\ref{Cijfull}]). 

Fig. \ref{fig:testcvellensB} shows the relative importance of the Poissonian and non-Poissonian lensing terms.
The upper panel shows with solid lines the non-Poissonian lensing term $C^{\rm lens.}_{ii}$ where
we have divided the redshift into bins of $\Delta z = 0.1$ each \cite{comments}.
The term $C^{\rm lens.}_{ii}$ depends on survey area, 
and we show from top to bottom the result for $1$, $5$, and $15$ square degrees.
The dotted line shows the Poissonian lensing term $2 (\sigma_i^{\rm Poiss.,\, lens})^2 /N$, where one can think of
$N=100$ as the number of SNe in a redshift bin.
For comparison, we show with dashed horizontal lines the 
Poissonian term 
due to intrinsic scatter $(\sigma^{\rm intr.})^2/100$, with the upper line
using $\sigma^{\rm intr.} = 0.15$ and the lower line using $\sigma^{\rm intr.} = 0.1$. The number of $N = 100$ is somewhat
arbitrary but it roughly corresponds to the number of SNe per redshift bin of $0.1$ in a survey like SNAP, which
also has an area of around $15$ square degrees (Table \ref{tabsurveys}).

What can we conclude from Fig. \ref{fig:testcvellensB}?
From the upper panel, one can see that lensing begins to be of comparable importance to the intrinsic scatter
when the redshift climbs above $1$ or so.
Moreover, unless the survey area is quite small (less than $\sim 1$ square degree), 
the non-Poissonian lensing term is small compared to the Poissonian
lensing term at redshifts where they matter. 
From the lower panel, we can see that there is a significant amount of lensing induced
correlations between different redshifts. However, because $C^{\rm lens}_{ii}$ is small to begin with
(compared to $2(\sigma^{\rm lens}_i)^2/N$), we do not expect these cross-redshift correlations to 
hugely impact dark energy errors from SN surveys. We have verified this to be the case.

The above conclusion about the relative importance of non-Poissonian and Poissonian lensing
terms is subject to changes in survey parameters, however.
Fig. \ref{fig:testcvellensC} is similar to Fig. \ref{fig:testcvellensB}, except that the number of
SNe per redshift bin is increased to $700$, as motivated by the ambitious SN survey JEDI. 
(The area is also slightly increased to $24$ square degrees from the
$15$ square degrees of SNAP, though that has a minor impact
relatively speaking.)
One can see that the non-Poissonian and the (boosted) Poissonian lensing terms are here of comparable
importance at the redshifts where they matter ($z \gsim 1$).

\subsection{Error Forecasts: Peculiar Motion versus Lensing}
\label{everything}

\begin{table}[htb]
\begin{center}
\begin{tabular}{|cccc|}\hline
Survey & No. of SNe & \, Area (sq. deg.) \, & \, Redshift Dist. \\ \hline \hline
DES & 1900 & 40 & 0.2 - 0.8 flat \\ \hline
ESSENCE & 200 & 12 & 0.2 - 0.8 flat \\ \hline
JEDI & 14000 & 24 & 0.1 - 1.7 \cite{kim} \\ \hline
SDSSII & 200 & 250 & 0.05 - 0.35 flat \\ \hline
SNAP & 2000 & 15 & 0.1 - 1.7 \cite{kim} \\ \hline
SNfactory & 300 & 20000 & 0.03 - 0.08 flat \\ \hline
SNLS & 600 & 4 & 0.2 - 0.8 flat \\
\hline
\end{tabular}
\end{center}
\caption{\label{tabsurveys} A summary of survey parameters we study in this paper.
Each set of parameters are supposed to mimic, but {\it not} necessarily exactly match, 
actual SN surveys that go by these names \cite{SNsurveys}.
For instance, the (flat) redshift distributions almost certainly differ from the actual ones -- they are
chosen for simplicity. For JEDI and SNAP, the redshift distribution is taken
from \cite{kim}, scaled to the appropriate number of SNe. 
For some surveys not included in this table, such
as the Carnegie Supernova Project and CfA Supernova Program at low redshifts and
the LSST and Pan-STARRS at high redshifts, see \cite{SNsurveys}.
}
\end{table}

We are finally ready to make error forecasts for several examples that 
approximate ongoing/proposed SN surveys. They are described in Table \ref{tabsurveys}.
Note that the adopted parameters by no means exactly match those of actual surveys that go by those
acronyms, but they should be close. 
Also, except for the SNfactory (see Fig. \ref{fig:testcvelB}), 
we assume the area listed for each survey is over a contiguous region, 
and, for simplicity, consists of a circular patch on the sky. (We will have more to say later on the implications
of the exact survey geometry.) See \cite{SNsurveys} for further details on these surveys, and others
we have not worked out explicitly. 
Our procedure, simply put, is: we put eq.
(\ref{sigPfull0b}) and eq. (\ref{sigPfull}) - (\ref{Wvellarge}) for
the large scale structure induced magnitude fluctuations into
eq. (\ref{tC}) and eq. (\ref{fisher}) for the Fisher matrix, and obtain the relevant dark energy errors.

%From SNlens/Code/TestCvelLensI2/testhistoB.ps
\Sfig{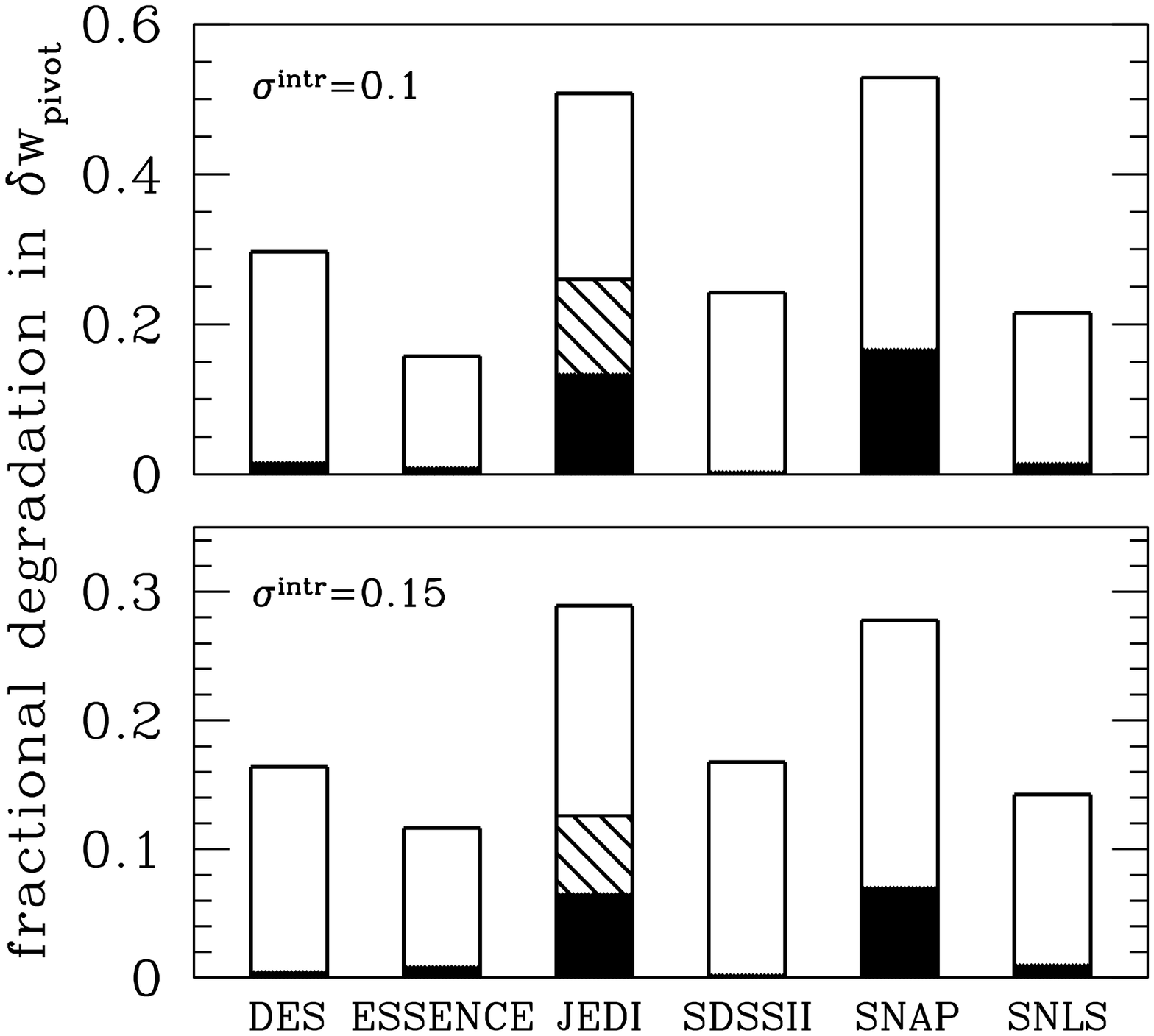}{The degradation in the marginalized error on the equation of state
$w_{\rm pivot}$ for several high redshift SN surveys, each used in conjunction with the 
SNfactory as a low redshift anchor. (The error on $w_{\rm pivot}$ is marginalized over
$w_a$, $\Omega_{\rm de}$ and $M$; because of our choice of the scale factor pivot,
this is identical with the error on a constant equation of state, marginalized over
$\Omega_{\rm de}$ and $M$; see \S \ref{prelim}.)
Degradation refers to the
fractional increase in the rms marginalized error on $w_{\rm pivot}$ due to large scale structure
induced fluctuations i.e. degradation $\equiv \delta w_{\rm pivot}/\delta w_{\rm pivot}^{\rm intr.} - 1$,
where $\delta w_{\rm pivot}^{\rm intr.}$ allows for only the intrinsic scatter while $\delta w_{\rm pivot}$
takes into account both the intrinsic scatter and large scale structure. 
The division of each histogram into black and white regions show how much of
the degradation is due to lensing (black) and how much due to peculiar motion (white).
The upper panel uses $0.1$ for the intrinsic scatter $\sigma^{\rm intr.}$, while the lower
panel uses $0.15$. In all cases, except JEDI, the lensing degradation is completely 
dominated by Poissonian lensing
fluctuations. For JEDI, the black-filled portion is the degradation due to Poissonian lensing fluctuations
and the black-hatched portion is the additional degradation due to non-Poissonian/coherent lensing
fluctuations. A prior of $\delta \Omega_{\rm de} = 0.03$ (flat universe) is assumed.}

%From SNlens/Code/TestCvelLensI2/testhistoBnoprior.ps
\Sfig{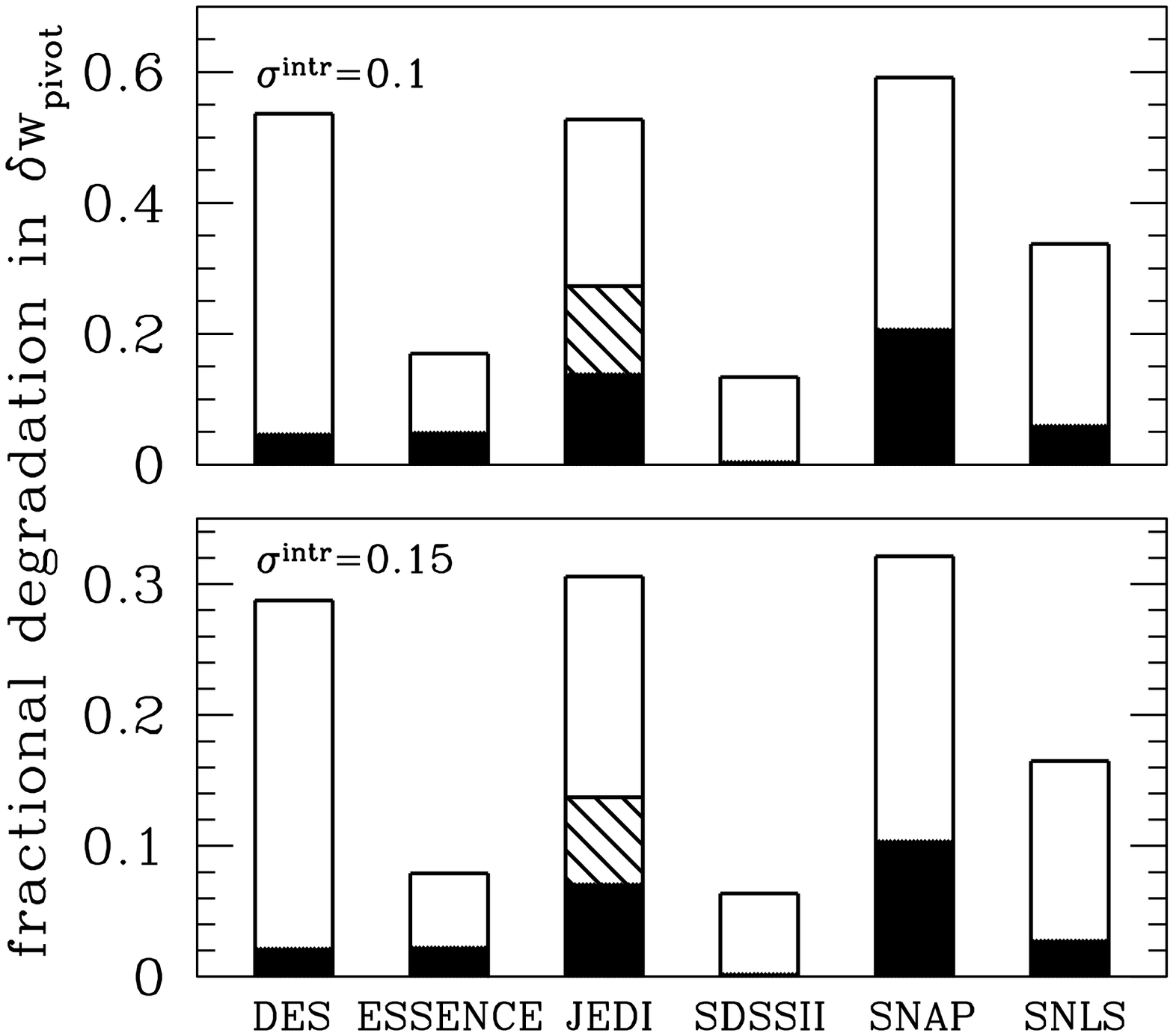}{Same as Fig. \ref{fig:testhistoB} except that no prior on $\Omega_{\rm de}$ is assumed,
though the universe is still flat.
}

%From SNlens/Code/TestCvelLensI2/Morecontours/contour.1.combo.ps
\Sfig{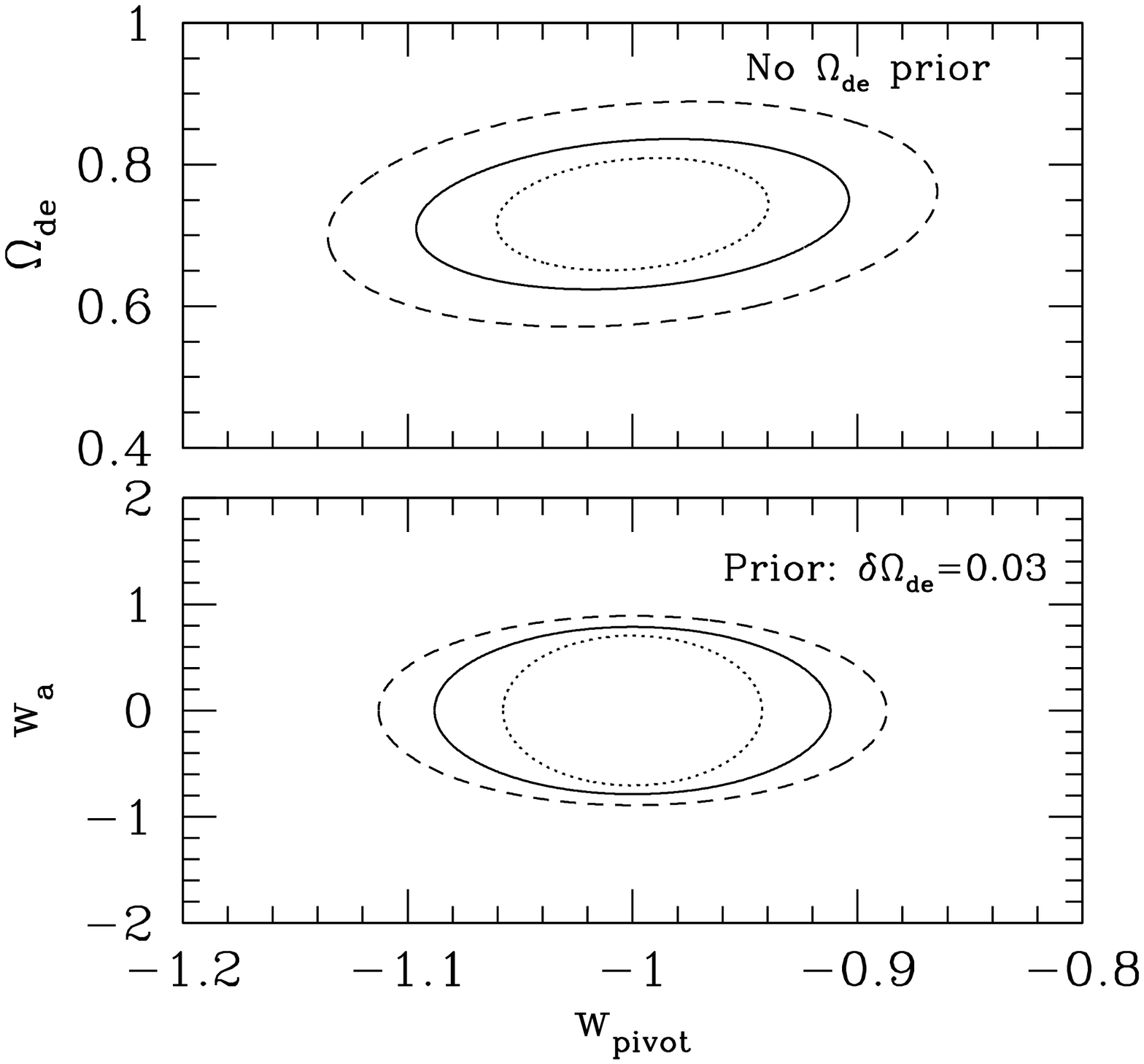}{Projected ($68.3 \%$) errors for SNAP $+$ SNfactory 
for three different combinations of sources of noise: 
intrinsic scatter only (dotted), intrinsic scatter $+$ large scale structure induced
fluctuations (solid), and intrinsic scatter $+$ large scale structure $+$ systematic error (dashed).
The upper panel shows the errors for $\Omega_{\rm de}$ and
$w_{\rm pivot}$ marginalizing over the derivative $w_a = -dw/da$ and the zero point $M$, assuming
a flat universe but no prior on $\Omega_{\rm de}$. The lower panel shows the errors for
$w_a$ and $w_{\rm pivot}$ marginalizing over the dark energy density
$\Omega_{\rm de}$ and the zero-point $M$, assuming a flat universe and a prior of $\delta\Omega_{\rm de} = 0.03$.
Here, $w=w_{\rm pivot} + w_a (a_{\rm pivot}-a)$, where $a_{\rm pivot}$ is chosen
to make the errors on $w_{\rm pivot}$ and $w_a$ uncorrelated (see \S \ref{prelim}).
Note that $\delta w_a \sim 2 \delta w'$, where $w'$ is another common parametrization of the evolution of
the equation of state: $w' = dw/dz$.
In both panels, $\sigma^{\rm intr.} = 0.1$ is used.
The degradation in errors due to large scale structure is primarily due
to coherent peculiar motion and Poissonian lensing fluctuations (see Fig. \ref{fig:testhistoB} 
\& \ref{fig:testhistoBnoprior}). The systematic error is assumed to be $\sigma^{\rm sys.} = 0.02 (1+z)/2.7$
\cite{sys}. 
It appears the degradation due to large scale structure
and that due to systematic error are comparable.
}

For readers interested in the bottom line, 
Fig. \ref{fig:testhistoB} \& \ref{fig:testhistoBnoprior} are in some sense the most important figures of this paper.
They show the degradation in the marginalized error on $w_{\rm pivot}$ for 
several high redshift SN surveys, each used in conjunction with the SNfactory as a low redshift anchor.
Degradation refers to the fractional increase in error. It is
defined as $\delta w_{\rm pivot}/\delta w_{\rm pivot}^{\rm intr.} - 1$, where
$\delta w_{\rm pivot}^{\rm intr.}$ is the rms error when only the intrinsic scatter is included, and
$\delta w_{\rm pivot}$ is the rms error when all sources of fluctuations (intrinsic $+$ 
lensing $+$ velocity) are included (full histogram, including both black and white portions), or
when only intrinsic $+$ lensing fluctuations are included (black portion of the histogram).
In other words, the white (black) portion of the histogram tells us the degradation due to peculiar motion (lensing).
In all cases other than JEDI, the lensing fluctuations are completely dominated by the Poissonian ones (the ones
that are customarily considered). For JEDI, the black-filled portion represents the degradation
when intrinsic $+$ Poissonian-lensing fluctuations are included, and the black-hatched portion
represents the additional degradation when non-Poissonian/coherent lensing fluctuations are taken
into account. The error on $w_{\rm pivot}$ here is obtained by marginalizing over
$w_a$, $\Omega_{\rm de}$ and $M$
(as discussed in \S \ref{prelim} and the Appendix \ref{app:fisher}, 
this is identical to the error on a constant equation of state, marginalized over
$\Omega_{\rm de}$ and $M$).
 A prior of rms $\delta \Omega_{\rm de} = 0.03$ is assumed
for Fig. \ref{fig:testhistoB}, while no such prior is used for Fig. \ref{fig:testhistoBnoprior}.
A flat universe is assumed in both cases.
The upper panel uses an intrinsic scatter of size $\sigma^{\rm intr.} = 0.1$, and the lower
panel uses $\sigma^{\rm intr.} = 0.15$. 
{\it An important conclusion: the general trend is for surveys
that have a higher statistical power to suffer more degradation,
and
in most cases, peculiar motion
dominates the degradation (white portions of the histograms).}

Fig. \ref{fig:contour.1.combo} shows the error contours for the example of a SNAP-SNfactory combination.
The dotted contours include the intrinsic scatter only, while the solid contours include
both the intrinsic scatter and large scale structure (velocity $+$ lensing) fluctuations.
The dashed contours are for the case where 
some level of systematic error ($\sigma^{\rm sys} = 0.02 (1+z)/2.7$ \cite{sys}) 
is included on top of intrinsic plus large scale structure fluctuations. One can see that a systematic
error of this size degrades the determination of $w_{\rm pivot}$ to a degree that is
comparable to large scale structure (see the end of this section for more discussion).
Note that this figure uses $\sigma^{\rm intr.} = 0.1$; for $\sigma^{\rm intr.} = 0.15$, the constraints would
be worse.

For readers interested in further details, 
Table \ref{table} displays error forecasts for all interesting 
parameters for various combinations of surveys, assumptions and priors.
There are several interesting points to be made.

\begin{itemize}

\item The prior of $\delta \Omega_{\rm de} = 0.03$, which is often assumed in the literature,
makes quite a difference to the projected errors on 
the equation of state $w_{\rm pivot}$ and its slope $w_a$ (eq. [\ref{linderparm}]), 
especially for the smaller surveys -- those with
a narrower redshift range and fewer SNe.

\item The addition of a low redshift survey like the SNfactory generally improves the errorbars
significantly, the more so for surveys with less statistical power.

\item As mentioned in \S \ref{prelim}, the error on $w_{\rm pivot}$, marginalized over
the slope $w_a$, dark energy density $\Omega_{\rm de}$ and the zero-point $M$, is
exactly equal to the error on a constant equation of state, marginalized over $\Omega_{\rm de}$
and $M$ (this is the result of an optimally chosen scale factor pivot).
This is relevant especially for the smaller surveys -- they produce only weak constraints
on the variation of $w$, which is why the error on $w$ generally projected for
these surveys assumes no variation with $z$. Our marginalized error on $w_{\rm pivot}$ can be
compared directly against these forecasts.
Note, however, that our error on $\Omega_{\rm de}$, marginalized over $w_{\rm pivot}$, 
$w_a$ and $M$, is generally larger
than the error on $\Omega_{\rm de}$, marginalized over a constant equation of state
and $M$.

\item Several surveys have a rather similar redshift coverage of $z \sim 0.2 - 0.8$: 
DES, ESSENCE and SNLS. On their own, without the addition of the SNfactory, one can see that
the large scale structure fluctuations make little difference to the errors (compare
'all' with 'intr'). This is because peculiar motion is only important
for $z \lsim 0.1$ and lensing is important only for $z \gsim 1$. 
It is with the addition of the SNfactory (which is useful because it improves the errors quite a bit) that
peculiar motion has an impact.

\item Large scale structure fluctuations generally have a larger impact (fractionally) on the
errors for larger surveys, a point already illustrated in Fig. \ref{fig:testhistoB} and
\ref{fig:testhistoBnoprior}.

\item For the most part,
we have so far focused on statistical errors. 
This is in part because the level of systematic errors is uncertain, depending
on the precise capabilities and operational details of the respective experiments.
To get some idea of
how systematic errors might change our conclusions, we try two different
prescriptions for the systematic error: one from \cite{kim}
$\sigma^{\rm sys.} = 0.02 \times (z/1.7)$, 
and the other from \cite{linderhut05} $\sigma^{\rm sys.} = 0.02 \times (1+z/2.7)$ (see \cite{sys} on
redshift binning).
They behave similarly at high $z$, but the latter implies a larger systematic error at low $z$.
We apply these to SNAP $+$ SNfactory in Table \ref{table}, with the former prescription
labeled as 'sys' and the latter labeled as 'sys2'. The general conclusion is that systematic
error contributes to a degradation that is comparable to large scale structure.
This is illustrated in Fig. \ref{fig:contour.1.combo}: compare the increase in error for $w_{\rm pivot}$
from the dotted to solid contours against that from the solid to dashed contours.

\end{itemize}

%%%%%%%%%%%%%%
%{\bf Will erase later: let's compare the SNAP error with the error in \cite{kim} to make sure they agree.
%I can compare my marginalized $\delta w_{\rm pivot}$ with their error on $w_0$ keeping $w'$ fixed at $0$.
%Their Fig. 1 innermost contour shows $w_0$ has an error of $< 10\%$. This seems consistent with our 
%$0.059 \times \sqrt{2.3)} 
%\sim 0.089$ for SNAP $+$ SNf (intr). (The factor of $\sqrt{2.3}$ is because in 2 dimensions, 
%the $68 \%$ contour corresponds
%to a $\chi^2$ of $2.3$.)
%Their Fig. 3 innermost contour shows $w_0$ has an error of about $0.1$, which is consistent with our 
%$0.66 \times \sqrt{2.3} \sim 0.1$
%for SNAP $+$ SNf (intr + sys).}
%%%%%%%%%%%%%%

%\onecolumngrid

%Table details, see SNlens/Code/guide
%New: these new runs (using new cosmological parameters) are in Home/SNlens2/Code/PRDRev/
\begin{table*}[tb]
\begin{center}
\begin{tabular}{|cccccccc|}\hline
Survey & $\delta w_{\rm pivot}$ (prior) & $\delta w_a$ (prior) & $\delta M$ (prior) & $\delta w_{\rm pivot}$ (no pr.) & $\delta w_a$ (no pr.) & $\delta\Omega_{\rm de}$ (no pr.) & $\delta M$ (no pr.)
\\ \hline \hline
DES + SNf (all) & 0.085/0.079 & 0.69/0.57 & 0.021/0.018 & 0.17/0.14 & 4.9/3.5 & 0.47/0.33 & 0.027/0.023 \\
%New: TestDESallI and TestDESallI.1
\hline
DES + SNf (intr) & 0.073/0.061 & 0.60/0.46 & 0.013/0.009 & 0.13/0.09 & 4.5/3.0 & 0.44/0.29 & 0.018/0.012 \\
%New: TestDES_nov and TestDES_nov.1
\hline
DES (all) & 0.11/0.01 & 1.4/1.0 & 0.064/0.044 & 0.36/0.25 & 17/12 & 1.4/0.9 & 0.23/0.15 \\
%New: TestDESallIhigh and TestDESallIhigh.1
\hline
DES (intr) & 0.11/0.10 & 1.4/1.0 & 0.063/0.042 & 0.35/0.23 & 17/11 & 1.3/0.9 & 0.22/0.15 \\
%New: TestDEShigh_nov and TestDEShigh_nov.1
\hline
ESSENCE+SNf (all) & 0.094/0.087 & 1.5/1.1 & 0.024/0.021 & 0.36/0.26 & 14.0/9.6 & 1.4/1.0 & 0.043/0.033 \\
%New: TestESSENCEallIredo and TestESSENCEallI.1redo
\hline
ESSENCE+SNf (intr) & 0.084/0.075 & 1.4/1.0 & 0.017/0.011 & 0.34/0.23 & 13.6/9.1 & 1.3/0.9 & 0.038/0.025 \\
%New: TestESSENCE_novredo and TestESSENCE_nov.1redo
\hline
ESSENCE (all) & 0.19/0.15 & 4.2/2.9 & 0.20/0.13 & 1.11/0.76 & 54/37 & 4.2/2.9 & 0.69/0.47 \\
%New: TestESSENCEallIhighredo and TestESSENCEallIhigh.1redo
\hline
ESSENCE (intr) & 0.18/0.14 & 4.1/2.8 & 0.19/0.13 & 1.1/0.72 & 52/35 & 4.1/2.7 & 0.68/0.45 \\
%New: TestESSENCEhigh_novredo and TestESSENCEhigh_nov.1redo
\hline
JEDI + SNf (all) & 0.041/0.032 & 0.46/0.41 & 0.013/0.011 & 0.042/0.033 & 0.76/0.60 & 0.042/0.033 & 0.017/0.013 \\
%New: TestJEDIallI and TestJEDIallI.1
\hline
JEDI + SNf (intr) & 0.032/0.021 & 0.41/0.32 & 0.010/0.007 & 0.032/0.022 & 0.60/0.40 & 0.033/0.022 & 0.011/0.008 \\
%New: TestJEDI_nov and TestJEDI_nov.1
\hline
JEDI (all) & 0.046/0.035 & 0.52/0.46 & 0.018/0.014 & 0.048/0.036 & 0.99/0.73 & 0.052/0.039 & 0.025/0.018 \\
%New: TestJEDIallIhigh and TestJEDIallIhigh.1
\hline
JEDI (intr) & 0.041/0.028 & 0.50/0.42 & 0.017/0.013 & 0.042/0.028 & 0.90/0.60 & 0.046/0.031 & 0.024/0.016 \\
%New: TestJEDIhigh_nov and TestJEDIhigh_nov.1
\hline
SDSSII + SNf (all) & 0.12/0.10 & 5.7/4.0 & 0.039/0.031 & 1.6/1.1 & 101/69 & 15/10 & 0.092/0.067 \\
%New: TestSDSSIIballI and TestSDSSIIballI.1
\hline
SDSSII + SNf (intr) & 0.105/0.081 & 5.4/3.6 & 0.031/0.021 & 1.5/1.0 & 99/66 & 15/10 & 0.085/0.057 \\
%New: TestSDSSIIb_nov and TestSDSSIIb_nov.1
\hline
SDSSII (all) & 0.17/0.12 & 8.0/5.4 & 0.081/0.056 & 2.3/1.6 & 145/97 & 21/14 & 0.20/0.14 \\
%New: TestSDSSIIballIhigh and TestSDSSIIballIhigh.1
\hline
SDSSII (intr) & 0.17/0.12 & 7.8/5.2 & 0.078/0.052 & 2.2/1.5 & 143/95 & 20/14 & 0.20/0.13 \\
%New: TestSDSSIIbhigh_nov and TestSDSSIIbhigh_nov.1
\hline
SNAP + SNf (all) & 0.069/0.058 & 0.57/0.52 & 0.018/0.015 & 0.079/0.063 & 1.5/1.2 & 0.09/0.07 & 0.022/0.019 \\
%New: TestCvelLensI2 and TestCvelLensI2.1
\hline
SNAP + SNf (intr) & 0.054/0.038 & 0.53/0.46 & 0.011/0.008 & 0.06/0.04 & 1.32/0.88 & 0.078/0.052 & 0.014/0.009 \\
%New: TestCvel2_nov and TestCvel2_nov.1
\hline
SNAP + SNf (all+sys) & 0.075/0.067 & 0.60/0.55 & 0.018/0.016 & 0.090/0.077 & 1.8/1.5 & 0.11/0.09 & 0.022/0.020 \\
%New: TestCvelLensI2sys and TestCvelLensI2sys.1
\hline
SNAP + SNf (intr+sys) & 0.061/0.049 & 0.56/0.50 & 0.012/0.008 & 0.073/0.056 & 1.6/1.2 & 0.10/0.08 & 0.015/0.010 \\
%New: TestCvel2sys_nov and TestCvel2sys_nov.1
\hline
SNAP + SNf (all+sys2) & 0.081/0.074 & 0.63/0.59 & 0.020/0.018 & 0.101/0.089 & 2.0/1.7 & 0.12/0.10 & 0.025/0.023 \\
%New: TestCvelLensI2sysB and TestCvelLensI2sysB.1
\hline
SNAP + SNf (intr+sys2) & 0.071/0.062 & 0.59/0.55 & 0.015/0.013 & 0.086/0.073 & 1.8/1.5 & 0.111/0.093 & 0.019/0.016 \\
%New: TestCvel2sysB_nov and TestCvel2sysB_nov.1
\hline
SNAP (all) & 0.101/0.078 & 0.70/0.62 & 0.038/0.028 & 0.121/0.087 & 2.6/1.9 & 0.14/0.10 & 0.066/0.046 \\
%New: TestCvelLensI2high and TestCvelLensI2high.1
\hline
SNAP (intr) & 0.096/0.069 & 0.68/0.60 & 0.037/0.026 & 0.110/0.074 & 2.4/1.6 & 0.122/0.082 & 0.063/0.042 \\
%New: TestCvel2high_nov and TestCvel2high_nov.1
\hline
SNLS + SNf (all) & 0.089/0.084 & 0.98/0.75 & 0.022/0.019 & 0.24/0.18 & 8.2/5.7 & 0.81/0.56 & 0.032/0.026 \\
%New: TestCFHTallIredo and TestCFHTallI.1redo
\hline
SNLS + SNf (intr) & 0.077/0.069 & 0.89/0.64 & 0.014/0.009 & 0.21/0.14 & 7.9/5.3 & 0.78/0.52 & 0.025/0.017 \\
%New: TestCFHT_novredo and TestCFHT_nov.1redo
\hline
SNLS (all) & 0.14/0.12 & 2.5/1.7 & 0.114/0.078 & 0.64/0.44 & 31/21 & 2.4/1.7 & 0.40/0.27 \\
%New: TestCFHTallIhighredo and TestCFHTallIhigh.1redo
\hline
SNLS (intr) & 0.13/0.11 & 2.4/1.6 & 0.11/0.07 & 0.62/0.42 & 30/20 & 2.4/1.6 & 0.39/0.26 \\
%New: TestCFHThigh_novredo and TestCFHThigh_nov.1redo
\hline
\end{tabular}
\end{center}
\caption{\label{table} Marginalized $1 \sigma$ errors for different combinations of surveys and assumptions.
The survey parameters are taken from Table \ref{tabsurveys} (SNf here stands for the SNfactory).
The descriptions in parentheses, 'all', 'intr', 'sys', 'sys2' refer to the sources of errors
that are included in the forecasts -- 'all' means including all sources of random fluctuations
intrinsic-scatter $+$ lensing $+$ velocity, 'intr' means including only the intrinsic scatter, and
'sys' and 'sys2' mean systematic error (see text for details). The errorbar on each parameter is obtained
by marginalizing over all the other parameters (altogether, there are four: $w_{\rm pivot}$,
$w_a$, $\Omega_{\rm de}$ and the zero-point $M$; see \S \ref{prelim}). 
Those errorbars with the description 'prior' assume a prior of rms $\delta \Omega_{\rm de} = 0.03$,
while those denoted with 'no pr.' assumes no such prior. 
For each entry, we give two numbers in the form $x/y$, where $x$ assumes the intrinsic scatter
has a size of $\sigma^{\rm intr.} = 0.15$ and $y$ assumes $\sigma^{\rm intr.} = 0.1$.
Note that in cases where the errorbars
are sufficiently large, the Fisher matrix analysis likely breaks down.
The above numbers are rounded-off at two significant figures.
A flat universe is assumed throughout.
}
\end{table*}

%\twocolumngrid

\section{Discussion}
\label{discuss}

Let us summarize the main lessons.

\begin{itemize}

\item Large scale structure induced magnitude fluctuations have a significant impact 
on dark energy measurements from
a whole array of ongoing or future SN surveys. For instance, the degradation, due to
large scale structure, in the error for 
the equation of state $w_{\rm pivot}$ 
ranges from $10 \%$ to $60 \%$ depending on surveys and assumptions
(Fig. \ref{fig:testhistoB}, \ref{fig:testhistoBnoprior}).
It appears difficult to measure from SNe alone the equation of state to better than about $7 - 10\%$
(depending on assumptions), unless one has a survey
considerably more ambitious than SNAP $+$ SNfactory (see Fig. \ref{fig:contour.1.combo}
and Table \ref{table}).

\item Of all possible large scale structure fluctuations, the dominant ones are due to peculiar motion
and gravitational lensing. Peculiar motion is important at $z \lsim 0.1$ (through a low 
redshift anchor such as the SNfactory) while lensing dominates
at $z \gsim 1$. The impact of peculiar motion is mainly through coherent/correlated large scale flows
(Fig. \ref{fig:testcvelB})
while the impact of lensing is mainly through Poissonian fluctuations (Fig. \ref{fig:testcvellensB}).
The Poissonian fluctuations can be reduced by increasing the number of SNe, while the coherent ones
can only be suppressed by increasing the survey area.
When a high redshift ($z \gsim 0.1$) survey is combined with a low redshift anchor ($z \lsim 0.1$),
we find that peculiar motion mostly dominates over lensing as a source of error.

\item What does the above mean for survey designs?
As has been emphasized in the literature (e.g. \cite{SNsurveys}), a low redshift anchor
such as the SNfactory is very useful for reducing the eventual dark energy errors from a high redshift
SN survey. For such a low redshift survey, one might have hoped to reduce the coherent peculiar motion induced
fluctuations by either increasing the survey area or moving it to a higher redshift. 
{\it All else being equal, neither will improve appreciably the precision on dark energy determination.}
The SNfactory already covers half of the sky; going to full sky will not reduce the errors 
significantly. For instance, combining SNAP with an all-sky version of SNfactory instead of the half-sky one that
we have been assuming, the marginalized error for $\delta w_{\rm pivot}$ would improve
by only about $1 \%$. 
Furthermore, moving SNfactory to a higher redshift, while useful in reducing
peculiar motion induced fluctuations, shortens the lever arm that the combination of
a high redshift survey and a low redshift anchor offers. The net effect is that moving the low redshift anchor
to a higher redshift actually does not reduce the error on the equation of state $w_{\rm pivot}$
appreciably (Fig. \ref{fig:testcveldwB}).

%%%%%%%%%
%{\bf Will erase later: for the full-sky runs, see TestCvelLensI2fullsky and TestCvelLensI2fullsky.1}
%%%%%%%%%

\item How about survey designs for a high redshift survey? For a high redshift survey (considered on its own) 
that does not extend beyond $z \sim 1$ (e.g. DES, ESSENCE and SNLS all cover roughly $z \sim 0.2 - 0.8$), 
neither peculiar motion nor lensing constitutes significant sources of errors. The only way to reduce
dark energy errors is to increase the number of SNe, and suppress systematic errors.
The precise survey area is of little importance for such a survey, as long as it is not too small (too small meaning
$1$ square degree or less, see Fig. \ref{fig:testcvellensB}).
(Of course, peculiar motion does play a role in the eventual errors once one combines
such a high redshift survey with a low redshift anchor, which as emphasized above, is generally a good idea.)
For a high redshift survey that extends beyond $z \sim 1$, gravitational lensing becomes
a non-negligible source of errors. But because lensing's impact is mainly through the Poissonian
fluctuations it introduces, increasing survey area (such as for SNAP) is not really necessary. 
The only instance in which a case can be made for increasing survey area is JEDI, which has
a sufficiently small Poissonian error (due to its large number of SNe) that coherent/correlated
lensing fluctuations actually play a role (see Fig. \ref{fig:testhistoB} and 
\ref{fig:testhistoBnoprior}).

\end{itemize}

Our investigations in this paper naturally raise a number of questions and issues,
some of which we address briefly here, and some require further research.

{\bf 1.} Perhaps the most natural and interesting question is whether {\it current} constraints
on dark energy, which typically come from some combination of high and low redshift
SNe ($z \gsim 0.1$ and $\lsim 0.1$), are already affected by peculiar motion.
The short answer is: not very much. This is because the current number of low redshift
SNe used (typically several 10's) is sufficiently small that the Poissonian error 
(due to simply intrinsic scatter)
is quite a bit larger than the coherent velocity error. 
This can be inferred from Fig. \ref{fig:testcvelB}: raising the dotted lines by
a factor of $\sim 10$ (due to dropping the number of SNe from $300$ as in the figure
to $\sim 30$) means the Poissonian intrinsic scatter constitutes a larger source
of error than coherent peculiar motion (solid lines).
Note that this argument assumes the existing low redshift SNe are selected from
a large area of the sky e.g. \cite{hamuy} (so that it is the lowest few solid lines
of Fig. \ref{fig:testcvelB} that is relevant). The conclusion could be quite different
if this assumption does not hold. We urge SN experiments to clearly state the survey
areas of their different samples (especially the low $z$ samples) when publishing their results.

{\bf 2.} In our forecasts for a selection of representative SN surveys (Table \ref{tabsurveys}), 
we have assumed simple geometries -- a contiguous circular region on the sky for the high
$z$ surveys, and two separate patches (one in the north and one in the south) for the SNfactory.
Realistic surveys are bound to be more complicated in shape, possibly with many holes or gaps.
The exact geometry affects the size of the coherent fluctuations but not the
Poissonian ones. Since for the most part the only coherent fluctuations we need
worry about are those due to peculiar motion which is important only at low $z$'s, 
it is mainly the exact geometry of something like the SNfactory that concerns us.
It is therefore worth repeating our calculations for the actual geometry of the SNfactory, including 
possible extra gaps for
instance. For this purpose, we have given sufficiently general expressions for the relevant window functions
($W^{\rm lens}_{ij}$ and $W^{\rm vel.}_{ij}$) in eq. (\ref{Cijfull}) and (\ref{cijvellarge}), and in
eq. (\ref{WvellargeSNf}) and (\ref{WvellargeGeneral}) in
Appendix \ref{app:poisson2}.
Note that gaps almost always increase the importance of coherent fluctuations because of the introduction
of high $k$ modes.
For the high redshift surveys that extend beyond $z \sim 1$, it would be useful to check
that a realistic survey geometry does not make the coherent lensing fluctuations much
more important (though we do not expect this to happen, as long as the survey area exceeds
$\sim 1$ square degree).
For high $z$ surveys that stay within the redshift range $0.1 \lsim z \lsim 1$, neither
velocity nor lensing fluctuations are expected to be important (unless the
number of SNe is much larger than what has been considered), and so the survey geometry has
a relatively minor impact.

{\bf 3.} We have largely ignored internal motion in our discussions of velocity induced fluctuations.
By internal motion we mean the motion of the SNe within galaxies, for instance due
to the virialized motion of the SN progenitors, or even due to the orbital motion of
the SN itself within the binary system that is its progenitor.
Such motion could contribute to the overall peculiar velocity of the SNe. 
Ignoring internal motion is partially justified
by the fact that in practice the redshifts are assigned based on the redshifts
of the host galaxies. (We thank the referee for emphasizing this to us.)
However, internal motion could still in principle modify the apparent
luminosity. The important point to keep in mind is that
such internal motion is not expected to be correlated between SNe in different galaxies, and so
our calculation of the coherent/correlated velocity fluctuations remains valid.
Internal motion can certainly increase the Poissonian velocity fluctuations. However, 
typical virialized motion is of the order of a few hundred km/s, similar to the typical
large scale flow velocity, and so the Poissonian velocity fluctuations remain subdominant
(Fig. \ref{fig:testcvelB}). Also, the orbital motion internal to the binary progenitor is too
slow to be of significance (see e.g. \cite{whelaniben}).

{\bf 4.} Our main focus in this paper is on statistical errors: from intrinsic scatter and from
large scale structure induced fluctuations. We have investigated the effect of systematic error
in some simple examples (see Table \ref{table}, entries for 'SNAP $+$ SNf (all $+$ sys)' and
'SNAP $+$ SNf (all $+$ sys2)', and the associated discussion at the end of \S \ref{everything};
see also Fig. \ref{fig:contour.1.combo}) -- 
we show that even in the presence of systematic error of the assumed magnitudes, large scale structure
fluctuations remain a non-negligible source of errors for dark energy measurements. 
It would obviously be useful to investigate this further and explore a wider range of
systematic errors suitable for each SN experiment.

{\bf 5.} An implicit assumption in our calculations is that the redshift measurements of
low $z$ SNe are sufficiently accurate for us to worry about their peculiar motion in the first place.
Existing low $z$ measurements typically report an accuracy of $\delta z \sim 0.001 - 0.002$
(e.g. \cite{hamuy96}). The spectral instrument of the SNfactory has a resolution of $1200$,
corresponding to $\delta z \sim 0.001$. 
(Note that the actual redshift accuracy is likely to be better than
the instrumental spectral resolution, so this is a conservative estimate.)
Translating into velocities, we are talking about
a velocity of $300 - 600$ km/s, or a magnitude fluctuation of $\delta m \sim 0.04 - 0.08$
(for $z \sim 0.05$). 
This is still smaller than the intrinsic magnitude scatter that
we assume: $\sigma^{\rm intr.} = 0.1$ or $0.15$. The redshift uncertainty adds
to the Poissonian scatter, and our range of $0.1$ to $0.15$ can be thought of as accounting
for this possibility already. It is important, however, that the redshift measurements
do not suffer from a systematic bias (that affects all SNe in the same way). 
From Fig. \ref{fig:testcvelB}, it can be seen that
a systematic bias of $\delta z \sim 0.0003$ or $100$ km/s (at $z \sim 0.05$) 
would have a comparable effect as coherent peculiar motion.

{\bf 6.} An interesting question is: to what extent can corrections be made for
the velocity and lensing fluctuations? 
For instance, one could imagine using galaxy weak lensing maps
to correct for the magnification of SNe. This has been shown to
be not viable, or not sufficiently accurate to be useful, by \cite{dalal}. 
This is because galaxy weak lensing maps typically tell us the magnification
on scales larger than are relevant for the Poissonian part of SN lensing. 
(These maps can be useful for correcting the non-Poissonian/coherent part of SN lensing,
but this part of lensing is not very important for most SN surveys anyway.)
More recently it was argued by \cite{gunnarsson} that corrections for 
(the Poissonian part of) SN lensing can
be made by modeling foreground galaxies as isothermal spheres or generalizations thereof.
One should keep in mind that the Poissonian lensing fluctuations are sensitive to
structures on relatively small scales ($k \gsim 10$ h/Mpc), and a smooth halo profile
does not necessarily capture all the relevant fluctuations. For instance, in
the case of strong lensing, substructures are often invoked to explain the observed
flux ratios \cite{nealchris}. Most of the high $z$ SNe will not be strongly lensed, but
a similar lesson applies here. Moreover, using the foreground
galaxies to make a magnification correction inevitably involves assumptions about
bias: how galaxies trace mass. 
As can be seen from Fig. \ref{fig:testcvellensB}, $\sigma^{\rm lens}/\sqrt{N} \sim 0.01$ for SNe
at $z \sim 1.5$ and $N = 100$ (per $\Delta z$ of $0.1$, as appropriate for SNAP for instance). 
For the lensing correction to be useful, it should therefore satisfy two criteria: first,
the correction should be more accurate than $0.1$ (in magnitude) per SN; second, 
it should not introduce a systematic bias that is larger than $0.01$ (in magnitude;
magnitude fluctuation $\sim$ magnification fluctuation). 
Even if the first can be achieved, the second seems challenging.

How about corrections for velocity fluctuations? Here, the situation is slightly different: 
what needs to be corrected is the coherent part (i.e. large scale), not the Poissonian part, of the fluctuations.
Roughly speaking, we need to know the low order multipoles of the peculiar flow
at the redshift of e.g. the SNfactory ($z \sim 0.055$). 
One option is to use peculiar velocity surveys (such as from the SNe themselves), but it should be kept
in mind that to disentangle the Hubble flow from say the monopole, one needs a survey that has
the same sky coverage as the SNfactory, but is deeper. This requires considerable resources.
Another option is to use the galaxy spatial distribution as a guide, i.e using mass conservation to relate
peculiar velocity to the galaxy overdensity. Such a procedure of course suffers from the
uncertain biasing relation between galaxies and mass. Note also that one needs a galaxy survey
that is deeper than the SNfactory to define the correct mean galaxy density.
(For a recent paper that examines the peculiar motions predicted by the PSCz survey, see \cite{john};
it focuses on peculiar flows at slightly lower redshifts than we need.)
Whether either option allows us to take out the effect of bulk flows to sufficient accuracy
is a question we would like to address in the future.

{\bf 7.} Another natural and interesting question is: to what extent can the noise here, due to lensing and peculiar
motion, be viewed as a useful signal? 
In the case of lensing, the issue is discussed in several recent papers \cite{williams,wang,menard,cooray,dodelson}.
In general, it is difficult for SNe to be competitive with galaxies as the sources for
weak gravitational lensing experiments. Consider for instance the measurement of the convergence
power spectrum: the shot-noise in the case of SNe is $(\sigma^{\rm intr.})^2/(4n)$ (the factor
of $4$ comes from $\delta m \sim 2 \kappa$ where $\kappa$ is convergence), while the shot-noise
in the case of galaxies is $(\sigma^\gamma)^2/(2n)$ (the factor of $2$ comes from the use of two components
of shear to estimate $\kappa$). 
Here the intrinsic magnitude scatter $\sigma^{\rm intr.}$ is roughly $0.1 - 0.15$, the
shape noise $\sigma^\gamma$ is about $0.3$, and the surface density $n$ is approximately $0.04$ per square 
arcminute for SNe
(taking numbers from SNAP), while $n \sim 30$ per square arcminute for a typical weak lensing galaxy survey.
The shot-noise from SNe is simply too big compared to that from galaxies. 
Nonetheless, the lensing of SNe is free from certain systematic errors that might affect the lensing of galaxies, 
such as intrinsic alignment, and so the SN method still provides a useful, though not terribly stringent, 
consistency test. 

How about the SN peculiar motion as a signal? 
For some of the earlier work on this issue, see e.g. \cite{hamuy96,riess95,riess97,idit}.
Among the different methods for measuring peculiar velocities (see \cite{strauss} for a review),
SNe Ia constitute the most accurate distance indicator 
on an object by object basis. For instance, 
SNe Ia yield distances with an error of $\sim 5 - 7\%$, while 
Tully-Fisher distances are typically uncertain
at the $15 - 20 \%$ level. On the other hand, 
Tully-Fisher galaxy catalogs 
(e.g. \cite{willick}) typically have significantly more objects than SN surveys, and
therefore have perhaps more statistical power.
Yet, SN surveys might suffer less from systematic errors that seem to have plagued at least 
some Tully-Fisher galaxy catalogs (e.g. \cite{davis}), and SN surveys generally go deeper.
It remains an interesting question to what extent competitive cosmological constraints can be obtained
from the peculiar motion of SNe. We hope to explore this in the future.

\acknowledgments

We thank Arlin Crotts and Zoltan Haiman for discussions.
We are grateful to Asantha Cooray, Josh Frieman,
Mike Hudson, Dragan Huterer, 
Saurabh Jha, Bob Kirshner, John Lucey, Ramon Miquel, David Polarski and Adam Riess
for comments on our manuscript and/or pointing out references to us.
Research for this work is supported in part by the DOE,
DE-FG02-92-ER40699, and by the the Columbia 
University Academic Quality Fund through ISCAP.

\appendix

\section{On Errorbars}
\label{app:fisher}

Let us briefly describe how to go from the Fisher matrix to errorbars by giving some examples.
Labeling the parameters $w_{\rm pivot}$, $w_a$, $\Omega_{\rm de}$ and $M$ as $p_1$ to $p_4$,
the (rms) errorbar on $w_{\rm pivot}$ marginalized over everything else with no prior is
given by $\sqrt{[F^{-1}]_{11}}$. 
If a prior on $\Omega_{\rm de}$ is desired, say
with rms $\delta \Omega_{\rm de} = 0.03$, the marginalized errorbar on $w_{\rm pivot}$
is equal to $\sqrt{[(F + G)^{-1}]_{11}}$, where $G$ is a diagonal matrix with all zero-entries
except for $G_{33} = 1/0.03^2$. 
On the other hand, the error on $w_{\rm pivot}$ keeping everything else fixed is
$\sqrt{1/F_{11}}$, which is generally smaller than the marginalized error $\sqrt{[F^{-1}]_{11}}$.
For instance, in this paper, we have occasion to study $\sqrt{1/F_{44}}$, the error on $M$ keeping
everything else fixed.

It is also useful to elaborate the nature of the pivot in the parametrization:
$w(a) = w_{\rm pivot} + w_a (a_{\rm pivot} - a)$. Suppose we start with choosing $a_{\rm pivot} = 1$,
and from the Fisher matrix, after appropriate marginalization over $\Omega_{\rm de}$ and $M$ (with or without prior), arrive
at a covariance matrix for the errors on $w_{\rm pivot}$ and $w_a$ of the form: 
$Q_{11} = \langle \delta w_{\rm pivot}^2 \rangle$,
$Q_{12} = Q_{21} = \langle \delta w_{\rm pivot} \delta w_a \rangle$ and $Q_{22} = \langle \delta w_a^2 \rangle$.
For instance, if we marginalize over both $\Omega_{\rm de}$ and $M$ with no prior, 
one can obtain $Q$ from inverting the full Fisher matrix: $Q_{11} = [F^{-1}]_{11}$, $Q_{12}=[F^{-1}]_{12}$
and $Q_{22} = [F^{-1}]_{22}$.
In general, the choice of $a_{\rm pivot} = 1$ would not lead to a vanishing $Q_{12}$.
Choosing a different $a_{\rm pivot}$ from the original $a_{\rm pivot} = 1$ corresponds to a simple
linear transformation on the parameters $w_{\rm pivot}$ and $w_a$. Our goal is to choose a new $a_{\rm pivot}$ such that
the errors on $w_{\rm pivot}$ and $w_a$ are uncorrelated. 
This can be accomplished by choosing $a_{\rm pivot} = 1 + Q_{12}/Q_{22}$.
With this choice, the resulting error on
$w_{\rm pivot}$ (marginalized over $w_a$) becomes $\sqrt{Q_{11} - Q_{12}^2/Q_{22}}$,
while the error on $w_a$ (marginalized over $w_{\rm pivot}$) remains the
same i.e. $\sqrt{Q_{22}}$. It can be shown that this marginalized error on 
$w_{\rm pivot}$ is exactly the same as the error on $w$ if $dw/da$ were fixed (to be zero, say)
\cite{wayne}.
This is easiest to see if there are only two parameters involved: $w_{\rm pivot}$ and $w_a$.
With our choice of $a_{\rm pivot}$, $Q$ is turned diagonal and so is its inverse (the Fisher matrix). 
In such a case, marginalization over $w_a$ is equivalent to fixing $w_a$ (to be zero for instance).
It is straightforward to generalize this to the case of more parameters (e.g. including $\Omega_{\rm de}$ and $M$).
In other words, the error on $w_{\rm pivot}$, marginalized over $w_a$, $\Omega_{\rm de}$ and $M$,
is exactly the same as the error on a constant equation of state, marginalized over $\Omega_{\rm de}$ and $M$.
This is a useful fact to know because some experiments quote errors
on $w$ assuming a vanishing $dw/da$.

For completeness, the following relation is useful: the background dark energy density scales with
redshift as $\rho \propto (1+z)^{3(1+w_{\rm pivot} + w_a)} {\,\rm exp\,}[-3 w_a z/(1+z)]$
for the parametrization $w(a) = w_{\rm pivot} + w_a (1 - a)$. This follows from
energy momentum conservation $d(\rho a^3) = - P d(a^3)$, and $P = w(a) \rho$.

\section{The Magnitude Covariance Matrix I -- Generalities}
\label{app:poisson}

Our goal is to derive eq. (\ref{tC}) for the magnitude covariance matrix $\tilde C_{ij}$,
as well as eq. (\ref{sigP}) and (\ref{Cij}) for its individual components.

The average magnitude $m_i$ from a redshift bin $i$ is:
\begin{eqnarray}
\label{mi}
m_i = {1 \over N_i} \sum_a m^0_a \Theta_{ai}
\end{eqnarray}
where $N_i$ is the number of SNe in that bin, $m^0_a$ denotes the magnitude
of individual SN labeled by $a$ (instead of the {\it averaged} magnitude as in $m_i$),
$\Theta_{ai}$ equals $1$ if SN $a$ falls within bin $i$ and vanishes otherwise.
The summation of $a$ is over all SNe in one's survey. 
Note $N_i = \sum_a \Theta_{ai}$.

From above, we have
\begin{eqnarray}
\label{dmdm}
\tilde C_{ij} = \langle \delta m_i \delta m_j \rangle
= {1\over N_i N_j} \sum_{a,b} \langle \delta m^0_a \delta m^0_b \rangle \Theta_{ai} \Theta_{bj} 
\end{eqnarray}
The summation over $a$ and $b$ can be split into two terms:
$a=b$ and $a\ne b$. The $a=b$ term gives:
\begin{eqnarray}
\label{aeqb}
{1\over N_i N_j} \sum_{a} \langle (\delta m^0_a)^2 \rangle \Theta_{ai} \Theta_{aj} 
= {\delta_{ij} \over N_i^2} \sum_{a} \langle (\delta m^0_a)^2 \rangle \Theta_{ai}
\end{eqnarray}
The magnitude variance of an individual SN, $\langle (\delta m^0_a)^2 \rangle$, 
has two contributions, one from intrinsic variations and the other from large scale structure
induced fluctuations. We usually treat the intrinsic part as independent of $a$. Therefore,
we have
\begin{eqnarray}
\label{pp}
\delta_{ij} {1\over N_i^2} \sum_{a} \langle (\delta m^0_a)^2 \rangle \Theta_{ai}
= \delta_{ij} { (\sigma^{\rm intr.})^2 + (\sigma_i^{\rm Poiss.})^2 \over N_i}
\end{eqnarray}
where $(\sigma^{\rm intr.})^2$ is intrinsic, and
$(\sigma_i^{\rm Poiss.})^2$ is due to structures. The latter is defined to be
\begin{eqnarray}
(\sigma_i^{\rm Poiss.})^2 = {1\over N_i} \sum_a \langle (\delta m^0_a)^2 \rangle^{\rm Poiss.} \Theta_{ai}
\end{eqnarray}
which can be approximated by taking the continuum limit:
\begin{eqnarray}
\label{sigPapp}
(\sigma_i^{\rm Poiss.})^2 = \left[ {5 \over {\rm ln\,} 10} \right]^2 
\int {dz \over \Delta z_i} \langle [\delta_{d_L}(z)]^2 \rangle
\end{eqnarray}
This last expression makes use of the fact that $\delta m^0 = (5/{\rm ln\,}10) \delta_{d_L}$
for the large scale structure fluctuations (eq. [\ref{dLm}]), and that $\langle [\delta_{d_L}]^2 \rangle$ depends on redshift
but not angular position. 

Eq. (\ref{pp}) reproduces exactly the Poissonian terms in eq. (\ref{tC}), while
eq. (\ref{sigPapp}) matches eq. (\ref{sigP}).

It remains to study the $a\ne b$ contribution to eq. (\ref{dmdm}).
Once again, making use of $\delta m^0 = (5/{\rm ln\,}10) \delta_{d_L}$ for the large
scale structure induced fluctuations, we can approximate 
this by taking the continuum limit:
\begin{eqnarray}
\label{SNlss}
&& {1\over N_i N_j} \sum_{a \ne b} \langle \delta m^0_a \delta m^0_b \rangle \Theta_{ai} \Theta_{bj} \\ \nonumber
&& = \left[{5 \over {\,\rm ln\,} 10}\right]^2 \int {dz {d^2 \theta} dz' {d^2 \theta'} \over \Delta z_i A_i \Delta z_j A_j}
\langle \delta_{d_L} (z, \bm{\theta}) \delta_{d_L} (z', \bm{\theta'}) \rangle
\end{eqnarray}
There is no need to zero out the $z=z'$ and $\bm{\theta} = \bm{\theta'}$ contributions in the above integrals
because they are vanishingly small. The expression above matches the non-Poissonian term $C_{ij}$ in eq. (\ref{tC}) and
(\ref{Cij}).

If one wishes, one can add a term, in addition to those in eq. (\ref{sigPapp}) and (\ref{SNlss}), to
$\langle \delta m_i \delta m_j \rangle$ accounting for systematic errors.

\section{Luminosity Distance in a Perturbed FRW Universe}
\label{app:dL}

Our aim is to derive an expression for the luminosity distance fluctuation that
is accurate to first order, and show that among all first order terms, those
in eq. (\ref{ddL2}) dominate in realistic applications.
The derivation follows closely the one given by Pyne \& Birkinshaw \cite{pyne1,pyne2}.
The final expression agrees with that of \cite{pyne2}
in the case of an Einstein de-Sitter universe. It also agrees with the earlier
result from Sasaki \cite{sasaki} who also gave an explicit expression for an Einstein de-Sitter universe.
It disagrees slightly with \cite{pyne2} for
a more general universe, but only in terms that are subdominant compared to the
lensing and velocity terms kept in eq. (\ref{ddL2}). The origin of this difference is discussed below.

%Derivation is taken from Notes 05 II.

For simplicity, we assume a flat universe.
The metric in a convenient gauge, the conformal Newtonian gauge, takes the form
\cite{ed}:
\begin{eqnarray}
\label{metric}
&& ds^2 = a^2 [- (1 + 2\phi) d\eta^2 + (1 - 2\phi) \sum_i (dx^i)^2] \\ \nonumber
&& \quad = a^2 [- (1 + 2\phi) d\eta^2 + (1 - 2\phi) (d\chi^2 + \chi^2 d\Omega^2)]
\end{eqnarray}
where $\phi$ is the scalar perturbation which
is also the gravitational potential, $\eta$ is the conformal time, $a(\eta)$ is the scale factor,
$x^i$ is the Cartesian comoving coordinate ($i = 1,2,3$),
$\chi$ is the radial comoving distance, and $d\Omega^2$ is the usual angular part of the metric.
We ignore vector and tensor fluctuations and assume zero anisotropic stress.

We will take the path of first working out the angular diameter distance, and then
using the general relation eq. (\ref{dLdA}) to obtain the luminosity distance.

The null geodesics can be most easily worked out by ignoring the
factor of $a^2$ in $ds^2$, since it is just an overall conformal factor.
For such a metric ($ds^2/a^2$), the affine connection components are
\begin{eqnarray}
\label{affine}
&& {\Gamma^0}_{00} = \phi_{,\eta} \quad 
{\Gamma^0}_{ij} = - \phi_{,\eta} \delta_{ij} \quad
{\Gamma^0}_{0i} = {\Gamma^i}_{00} = \phi_{,i} \\ \nonumber
&& {\Gamma^i}_{0j} = - \phi_{,\eta} \delta_{ij} \quad
{\Gamma^i}_{jk} = - \delta_{ij} \phi_{,k} - \delta_{ik} \phi_{,j} + \delta_{jk} \phi_{,i}
\end{eqnarray}
We use Latin indices to denote the spatial components.
Note that even though we work out the null geodesics in the rescaled metric $ds^2/a^2$,
all subsequent manipulations (i.e. subsequent to eq. [\ref{detadx}]) are done
with the original metric $ds^2$.

Let $\lambda$ be the affine parameter, and let us split the photon path
into background and first order pieces:
\begin{eqnarray}
\label{split}
\eta (\lambda) = \bar\eta (\lambda) + \delta \eta (\lambda) \, , \,
x^i (\lambda) = {\bar x}^i (\lambda) + \delta x^i (\lambda)
\end{eqnarray}

The zero-order background solution is just a straight line
\begin{eqnarray}
\label{straight}
\bar\eta (\lambda) = \eta_0 - \lambda \, , \,
{\bar x}^i (\lambda) = \lambda {n}^i 
\end{eqnarray}
where $\eta_0$ is the conformal time today, and ${n}^i$ is a
unit vector pointing away from the observer. 
Note the slightly strange choice of the sign for $\lambda$: it increases
as one traverses backward in time along the path of the photon. 
The choice is perfectly acceptable since the geodesic equation is invariant
under a sign flip for $\lambda$.

%notes 05 II p. 8

The first order geodesic equation can be integrated to give
\begin{eqnarray}
\label{detadx}
&& \delta \eta (\lambda) = 2 \int_0^\lambda d\lambda' \phi
+ 2\int_0^\lambda d\lambda' (\lambda - \lambda') \phi_{,\eta}
\\ \nonumber
&& \delta x^i (\lambda) = - 2\int_0^\lambda d\lambda' (\lambda - \lambda') [\partial_i - {n}^i {n}^j \partial_j] \phi
\\ \nonumber
&& \quad \quad - 2\int_0^\lambda d\lambda' (\lambda-\lambda') \phi_{,\eta} {n}^i
\end{eqnarray}
where we have chosen the boundary conditions $\delta x^i(0)=\delta\eta(0)=d\delta x^i/d\lambda (0)=0$,
with $\lambda=0$ denoting the location of the observer.
The condition on the photon direction at the observer ($dx^i/d\lambda (0) = d{\bar x}^i/d\lambda (0) = {n}^i$) 
implies $d\delta\eta/d\lambda - 2\phi$ vanishes at the observer by nullness of the geodesic.
Note for objects with Latin (spatial) indices, we are being cavalier about
their placement upstairs or downstairs: e.g. $n_i = n^i$ is the $\mu=i$-th component of $n^\mu$.

Note that the photon 4-momentum in the original metric $ds^2$ is given by
\begin{eqnarray}
\label{4mom}
k^\mu = {1\over a^2} {dx^\mu \over d\lambda}
\end{eqnarray}
The factor of $1/a^2$ is necessary because the affine parameter $\lambda$ was chosen in 
the rescaled metric $ds^2/a^2$. See \cite{wald} for a proof.

We are interested in a cone of light emanating from the observer back towards the emitter
i.e. the observer is located at the tip of the cone. Suppose the center of this cone
points in the (fiducial) direction ${\bf {\bar n}}$ (at the observer). A photon moving along this direction
follows a geodesic described by the above equations (\ref{straight}) and (\ref{detadx}), with
${\bf n} = {\bf {\bar n}}$. Consider a photon moving along the surface of this cone i.e.
its path direction at the observer differs slightly from the fiducial direction,
say ${\bf n} = {\bf {\bar n}} + \delta {\bf n}$. Such a photon follows a geodesic also
described by eq. (\ref{straight}) and (\ref{detadx}), but with ${\bf n} = {\bf {\bar n}} + \delta {\bf n}$.

Our strategy is to work out $x^\mu(\lambda)$ for these two paths which
tells us how the light cone gets deformed as $\lambda$ moves away from zero (the observer)
and allows us to compute the angular diameter distance.

Let's define the deviation $w^\mu$:
\begin{eqnarray}
\label{deviation}
&& w^\mu (\lambda, {\bf {\bar n}}) \equiv x^\mu (\lambda, {\bf {\bar n}} + \delta {\bf n}) - x^\mu (\lambda, {\bf {\bar n}})
\\ \nonumber
&& = \left[ {\partial x^\mu (\lambda, {\bf n}) \over \partial n^j} \right]_{{\bf n} = {\bf \bar n}} \delta n^j
\end{eqnarray}
where we have added the argument ${\bf n}$ to $x^\mu$ to remind ourselves that
these are geodesics that start out at the observer with a particular direction.

We are interested in the area spanned by $w^\mu$ at the emitter. 
More precisely, it is the area in a two-dimensional space perpendicular to both
$k^\mu$ (for the fiducial ray) at the emitter and the four velocity of the emitter.
Let us define a projection operator ${H^{\mu}}_\nu$ \cite{pyne1} that projects onto this space:
\begin{eqnarray}
\label{wperp}
w_\perp^\mu = {H^\mu}_\nu w^\nu
\end{eqnarray}

Suppose for simplicity that ${\bf \bar n} = (0,0,1)$ i.e. points in the $x^3$ direction.
Then $\delta {\bf n}$ points in the $x^1 - x^2$ plane. One can parametrize it by
\begin{eqnarray}
\label{cone}
\delta {\bf n} = (\epsilon {\,\rm sin}\psi, \epsilon {\,\rm cos}\psi, 0)
\end{eqnarray}
where $\psi$ can take any value from $0$ to $2\pi$, and $\epsilon$ basically
defines the size of the light cone at the observer's position
(though the observer sees a light cone with a different size if she/he has
a non-zero peculiar velocity; see below).

Now, consider how $w_\perp^\mu$ varies as we vary $\psi$:
\begin{eqnarray}
\label{conedeformed}
&& w_\perp^\mu = {H^\mu}_\nu \left[ {\partial x^\nu (\lambda, {\bf n}) \over \partial n^1} \right]_{{\bf n} = {\bf \bar n}} 
\epsilon {\,\rm sin}\psi \\ \nonumber 
&& \quad \quad
+ {H^\mu}_\nu \left[ {\partial x^\nu (\lambda, {\bf n}) \over \partial n^2} \right]_{{\bf n}= {\bf \bar n}} \epsilon {\,\rm cos}\psi
\end{eqnarray}
In the two-dimensional space where $w_\perp^\mu$ lives, this traces
out an ellipse. In other words, $w_\perp^\mu w_{\perp\mu}$ as a function of $\psi$ can
be rewritten in the form $p^2 {\,\rm sin}^2 (\psi - \psi_0) + q^2 {\,\rm cos}^2 (\psi - \psi_0)$
where $p$ and $q$ are the lengths of the major and minor axes. The area of such an ellipse is $\pi p q$.
Relating $p$ and $q$ back to the quantities in eq. (\ref{conedeformed}), it can
be shown that the area of the light beam at the emitter is
\begin{eqnarray}
\label{areaemit}
&& \delta A_e = \pi \epsilon^2 
\Big( [w_\perp^\mu]_{1} [w_{\perp\mu}]_{1} [w_\perp^\nu]_{2} [w_{\perp\nu}]_{2} \\ \nonumber
&& \quad -  ([w_\perp^\mu]_{1} [w_{\perp\mu}]_{2})^2 \Big)^{1/2}
\end{eqnarray}
where we have defined
\begin{eqnarray}
\label{wperpmore}
[w_\perp^\mu]_1 \equiv {H^\mu}_\nu \left[ {\partial x^\nu (\lambda, {\bf n}) \over \partial n^1} \right]_{{\bf n} = {\bf \bar n}} 
\\ \nonumber
[w_\perp^\mu]_2 \equiv {H^\mu}_\nu \left[ {\partial x^\nu (\lambda, {\bf n}) \over \partial n^2} \right]_{{\bf n} = {\bf \bar n}} 
\end{eqnarray}

Eq. (\ref{areaemit}) and (\ref{wperpmore}), together with the geodesics
worked out earlier, allow us to compute the light beam area at the emitter.
One fact simplifies the calculation a bit: the quantities
$[w_\perp^\mu]_1 [w_{\perp\mu}]_1$, 
$[w_{\perp}^{\mu}]_{2} [w_{\perp\mu}]_{2}$ and 
$[w_\perp^\mu]_1 [w_{\perp\mu}]_{2}$ can be computed accurately to
first order by just using the zero-order ${H^\mu}_\nu$.
The zero-order ${H^\mu}_\nu$ for our choice of
coordinates is simple, it is diagonal: ${H^\mu}_\nu = {\,\rm diag} (0,1,1,0)$.
To see this, note that the only possible place where we might
care about the first order part of ${H^\mu}_\nu$ is when it is
contracted with the zero order part of $[{\partial x^\nu (\lambda, {\bf n}) /\partial n^1}]$ or
$[{\partial x^\nu (\lambda, {\bf n}) /\partial n^2}]$, which is simple: $(0,\lambda,0,0)$ or 
$(0,0,\lambda,0)$ (see eq. [\ref{straight}]). This means the only part of
the first order piece of ${H^\mu}_\nu$ that we care about is ${H^\mu}_1$ or ${H^\mu}_2$. 
Moreover, in the quantities of interest such as
$[w_\perp^\mu]_1 [w_{\perp\mu}]_1 = {H^\mu}_\nu [{\partial x^\nu (\lambda, {\bf n}) /\partial n^1}]
{H^\alpha}_\beta [{\partial x^\beta (\lambda, {\bf n}) /\partial n^1}] g_{\mu\alpha}$,
suppose we are interested in the contribution from the first order piece of the first ${H^\mu}_\nu$ --
then only the zero-order pieces of the other terms need to be considered, and to zero-order,
both ${H^{\alpha}}_\beta$ and $g_{\mu\alpha}$ are diagonal (and ${H^\alpha}_\beta = {\,\rm diag\,}(0,1,1,0)$).
The upshot is that the only first order piece of ${H^\mu}_\nu$ that we need to worry about is
${H^1}_1$, ${H^2}_2$, ${H^1}_2$ or ${H^2}_1$. It can be shown explicitly that the first
order corrections to these vanish. Further discussions can be found in \cite{pyne1}.

%notes 05 I p. 120 - 122 has more details.

Putting everything together, we can substitute the zero-order ${H^\mu}_\nu$ and the
geodesic solution from eq. (\ref{straight}) and (\ref{detadx}) into eq. (\ref{wperpmore}), and
get:
%notes 05 II p. 31
\begin{eqnarray}
\label{wperpfull}
&& [w^1_\perp]_1 = \lambda + 2\int_0^\lambda d\lambda' (\lambda - \lambda') \phi_{,3} \\ \nonumber
&& \quad - 2 \int_0^\lambda d\lambda' (\lambda - \lambda') \lambda' \phi_{,11}
- 2 \int_0^\lambda d\lambda' (\lambda - \lambda') \phi_{,\eta} \\ \nonumber
&& [w^2_\perp]_1 = [w^1_\perp]_2 = -2\int_0^\lambda d\lambda' (\lambda - \lambda')\lambda' \phi_{,12} \\ \nonumber
&& [w^2_\perp]_2 = \lambda + 2\int_0^\lambda d\lambda' (\lambda - \lambda') \phi_{,3} \\ \nonumber
&& \quad - 2 \int_0^\lambda d\lambda' (\lambda - \lambda') \lambda' \phi_{,22}
- 2 \int_0^\lambda d\lambda' (\lambda - \lambda') \phi_{,\eta}
\end{eqnarray}
One useful tip in deriving the above is to remember that $\phi$ in the integrands
is not only an explicit function of $\lambda'$, but also an implicit function of ${\bf n}$
(i.e. $\partial \phi(\lambda')/\partial n^i = \lambda' \phi_{,i} (\lambda')$ to first order).
The result above can be put into eq. (\ref{areaemit}) to obtain the area at emission:
%notes 05 II p. 33
\begin{eqnarray}
\label{area}
&& \delta A_e = \pi \epsilon^2 a_e^2 (1-2\phi_e) \lambda_e^2 \\ \nonumber
&& \quad \Big( 1 - 4 \int_0^{\lambda_e} d\lambda' {\lambda_e - \lambda' \over \lambda_e} [-\phi_{,3} + \phi_{,\eta}]
\\ \nonumber
&& \quad - 2 \int_0^{\lambda_e} d\lambda' {(\lambda_e - \lambda')\lambda' \over \lambda_e}
[\phi_{,11} + \phi_{,22}] \Big)
\end{eqnarray}
where $\lambda_e$ is the affine parameter at the emitter, and
all quantities with a subscript $e$ are evaluated at emission.

To find the angular diameter distance, we also need to know the
solid angle of the beam according to the observer.
If the observer does not have peculiar motion, the answer is simple: from eq. (\ref{cone}), 
the solid angle is obviously $\pi \epsilon^2$. If the observer has peculiar motion,
the solid angle can be obtained by performing a simple boost:
\begin{eqnarray}
\label{solidangle}
\delta \Omega_0 = \pi \epsilon^2 (1 - 2 v_0^3)
\end{eqnarray}
where $v_0^3$ is the line of sight component of the observer's peculiar velocity i.e.
$dx^3/d\eta$. 

With the above two equations, all ingredients are in place to write
down the angular diameter distance and therefore also the luminosity distance
(eq. [\ref{dLdA}]):
\begin{eqnarray}
\label{dLinterm}
&& d_L(\lambda_e) =  a_e \lambda_e (1+z_e)^2 \Big( 1 
- \phi_e + v_0^3 \\ \nonumber
&& \quad - 2 \int_0^{\lambda_e} d\lambda' {\lambda_e - \lambda' \over \lambda_e} [-\phi_{,3} + \phi_{,\eta}] \\ \nonumber
&& \quad - \int_0^{\lambda_e} d\lambda' {(\lambda_e - \lambda')\lambda' \over \lambda_e}
[\phi_{,11} + \phi_{,22}] \Big)
\end{eqnarray}

We also need an expression for $1 + z_e$, the observed redshift of the emitter.
This can be obtained from the ratio $[k^\mu u_\mu]_e/[k^\nu u_\nu]_0$, where
$k^\mu$ is the photon momentum in eq. (\ref{4mom}) and $u^\mu$ is the 4-velocity of
the emitter/observer: $u^\mu = dx^\mu/d\tau = [(1-\phi)/a , {\bf v}/a]$.
Therefore
\begin{eqnarray}
\label{zlambdae}
1 + z(\lambda_e) = {1 \over a_e} \left[1 - [\phi - v^3]^e_0 - 2 \int_0^{\lambda_e} \phi_{,\eta} d\lambda' \right]
\end{eqnarray}
where $[Q]^e_0$ for some quantity $Q$ is defined to be $Q_e - Q_0$, the difference at emission and observation.

Ultimately, we are interested in the luminosity distance fluctuation
$\delta_{d_L} = [d_L (\lambda_e + \delta\lambda_e) - \bar d_L (\lambda_e)]/\bar d_L (\lambda_e)$, 
where $\bar d_L$ is the luminosity distance in an unperturbed universe, and $\delta\lambda_e$ is
chosen such that $1+z(\lambda_e + \delta\lambda_e) = 1+\bar z(\lambda_e)$ with $\bar z$ being the
redshift in the unperturbed universe (see Fig. \ref{fig:hubble}).

From eq. (\ref{zlambdae}), it can be shown that to first order
\begin{eqnarray}
\label{zdlambdae}
&& 1 + z(\lambda_e + \delta\lambda_e) = [1 + \bar z(\lambda_e)] \\ \nonumber
&& \Big( 1 + \left[{a' \over a}\right]_e (\delta\lambda_e - 2\int_0^{\lambda_e} \phi d\lambda' - 
2\int_0^{\lambda_e} (\lambda_e - \lambda') \phi_{,\eta} d\lambda' ) \\ \nonumber
&& - [\phi - v^3]^e_0 - 2 \int_0^{\lambda_e} \phi_{,\eta} d\lambda' \Big)
\end{eqnarray}
To derive this, it is useful to remember that the scale factor for affine parameter 
$\lambda_e + \delta\lambda_e$ is
equal to $a$ evaluated at the time $\bar\eta(\lambda_e + \delta\lambda_e) + \delta\eta(\lambda_e)$,
accurate to first order. The scale factor evaluated at $\bar\eta(\lambda_e)$ is of course equal to
$1/[1+\bar z (\lambda_e)]$. 

Similarly, from eq. (\ref{dLinterm}), one can show that
\begin{eqnarray}
\label{dLdlambdae}
&& d_L (\lambda_e + \delta\lambda_e) 
= a[\bar\eta (\lambda_e)] [1+z(\lambda_e +\delta\lambda_e)]^2 \lambda_e \\ \nonumber
&& \Big(1 - \left[{a' \over a}\right]_e (\delta\lambda_e - 2\int_0^{\lambda_e} \phi d\lambda' - 
2\int_0^{\lambda_e} (\lambda_e - \lambda') \phi_{,\eta} d\lambda' ) \\ \nonumber
&& - \phi_e + v_0^3 + {\delta\lambda_e \over \lambda_e} 
- 2 \int_0^{\lambda_e} d\lambda' {\lambda_e - \lambda' \over \lambda_e} [-\phi_{,3} + \phi_{,\eta}] \\ \nonumber
&& - \int_0^{\lambda_e} d\lambda' {(\lambda_e - \lambda')\lambda' \over \lambda_e} [\phi_{,11} + \phi_{,22}] \Big)
\end{eqnarray}

Choosing $\delta\lambda_e$ such that $1 + z(\lambda_e + \delta\lambda_e) = 1 + \bar z (\lambda_e)$
in eq. (\ref{zdlambdae}), and noting that 
$\bar d_L (\lambda_e) = a[\bar\eta(\lambda_e)] [1 + \bar z(\lambda_e)]^2 \lambda_e$,
we finally have from eq. (\ref{dLdlambdae}):
%notes 05 II p. 51
\begin{eqnarray}
\label{ddLorig}
&& \delta_{d_L} (z) = v_e^3 - \phi_0 - 2\phi_e - 2\int_0^{\lambda_e} d\lambda' \phi_{,\eta} \\ \nonumber
&& + {a_e \over \lambda_e a'_e} \left[ [\phi-v^3]^e_0 + 2\int_0^{\lambda_e} d\lambda' \phi_{,\eta} \right]
+ 4 \int_0^{\lambda_e} {d\lambda' \over \lambda_e} \phi \\ \nonumber
&& - \int_0^{\lambda_e} d\lambda' { (\lambda_e - \lambda')\lambda' \over \lambda_e} [\phi_{,11} + \phi_{,22}] \\ \nonumber
&& + 2\int_0^{\lambda_e} d\lambda' { \lambda_e - \lambda' \over \lambda_e} \phi_{,\eta}
\end{eqnarray}
where $z$ is the observed redshift in the perturbed universe.
All quantities with subscript $e$ are to be evaluated at $\lambda_e$; evaluating them
at $\lambda_e + \delta\lambda_e$ instead would only make a difference to $\delta_{d_L}$ to second order.

It is useful to rewrite the expression above in the following form:
%see notes 05 II p. 72
\begin{eqnarray}
\label{ddL}
&& \delta_{d_L} (z, {\bf n}) = {\bf v_e} \cdot {\bf n} 
- {1\over \chi_e} \left[{a\over a'}\right]_e ({\bf v_e} \cdot {\bf n} - {\bf v_0} \cdot {\bf n}) \\ \nonumber
&& - \int_0^{\chi_e} d\chi {(\chi_e - \chi)\chi \over \chi_e} \nabla^2 \phi(\chi) \\ \nonumber
&& - \phi_e + {1\over \chi_e} \left[{a\over a'}\right]_e (\phi_e - \phi_0) + 2\int_0^{\chi_e} {d\chi\over\chi_e} \phi(\chi) \\ \nonumber
&& + \int_0^{\chi_e} d\chi {(\chi_e - \chi)\chi \over \chi_e} \phi''(\chi) - 2\int_0^{\chi_e} d\chi {\chi_e - \chi \over \chi_e} \phi'(\chi)
\\ \nonumber
&& + {2\over \chi_e} \left[{a\over a'}\right]_e \int_0^{\chi_e} d\chi \phi'(\chi)
\end{eqnarray}
which is accurate to first order in fluctuations, and we have switched from 
${\bf \bar n}$ pointing in the $x^3$ direction to allow it to point anywhere
(and we have relabeled ${\bf \bar n} \rightarrow {\bf n}$).
Note that $\phi' \equiv \phi_{,\eta}$, $\phi'' \equiv \phi_{,\eta\eta}$ and $a' \equiv a_{,\eta}$.
The argument $\chi$ (radial comoving distance) of $\phi(\chi)$ is supposed to remind us that 
$\phi$ in general depends on position, as well as, implicitly, time i.e. under the line of sight integrals,
$\phi(\chi)$ should be evaluated at position $\chi$ and at a time when a photon is supposed to reach $\chi$.
Symbols with subscript $e$ denote evaluation at the time of photon emission and symbols
with subscript $0$ denote evaluation today.
One manipulation we have done is to replace $\phi_{,11} + \phi_{,22}$ in the integrands
by $\nabla^2 \phi - \phi_{,33}$ and use $\partial/\partial x^3 = d/d\lambda' + \partial/\partial\eta$. 
Also, we have replaced $\lambda_e$ by $\chi_e$ where $\chi_e$ is the zero-order comoving distance to
the emitter. Doing so is justified because fluctuations in $\chi_e$ make a difference
to the final expression above only to second order. To be precise:
\begin{eqnarray}
\label{chie}
\chi_e = \int_0^{z} dz'/H(z')
\end{eqnarray}
where $z$ is the {\it observed} redshift in the perturbed universe (same $z$ as the argument of
$\delta_{d_L}$ in eq. [\ref{ddL}]).

%notes 05 III p. 47

The above expressions (eq. [\ref{ddLorig}] \& [\ref{ddL}]) are consistent with eq. (5.16) of Sasaki \cite{sasaki}, who
gave an explicit expression for an Einstein-de-Sitter universe (adopting $\delta\eta_0 = 0$
in \cite{sasaki}). Pyne \& Birkinshaw's \cite{pyne2} eq. (32) can be most readily compared with our eq. (\ref{ddLorig}).
The two are identical, except that \cite{pyne2} did not have the last term on the
right hand side of our expression:
\begin{eqnarray}
2 \int_0^{\lambda_e} d\lambda' {\lambda_e - \lambda' \over \lambda_e} \phi_{,\eta} \\ \nonumber
\end{eqnarray}
It takes a little bit of work to trace back the origin of the difference.
Everything up to (and including) the expressions for $d_L$ and the redshift, eq. (\ref{dLinterm}) and (\ref{zlambdae}),
appears to be consistent with \cite{pyne2} (e.g. our eq. [\ref{area}] \& [\ref{solidangle}] are consistent
with their eq. 2 \& 4).
Where we start to differ appears to be in working out how 
$a[\eta(\lambda_e + \delta\lambda_e)] (\lambda_e + \delta\lambda_e)$ is related to
$a[\bar\eta(\lambda_e)]\lambda_e$ (see eq. [9] of \cite{pyne2}). We believe the correct relation (to first order) is
\begin{eqnarray}
\label{differus}
&& a[\eta(\lambda_e + \delta\lambda_e)] (\lambda_e + \delta\lambda_e) \\ \nonumber
&& = a[\bar\eta(\lambda_e + \delta\lambda_e) + \delta\eta (\lambda_e)] \lambda_e (1 + \delta\lambda_e/\lambda_e)
\\ \nonumber
&& = a[\bar\eta(\lambda_e)]\lambda_e \Big(1 + \left[a' \over a\right]_e [\delta\eta(\lambda_e) - \delta\lambda_e]
+ {\delta\lambda_e \over \lambda_e} \Big)
\end{eqnarray}
The combination $\delta\eta(\lambda_e) - \delta\lambda_e$ is equivalent to
$\eta(\lambda_e + \delta\lambda_e) - \bar\eta (\lambda_e)$, which was referred to as
$\delta\eta$ in \cite{pyne2}, let us denote this as $\delta\eta_{PB}$. 
Eq. (9) of \cite{pyne2} can be written as
\begin{eqnarray}
\label{PB}
&& a[\eta(\lambda_e + \delta\lambda_e)] (\lambda_e + \delta\lambda_e) \\ \nonumber
&& = a[\bar\eta(\lambda_e)]\lambda_e \Big(1 + \left[a' \over a\right]_e \delta\eta_{PB} + {\delta z_{PB} \over \lambda_e}\Big)
\end{eqnarray}
where $\delta z_{PB}$ was defined in eq. (6) of \cite{pyne2} as
$\delta z_{PB} = x^3 (\lambda_e + \delta\lambda_e) - {\bar x}^3 (\lambda_e)
= {\bar x}^3 (\lambda_e + \delta\lambda_e) - {\bar x}^3 (\lambda_e) + \delta x^3(\lambda_e)$.
Using eq. (\ref{straight}) and (\ref{detadx}), we can see that
$\delta z_{PB} = \delta\lambda_e + \delta x^3 (\lambda_e)$ and
$\delta x^3 (\lambda_e)= -2\int_0^{\lambda_e} d\lambda' (\lambda_e-\lambda') \phi_{,\eta}$. 
The issue is that \cite{pyne2} appears to have wrongly used $\delta z_{PB}$ instead of
$\delta\lambda_e$ as it should be (compare eq. [\ref{differus}] and [\ref{PB}]), and the two
differ by $\delta x^3(\lambda_e)$, which leads to exactly the difference of
$2\int_0^{\lambda_e} d\lambda' (\lambda_e-\lambda')/\lambda_e \phi_{,\eta}$ in $\delta_{d_L}$.
The source of the error in \cite{pyne2} seems to be confusion in the definition of
what they refer to as $r_e$. According to an earlier paper by the same authors \cite{pyne1}, $r_e$ should be equated with
$r(\lambda_e)$ where $r$ is the comoving radial distance in an unperturbed, rather than a perturbed, universe (see
eq. [6] and [29] of \cite{pyne1}). (Note that our $\lambda$ convention is opposite in sign from that in \cite{pyne1}:
we have $\bar x^i = \lambda n^i$ while they effectively have $\bar x^i = - \lambda n^i$, setting their $\lambda_0 = 0$.)

%notes 05 III p. 29

Eq. (\ref{ddL}) has quite a number of terms. 
They can be loosely divided into four categories: peculiar motion (first line), gravitational lensing (second line),
gravitational redshift (third line) and integrated Sachs-Wolfe (fourth and fifth lines).
This division is loose in a number of ways. For instance, some of what we call the integrated Sachs-Wolfe terms (those
that involve integrals over time derivatives of $\phi$) can actually be physically ascribed to gravitational lensing.
We call the second line the gravitational lensing term mainly because the literature often uses it as
the only contribution to the lensing convergence.
The first two terms of the third line, what we call gravitational redshift terms,
obviously mirror those for peculiar motion i.e. Doppler shift, but
the last term of the third line has no Doppler analog.

Fortunately, for most applications we can confine our attention to the peculiar motion and lensing terms (first two lines).
This is because we are generally interested in fluctuations on scales smaller than the horizon
-- the high redshift SN surveys generally cover a small fraction of the sky while the low redshift
surveys, even though they cover a significant fraction of the sky, do not extend out to a sufficient depth to
be sensitive to horizon scale fluctuations. 
In other words,
$H/k$ is a small number where $H$ is the Hubble constant, and $k$ is the wavenumber of interest
(strictly speaking, the correct quantity to look at is $aH/k$, but here $a \sim 1$).
For instance, the term involving $\phi''$ is obviously smaller than the lensing term involving
$\nabla^2 \phi$ i.e. $\phi''/\nabla^2 \phi \sim (H/k)^2 \ll 1$. The rate of change
of $\phi$, if non-zero, should be of the order of the inverse Hubble time $\sim H$. 
Similarly, terms like $\int d\chi \phi'$, $\int d\chi \phi' (\chi_e - \chi)/\chi_e$, or
$\int \phi d\chi/\chi_e$ are at most of the order of $\phi$, which in turn is smaller than $v$. 
The latter holds because $v \sim (k/H) \phi$ (for $a \sim 1$), from gravitational instability.

This is why for the purpose of this paper, we can 
consider only the peculiar motion and lensing contributions to
the luminosity distance fluctuation. This justifies the use of eq. (\ref{ddL2}).

\section{The Magnitude Covariance Matrix II -- Explicit Expressions in Terms of the Mass Power Spectrum}
\label{app:poisson2}

Our main task here is to derive explicit expressions for 
the lensing and velocity contributions to the (binned) magnitude covariance matrix, including
both the Poissonian and non-Poissonian terms  i.e. eq. (\ref{sigPfull0}), (\ref{sigPfull}), (\ref{Cijfull}), (\ref{cijvellarge})
and (\ref{Wvellarge}). At the end of this Appendix, we will also give expressions for the velocity
window function that are useful for more general survey geometries.

Let us work in the plane-parallel (small angle) approximation first. We will generalize some of the relevant
terms to large angles later.

The starting point is the luminosity distance fluctuation $\delta_{d_L}$ given in
eq. (\ref{ddL2}). We substitute this into eq. (\ref{sigP}) and (\ref{Cij}) to compute
$(\sigma_i^{\rm Poiss.})^2$ and $C_{ij}$. 
Let us first study the lensing contributions. The expressions are, up to normalization factors,
identical to those worked out in the lensing community for convergence variance (e.g. \cite{kaiser,scottsbook}), and so
our derivation here is very brief.
From eq. (\ref{ddL2}), using Limber's approximation
and Poisson's equation,
it can be shown that
\begin{eqnarray}
\label{lenscheck}
&& \langle \delta_{d_L} (z_i, \bm{\theta}) \delta_{d_L} (z_j, \bm{\theta'}) \rangle_{\rm lens}
= (3 H_0^2 \Omega_m/2)^2 \\ \nonumber
&& \quad \int_0^{{\rm min.}(\chi_i,\chi_j)} {d\chi \over a^2}  {(\chi_i - \chi)\chi \over \chi_i}
{(\chi_j - \chi)\chi \over \chi_j} \\ \nonumber
&& \quad \int {d^2 k_\perp \over (2\pi)^2} P(k=k_\perp, a) e^{-i {\bf k_\perp} \cdot \chi (\bm{\theta} - \bm{\theta'})} 
\end{eqnarray}
where $\chi_i$ and $\chi_j$ are the radial comoving distances (eq. [\ref{chie}]) to redshifts $z_i$ and $z_j$ respectively,
and $P(k,a)$ is the mass power spectrum at a scale factor $a$ corresponding to when the photon is at distance $\chi$.
Here ${\bf k_\perp}$ is the projection of ${\bf k}$ onto the plane perpendicular to the line of sight.

It is simple to set $\bm{\theta} = \bm{\theta'}$ and $i=j$ in the above expression and show 
from eq. (\ref{sigP}) that 
\begin{eqnarray}
&& (\sigma_i^{\rm Poiss.,\, lens})^2 = 
\left[{5 \over {\,\rm ln} 10}\right]^2
(3 H_0^2 \Omega_m/2)^2 \\ \nonumber
&& \int_0^{\chi_i} {d\chi \over a^2}  \left[{(\chi_i - \chi)\chi \over \chi_i}\right]^2
\int {d^2 k_\perp \over (2\pi)^2} P(k=k_\perp, a) 
\end{eqnarray}
provided the redshift bin width is not too large.
This reproduces the lensing part of the Poissonian variance in eq. (\ref{sigPfull}). 

We can similarly substitute eq. (\ref{lenscheck}) into eq. (\ref{Cij}) to obtain
the lensing term in the non-Poissonian correlation matrix in eq. (\ref{Cijfull}).
In doing so, it is useful to remember that eq. (\ref{lenscheck}) is a slow
function of $z_i$ and $z_j$, and the averages over redshifts within the respective bins in 
eq. (\ref{Cij}) can safely be replaced by the middle values, provided the redshift
bins are not too large. It is also useful to note that for a survey that spans a (not too large) circle
on the sky with angular radius $\theta_i^{\rm max.}$ (area $A_i$):
\begin{eqnarray}
\label{J0J1}
&& \int {d^2 \theta \over A_i} e^{-i {\bf k_\perp} \cdot \chi \bm{\theta}} \\ \nonumber
&& = {1\over \pi (\theta_i^{\rm max.})^2} \int_0^{\theta_i^{\rm max.}} \theta d\theta \int_0^{2\pi} d\alpha
e^{-i k_\perp \chi \theta {\,\rm cos}\alpha} \\ \nonumber
&& = {2 \over (k_\perp \chi \theta_i^{\rm max.})^2} \int_0^{k_\perp \chi \theta_i^{\rm max.}} dx x J_0(x)
= {2 J_1 (k_\perp \chi \theta_i^{\rm max.}) \over k_\perp \chi \theta_i^{\rm max.}}
\end{eqnarray}

The velocity analogs of the above can be derived by using mass conservation (to linear order):
\begin{eqnarray}
\delta' = -{\bf \nabla} \cdot {\bf v}
\end{eqnarray}
where $\delta$ is the density fluctuation $\delta\rho/\rho$.
This tells us that in linear theory and
Fourier space: $v^j ({\bf k}) = i (D'/D) (k^j / k^2) \delta ({\bf k})$, where $D$ is the 
linear growth factor and $D'$ is its derivative with respect to conformal time.
Therefore, we have
\begin{eqnarray}
\label{vv}
&& \langle v^3 ({\bf x_1}) v^3 ({\bf x_2}) \rangle
= \\ \nonumber
&& \quad \int {d^3 k \over (2\pi)^3} {(k_z)^2 \over k^4} D'_1 D'_2 P(k, a=1) 
e^{-i {\bf k} \cdot ({\bf x_1} - {\bf x_2})}
\end{eqnarray}
where we have adopted the $z$ (or third) direction
as the line of sight direction. 
The derivatives of the growth factor $D'_1$ and $D'_2$
are evaluated at redshifts corresponding to the positions
${\bf x_1}$ and ${\bf x_2}$.

The velocity two-point correlation above, together with 
the relation between ${\bf v}$ and $\delta_{d_L}$ in eq. (\ref{ddL2}), 
and the relation between $\sigma_i^{\rm Poiss.}$ and $\delta_{d_L}$ in eq. (\ref{sigP}), 
gives us the velocity contribution to the Poissonian magnitude variance, if
we set ${\bf x_1} = {\bf x_2}$:
\begin{eqnarray}
\label{sigPfullapp}
&& (\sigma_i^{\rm Poiss.,\, vel.})^2 =
\left[{5 \over {\,\rm ln \,} 10}\right]^2
(1 - {a_i \over a'_i \chi_i})^2 \\ \nonumber
&& \quad \int {d^3 k \over (2\pi)^3} {(k_z)^2 \over k^4} (D'_i)^2
P(k, a=1)
\end{eqnarray}
which is consistent with eq. (\ref{sigPfull}). Note that
as in the case of lensing, we assume that the redshift bin is sufficiently narrow that
factors like $a$, $\chi$, $D$ and so on do not vary much across the bin, and so one can use
their values at the center of the bin (i.e. at $z=z_i$) to substitute for what should strictly speaking
be bin-averages. 
In deriving the above, we have ignored the $v_0^3$ term in the luminosity
distance fluctuation (eq. [\ref{ddL2}]):
$\delta_{d_L} = v^3_e - (a/a'/\chi)_e (v^3_e - v^3_0)$ (ignoring the lensing term for now).
This is justified to the extent that {\it our} peculiar motion is fairly well constrained
by the microwave background dipole, and so one could eliminate its influence on
luminosity distance measurements. 
The other extreme would be to allow $v^3_0$ to be a stochastic variable just like
$v^3_e$ in which case the velocity contributions to the magnitude variance
would be even higher than our estimate.
There is one more subtlety: even in the case where $v_0^3$ is known precisely,
our computation of $(\sigma_i^{\rm Poiss.,\, vel.})^2$ is strictly speaking only
approximate. We have essentially computed
$[5/{\,\rm ln\,}10]^2 [1 - (a/a'/\chi)_e]^2 \langle (v^3_e)^2 \rangle$, whereas the correct thing to do
is to compute $[5/{\,\rm ln\,}10]^2 [1 - (a/a'/\chi)_e]^2 [\langle (v^3_e)^2 | v^3_0 \rangle
- \langle v^3_e | v^3_0 \rangle^2 ]$, because there is some correlation between
$v^3_0$ and $v^3_e$ ($\langle v_e^3 | v_0^3 \rangle$ is the
expectation value of $v_e^3$ given $v_0^3$). If the emitter and the observer are sufficiently
far apart, $\langle (v^3_e)^2 | v^3_0 \rangle \sim \langle (v^3_e)^2 \rangle$.
The term $\langle v^3_e | v^3_0 \rangle^2$ for Gaussian random fluctuations
is equal to $[ \langle v^3_e v^3_0 \rangle / \langle (v^3_0)^2 \rangle ]^2 (v^3_0)^2$.
Approximating $(v^3_0)^2$ by $\langle (v^3_0)^2 \rangle$, one can see that
our computation of $(\sigma_i^{\rm Poiss.,\, vel.})^2$ should be fairly accurate if 
$\langle v^3_e v^3_0 \rangle^2 / [\langle (v^3_0)^2 \rangle \langle (v^3_e)^2 \rangle]
\ll 1$. 
We have checked that in all cases we have considered in this paper, the corrections due
to the correlation between ${\bf v_e}$ and ${\bf v_0}$ are negligible.
Similar, but slightly modified, arguments apply to the non-Poissonian velocity terms 
(e.g. the monopole and so on) considered below.
The relevant quantity in that case is
$\langle [{\bf v} ({\bf x_i}) \cdot \hat {\bf x_i}] [{\bf v_0} \cdot \hat {\bf x_i}] \rangle 
\langle [{\bf v} ({\bf x_j}) \cdot \hat {\bf x_j}] [{\bf v_0} \cdot \hat {\bf x_j}] \rangle /
[\langle [{\bf v} ({\bf x_i}) \cdot  \hat {\bf x_i}] [{\bf v} ({\bf x_j})
\cdot \hat {\bf x_j}] \rangle \langle {\bf v_0} \cdot {\bf v_0} \rangle]$, where ${\bf x_i}$ and ${\bf x_j}$
are to be averaged over redshift bins $i$ and $j$ respectively; we have checked explicity
that this is also $\ll 1$ for all cases of interest in this paper.
%notes 05 III p. 119.

The velocity contribution to the non-Poissonian correlation matrix $C_{ij}^{\rm vel.}$ can be
worked out in an analogous manner, and one arrives at eq. (\ref{Cijfull}).
As before, we have made use of the fact that quantities like $a$, $D$, $\chi$ and so on
vary slowly across the redshift bin (but not quantities such as 
$e^{i {\bf k} \cdot {\bf x}}$).
The window function given in
eq. (\ref{Cijfull}) is for volumes that are top-hat in the z-direction
and circular in the transverse direction. Note that if $i\ne j$, 
strictly speaking the window function $W_{ij}^{\rm vel.}$ should have
an imaginary piece proportional to $i {\,\rm sin\,}[k_z (\chi_i - \chi_j)]$, but
such a piece doesn't contribute to the Fourier integral because it is odd in $k_z$.

The reader might wonder whether there should also be a cross-term that involves
the product of velocity-induced and lensing-induced fluctuations.
Such a term is in principle possible, but the lensing projection sends $k_z$ to zero
(Limber approximation) while the velocity contribution is proportional to $k_z$, making
the cross-term identically zero.
(A similar statement holds even in the large angle case studied below: 
essentially, the cross velocity-lensing term would involve an integral
of $j_\ell (k\chi)$ over a large distance $\chi$, making it very small.)

Finally, for realistic applications, we need to generalize the above discussion for velocities
to large angles. This is because low redshift SN surveys, where peculiar motion (but not lensing) is important,
typically cover a significant fraction of the sky.
A more general version of eq. (\ref{vv}) is
\begin{eqnarray}
\label{vvlargeangle}
&& \langle [{\bf v} ({\bf x_1}) \cdot {\bf \hat x_1}] [{\bf v} ({\bf x_2}) \cdot {\bf \hat x_2}] \rangle
= \\ \nonumber
&& \int {d^3 k \over (2\pi)^3} { ({\bf k} \cdot {\bf \hat x_1}) ({\bf k} \cdot {\bf \hat x_2}) \over k^4} D'_1 D'_2 P(k, a=1) 
e^{-i {\bf k} \cdot ({\bf x_1} - {\bf x_2})}
\end{eqnarray}
where ${\bf \hat x_1}$ and ${\bf \hat x_2}$ are unit vectors pointing towards the positions
${\bf x_1}$ and ${\bf x_2}$. 
It is straightforward to show that eq. (\ref{cijvellarge}) follows from the above expression plus
eq. (\ref{ddL2}) and eq. (\ref{Cij}).
It takes a little more work to derive the velocity window function $W^{\rm vel.}_{ij}$ given in 
eq. (\ref{Wvellarge}) for a circularly symmetric survey. The definition given in eq. (\ref{cijvellarge})
can be written as:
\begin{eqnarray}
&& W^{\rm vel.}_{ij} ({\bf k}) 
= \\ \nonumber
&& {1\over \chi_i^2 \Delta\chi_i \Delta\Omega_i} \int_{\chi_i - \Delta\chi_i/2}^{\chi_i + \Delta\chi_i/2} 
\chi^2 d\chi d\Omega 
e^{-i {\bf k} \cdot {\bf x}}  {\bf \hat k} \cdot {\bf \hat x} \\ \nonumber
&& {1\over \chi_j^2 \Delta\chi_j \Delta\Omega_j} \int_{\chi_j - \Delta\chi_j/2}^{\chi_j + \Delta\chi_j/2}  
{\chi'}^2 d\chi' d\Omega' 
e^{i {\bf k} \cdot {\bf x'}} {\bf \hat k} \cdot {\bf \hat x'}
\end{eqnarray}
where $\Delta\chi_i$, $\Delta\chi_j$ are the radial widths and
$\Delta\Omega_i$ and $\Delta\Omega_j$ are the solid angles.

A useful expansion of the plane wave is:
\begin{eqnarray}
e^{-i {\bf k} \cdot {\bf x}} = \sum_{\ell, m} 4\pi (-i)^\ell j_\ell (k\chi)
Y_{\ell m} ({\bf \hat k}) Y_{\ell m}^* ({\bf \hat x})
\end{eqnarray}
where $j_\ell$ is the spherical Bessel function and the $Y_{\ell m}$ are spherical harmonics.
From this, one can see that
\begin{eqnarray}
e^{-i {\bf k} \cdot {\bf x}} {\bf \hat k} \cdot {\bf \hat x} = \sum_{\ell, m} 4\pi (-i)^{\ell-1}
j'_\ell (k\chi) Y_{\ell m} ({\bf \hat k}) Y_{\ell m}^* ({\bf \hat x})
\end{eqnarray}
where $j'_\ell$ is the derivative of $j_\ell$ with respect to its argument.

Looking at eq. (\ref{cijvellarge}), one can see that eventually we integrate
$W^{\rm vel.}_{ij} ({\bf k})$ over ${\bf k}$ with other functions of $k$ (but not of ${\bf \hat k}$).
In other words, we can replace $W^{\rm vel.}_{ij} ({\bf k})$ by its average over
the solid angle of the wave vector.
One can then take advantage of the fact that
$\int d\Omega_k Y_{\ell m}^* ({\bf \hat k}) Y_{\ell' m'} ({\bf \hat k}) = \delta_{\ell \ell'} \delta_{m m'}$.
The integration over the solid angle $d\Omega$ of the survey can be done quite easily:
%notes 05 III p. 83
\begin{eqnarray}
&& \int d\Omega Y_{\ell m} ({\bf \hat x}) = 
\int d\Omega Y_{\ell 0} ({\bf \hat x}) \delta_{m 0}
\\ \nonumber
&& = 2\pi \sqrt{2\ell + 1 \over 4\pi} \int_0^{\theta^{\rm max.}_i} d\theta {\,\rm sin}\theta P_\ell ({\,\rm cos}\theta)
\delta_{m0}
\end{eqnarray}
The solid angle $\Delta\Omega_i$ equals $2\pi (1 - {\,\rm cos \,}\theta^{\rm max.}_i)$.
Putting all these together yields eq. (\ref{Wvellarge}) for the velocity window function
$W_{ij}^{\rm vel.}$, which is valid for a contiguous, circularly symmetric survey.

A slightly more general situation is one where the survey is divided into several different patches,
but each patch retains azimuthal symmetry (i.e. each patch spans from $\phi = 0$ to $\phi = 2\pi$; 
we assume there is no danger of confusing the azimuthal angular coordinate $\phi$ with the metric fluctuation $\phi$). 
This is precisely the kind of geometry we have adopted for the SNfactory, which
consists of two patches, one centered
at the north pole and the other at the south pole. For a survey like this, generalizing
eq. (\ref{Wvellarge}) a little bit, the velocity window function should be:
\begin{eqnarray}
\label{WvellargeSNf}
&& W^{\rm vel.}_{ij} (k) = 
\sum_{\ell=0}^\infty (2\ell + 1)  \\ \nonumber 
&& \,\, \left[ \int_{\chi_i - \Delta \chi_i/2}^{\chi_i + \Delta \chi_i} {d\chi\over \Delta\chi_i} j'_\ell (k\chi)\right]
\left[ \int_{\chi_j - \Delta \chi_j/2}^{\chi_j + \Delta \chi_j} {d\chi\over \Delta\chi_j} j'_\ell (k\chi)\right]
\\ \nonumber
&& \,\, \left[ {\int_i d\theta {\,\rm sin}\theta P_\ell ({\,\rm cos}\theta) 
\over \int_i d\theta {\,\rm sin}\theta}
\right] \left[ {\int_j d\theta' {\,\rm sin}\theta' P_\ell ({\,\rm cos}\theta') 
\over \int_j d\theta' {\,\rm sin}\theta'}
\right]
\end{eqnarray}
where the symbol $\int_i$ denotes integration over the patches that belong to
the $i$th redshift bin. For instance, if one has a survey that covers all sky aside from
a galactic cut of $\pm 30^0$, $\int_i$ should represent an integration of $\theta$ from
$0$ to $\pi/3$, and from $2\pi/3$ to $\pi$. Eq. (\ref{WvellargeSNf}), 
in place of eq. (\ref{Wvellarge}), is what we
use to obtain predictions for the SNfactory.

Generalizing further, suppose one has a survey where for each angle $\theta$, the azimuthal
angle $\phi$ spans $\phi_{\rm min.} (\theta)$ to $\phi_{\rm max.} (\theta)$. We will further
give $\phi_{\rm min.}$ and $\phi_{\rm max.}$ superscripts $i$ or $j$ to denote the fact that
these ranges can even vary depending on the redshift bin under consideration.
It can be shown that the velocity window function is given by:
\begin{eqnarray}
\label{WvellargeGeneral}
&& W^{\rm vel.}_{ij} (k) = 
\sum_{\ell=0}^\infty (2\ell + 1)  {\Delta \Omega_i}^{-1}  {\Delta \Omega_j}^{-1}  \\ \nonumber 
&& \,\, \left[ \int_{\chi_i - \Delta \chi_i/2}^{\chi_i + \Delta \chi_i} {d\chi\over \Delta\chi_i} j'_\ell (k\chi)\right]
\left[ \int_{\chi_j - \Delta \chi_j/2}^{\chi_j + \Delta \chi_j} {d\chi\over \Delta\chi_j} j'_\ell (k\chi)\right]
\\ \nonumber
&& \,\, 
\Big( \Big[ 
\int_i d\theta {\,\rm sin}\theta P_\ell ({\,\rm cos}\theta) (\phi_{\rm max.}^i - \phi_{\rm min.}^i)
\\ \nonumber 
&& \quad \int_j d\theta' {\,\rm sin}\theta' P_\ell ({\,\rm cos}\theta') (\phi_{\rm max.}^j - \phi_{\rm min.}^j)
\Big] \\ \nonumber
&& \quad + \quad 2 \sum_{m=1}^{\ell} {(\ell - m)! \over (\ell + m)!} {1\over m^2} \times \\ \nonumber 
&& \quad \Big[
\int_i d\theta {\,\rm sin}\theta P_\ell^m ({\rm cos}\theta) ({\,\rm sin}m\phi_{\rm max.}^i - {\,\rm sin}m\phi_{\rm min.}^i) 
\\ \nonumber
&& \quad 
\int_j d\theta' {\,\rm sin}\theta' P_\ell^m ({\rm cos}\theta') ({\,\rm sin}m\phi_{\rm max.}^j - {\,\rm sin}m\phi_{\rm min.}^j) 
\\ \nonumber
&& \quad + 
\int_i d\theta {\,\rm sin}\theta P_\ell^m ({\rm cos}\theta) ({\,\rm cos}m\phi_{\rm max.}^i - {\,\rm cos}m\phi_{\rm min.}^i) 
\\ \nonumber
&& \quad 
\int_j d\theta' {\,\rm sin}\theta' P_\ell^m ({\rm cos}\theta') ({\,\rm cos}m\phi_{\rm max.}^j - {\,\rm cos}m\phi_{\rm min.}^j) 
\Big]
\Big)
\end{eqnarray}
where
\begin{eqnarray}
\Delta \Omega_i = \int_i d\theta {\,\rm sin}\theta (\phi_{\rm max.}^i - \phi_{\rm min.}^i) \\ \nonumber
\Delta \Omega_j = \int_j d\theta {\,\rm sin}\theta (\phi_{\rm max.}^j - \phi_{\rm min.}^j) 
\end{eqnarray}
and $P_\ell^m$ denotes the associated Legendre polynomial.

\newcommand\spr[3]{{\it Physics Reports} {\bf #1}, #2 (#3)}
\newcommand\sapj[3]{ {\it Astrophys. J.} {\bf #1}, #2 (#3) }
\newcommand\sapjs[3]{ {\it Astrophys. J. Suppl.} {\bf #1}, #2 (#3) }
\newcommand\sprd[3]{ {\it Phys. Rev. D} {\bf #1}, #2 (#3) }
\newcommand\sprl[3]{ {\it Phys. Rev. Letters} {\bf #1}, #2 (#3) }
\newcommand\np[3]{ {\it Nucl.~Phys. B} {\bf #1}, #2 (#3) }
\newcommand\smnras[3]{{\it Monthly Notices of Royal
        Astronomical Society} {\bf #1}, #2 (#3)}
\newcommand\splb[3]{{\it Physics Letters} {\bf B#1}, #2 (#3)}

\newcommand\apjs{Astrophys. J. Suppl.}
\newcommand\aj{Astron. J.}
\newcommand\mnras{Mon. Not. R. Astron. Soc.~}
\newcommand\apjl{Astrophys. J. Lett.~}
\newcommand\etal{{\it et al.}}


\begin{thebibliography}{99}

%\bibitem{inflation}
%         A. H. Guth, Phys. Rev. D {\bf 23}, 347 (1981);
%         A.~D.~Linde, Phys.\ Lett. {\bf B108} 389 (1982);
%         A.~Albrecht \& P.~J.~Steinhardt, Phys. Rev. Lett {\bf48}, 1220 (1982).

\bibitem{evidence}
  For instance: for evidence from flatness from CMB, combined with matter density from
  galaxy surveys, see e.g. D.~N. Spergel \etal, \apjs 148 175 (2003);
  for evidence from the integrated Sachs-Wolfe effect, see e.g. 
  P. Fosalba, E. Gaztanaga, F. J. Castander, Astrophys. J. Lett. 597, 89 (2003),
  R. Scranton \etal, submitted to Phys. Rev. Lett. [astro-ph/0307335],
  M. R. Nolta \etal, \apj, 608, 10 (2004).

%\cite{Riess:1998}
\bibitem{Riess:1998}
  A.~G.~Riess \etal  [High-Z Supernova Search Team],
  %``Observational Evidence from Supernovae for an Accelerating Universe and a
  %Cosmological Constant,''
  Astron.\ J.\  116, 1009 (1998).
  %[arXiv:astro-ph/9805201].
  %%CITATION = ASTRO-PH 9805201;%%

%\cite{Perlmutter:1999}
\bibitem{Perlmutter:1999}
  S.~Perlmutter \etal  [Supernova Cosmology Project],
  %``Measurements of Omega and Lambda from 42 High-Redshift Supernovae,''
  Astrophys.\ J.\  517, 565 (1999).
  %[arXiv:astro-ph/9812133].
  %%CITATION = ASTRO-PH 9812133;%%

\bibitem{SNsurveys}
  An incomplete list includes:
  Carnegie Supernova Project (W. L. Freedman (2004) [astro-ph/0411176]; see also 
  {\tt http://csp1.lco.cl/$\sim$cspuser1/CSP.html}),
  CfA Supernova Program (S. Jha \etal (2005) [astro-ph/0509234]),
  DES Dark Energy Survey (Annis \etal, from {\tt http://decam.fnal.gov}, {\tt http://www.darkenergysurvey.org}),
  ESSENCE ({\tt http://www.ctio.noao.edu/$\sim$wsne}, T. Matheson \etal, \aj, 129, 2352 
 (2005), Sollerman \etal, (2005) [astro-ph/0510026]),
  JEDI (A. Crotts \etal, white paper submitted to the Dark Energy Task Force (2005) [astro-ph/0507043]),
  LOTOSS ({\tt http://astron.berkeley.edu/$\sim$bait/lotoss.html}),
  LSST ({\tt http://www.lsst.org}),
  Nearby Supernova Factory (G. Aldering \etal, Proceedings of the SPIE, 4836, 61 
 (2002), W. M. Wood-Vasey \etal, New Astronomy Review, 48, 637 (2004),
  W. M. Wood-Vasey, PhD dissertation [astro-ph/0505604]),
  Pan-STARRS (Article by J. Tonry at {\tt http://pan-starrs.ifa.hawaii.edu/public/home.html}),
  SDSS II (M. Sako \etal, Proceedings of the 22nd Texas Symposium (2004) [astro-ph/0504455]),
  SNAP (G. Aldering \etal, submitted to PASP (2004) [astro-ph/0405232], G. Aldering, Proceedings to the Wide-Field  
  Imaging from Space Conference (2005) [astro-ph/0507426]),
  SNLS as part of CFHTLS (P. Antilogus \etal, from {\tt http://cfht.hawaii.edu/SNLS/}, 
  Astier \etal (2005) [astro-ph/0510447]).

\bibitem{gravity}
	e.g. G. Dvali, G. Gabadadze \& M. Porrati, Phys. Lett. B 485, 208 (2000),
	C. Deffayet, G. Dvali \& G. Gabadadze, \prd 65, 4023 (2002),
	J. Khoury~\& A.~Weltman \prd 69, 044026 (2004),
	S. M. Carroll, V. Duvvuri, M. Trodden, M. S. Turner \prd 70, 3528 (2004).
  
\bibitem{josh0}
	J. A. Frieman, Comments on Astrophysics 28, 323 (1997).

%\bibitem{hutererturner}
%	D. Huterer \& M. S. Turner, \prd 60, 1301 (1999).

\bibitem{hutererturner}
	D. Huterer \& M. S. Turner, \prd 64, 123527 (2001).

\bibitem{josh}
	J. A. Frieman, D. Huterer, E. V. Linder \& M. S. Turner, \prd 67, 083505 (2003).

\bibitem{linder}
	E. V. Linder \& D. Huterer, \prd 67, 081303 (2003).

\bibitem{kim}
	A. G. Kim, E. V. Linder, R. Miquel \& N. Mostek, \mnras 347, 909 (2004) [astro-ph/0304509].

\bibitem{hutererkim}	
	D. Huterer, A. Kim, L. M. Krauss \& T. Broderick, \apj 615, 595 (2004).

\bibitem{hl}
	D. E. Holz~\& E. V. Linder, \apj 631, 678 (2005)
%D. E. Holz, E. V. Linder, submitted to ApJ (2004) [astro-ph/0412173].

\bibitem{sigintr}
	A magnitude spread of $0.15$ (rms) is probably closer to what is generally assumed/found
	(e.g. \cite{Riess:1998,Perlmutter:1999}), but
	a spread of $0.1$ is sometimes used (e.g. \cite{hl}).
	A recent survey, the SNLS (see \cite{SNsurveys}), found $\sigma^{\rm intr.} = 0.13$.
        It is conceivable that,
        with large samples of well measured SNe in the future, hitherto unknown correlations
	can be exploited to further standardize SNe, much like how the Phillips relation 
	\cite{phillips} is currently used.
	We thank Arlin Crotts for discussions on this point.

\bibitem{phillips} M. M. Phillips, \apjl 413, L105 (1993).	

\bibitem{sasaki} M. Sasaki, \mnras 228, 653 (1987).

\bibitem{pyne1} T. Pyne~\& M.~Birkinshaw, \apj 458, 46 (1996).

\bibitem{pyne2} T. Pyne \& M.~Birkinshaw, \mnras 348, 581 (2004).

\bibitem{sss99} N. Sugiura, N. Sugiyama \& M. Sasaki, Progress of Theoretical Physics Vol. 101 No. 4, 903 (1999).

\bibitem{strauss}
	M. A. Strauss \& J. A. Willick, Physics Reports 261, 271 (1995).

\bibitem{shi}
	X. Shi, \apj 486, 32 (1997).

\bibitem{shiturner}
	X. Shi \& M. S. Turner, \apj 493, 519 (1998).


\bibitem{idit}
  I.~Zehavi, A.~G.~Riess, R.~P.~Kirshner and A.~Dekel,
  %``A Local Hubble Bubble from SNe Ia?,''
  Astrophys.\ J.\  {\bf 503}, 483 (1998).
  %[arXiv:astro-ph/9802252].
  %%CITATION = ASTRO-PH 9802252;%%


\bibitem{cooray} A. Cooray, D. Holz \& D. Huterer, \apjl 637, 77 (2006) [astro-ph/0509579],
	A. Cooray, D. Huterer \& D. Holz, \prl 96, 021301 (2006) [astro-ph/0509581].

\bibitem{dodelson} S. Dodelson \& A. Vallinotto, preprint (2005) [astro-ph/0511086].

\bibitem{bdg} C. Bonvin, R. Durrer \& M. A. Gasparini, \prd 73, 023523 (2006) [astro-ph/0511183].

\bibitem{schucker} T. Schucker, I. ZouZou, preprint (2005) [astro-ph/0511521].

\bibitem{polarski} M. Chevallier \& D. Polarski, Int. J. Mod. Phys. D10, 213 (2001).

\bibitem{linder0} E. V. Linder, \prl 90, 130 (2003).

\bibitem{scottsbook} S. Dodelson, Modern Cosmology, Academic Press (2003). It reviews, among other things,
	the Fisher matrix and weak gravitational lensing.

\bibitem{lie}
	It is not quite accurate to say there is no prior: there is always some hidden prior. For instance,
	we assume the universe is flat. When we say 'no prior', we mean no prior on the parameters
	$w_{\rm pivot}$, $w_a$, $\Omega_{\rm de}$ and $M$.

\bibitem{wayne}
	A nice review of some of these issues can be found in W. Hu, Proceedings of Observing Dark Energy,
	edited by S. C. Wolff and T. R. Lauer (2004) [astro-ph/0407158].

\bibitem{wmap3}
	We adopt values for cosmological parameters that come from combining the
	WMAP three-year data and the SDSS survey: Spergel \etal , preprint, (2006) [astro-ph/0603449].

\bibitem{binning}
	The choice of binning is somewhat arbitrary. The important thing is to 
	avoid choosing bin sizes so large that one loses information.
	Beyond that, the final errorbars
	on parameters of interest are not sensitive to the precise choice
	of binning.
	When analyzing actual data, binning is in fact not strictly necessary.
	For us, it is just a convenient calculational device.

\bibitem{pfrw}
	It is impractical to list all papers on this subject. A recent paper that contains some
	relevant references is E. W. Kolb, S. Mattarese, A. Notari \& A. Riotto, \prd 71, 023524 (2005).

\bibitem{peebles}
	P. J. E. Peebles, Principles of Physical Cosmology, Princeton University Press (1993). See eq. 14.54 and 14.51.

\bibitem{diffway}
	The reader might wonder if it is possible to derive eq. (\ref{dLshift}) differently, using
	instead $F(z) = [L/(4\pi (1+z)^2)] [\delta\Omega_e /\delta A_0]$ \cite{peebles}, which like eq. (\ref{peebles})
        holds regardless of whether the universe is homogeneous.
	This implies $d_L (z) = (1+z) \sqrt{\delta A_0/\delta\Omega_e}$. Peculiar motion causes the
	redshift to change according to eq. (\ref{shiftshift}), and $\delta\Omega_e \rightarrow \delta\Omega_e
	(1 - 2 {\bf v_e} \cdot {\bf n})$. The net result is once again eq. (\ref{dLshift}).

\bibitem{asymmetry}
	The reader might wonder why the expression looks asymmetric:
	it depends separately on ${\bf v_e}$ and ${\bf v_0}$
	instead of depending only on their difference.
	Perhaps the best way to see this is through a simplified example.
	Consider light propagation in Minkowski space (no expansion). 
	A detector receives a bunch of photons at $(t,x,y,z)=(0,0,0,0)$ which were emitted at $(-d,d,0,0)$.
	Clearly the detected flux $F = L/(4\pi d^2)$. 
	Now consider a frame which is moving at velocity $-v$ in the $x$
	direction with respect to the above. The detection event and emission
	event are transformed to first order in $v$ to $(t',x',y',z')=(0,0,0,0)$
	and $(-d+vd,-vd+d,0,0)$. Let us use the relation in \cite{diffway}:
	$F = [L/(4\pi (1+z)^2)][\delta\Omega_e/\delta A_0]$, to work out the
	detected flux, which we know must equal the original: $L/(4\pi d^2)$.
	In the primed frame, the photons suffer no redshift because
	both the detector and the emitter are moving at the same velocity.
	Naively, one might think that $[\delta\Omega_e/\delta A_0]$ should
	just equal $1/(d-vd)^2$, since $d-vd$ is the actual distance 
	the photons have traveled between the emitter and the detector in the primed frame. 
	However, it is crucial to remember that $\delta\Omega_e \rightarrow
	\delta\Omega_e (1-2v)$. The reduction in distance and the 
	reduction in solid angle precisely cancel to give us back $F$ in the unprimed frame.
	More generally, one can see from the expression for $F$ in \cite{diffway} that
	the asymmetry in the general expression for $d_L$ originates from
	the fact that $\delta\Omega_e$ is affected only by the emitter's
	but not the observer's velocity.


\bibitem{kaiser}
	N. Kaiser, \apj 388, 272 (1992).

\bibitem{ned}
	For a recent discussion, see
	E. L. Wright, Lectures at the Frontiers of the Universe: Cosmology 2003 (2003)
	[astro-ph/0401001].

\bibitem{ehu}
%	G. Efstathiou, J. R. Bond \& S. D. M. White, \mnras 258, 1 (1992).
	D. J. Eisenstein \& W. Hu, \apj 511, 5 (1997).

\bibitem{bhuv}
	B. Jain \& U. Seljak, \apj 484, 560 (1997).

\bibitem{scott}
	S. Dodelson, E. W. Kolb, S. Matarrese, A. Riotto \& P. Zhang, submitted to \prd (2005) [astro-ph/0503160].

\bibitem{smith}
	R. E. Smith \etal, \mnras 341, 1311 (2003).

\bibitem{bin}
	We have checked that our results in this paper are insensitive to the choice of redshift binning.
	For instance, the rms error on $M$ (Fig. \ref{fig:testM}) from a low redshift survey like
	the SNfactory ($z = 0.03 - 0.08$) changes by less than a percent when we change the binning:
	from treating it as one redshift bin to dividing it into 5 bins.

\bibitem{area}
	Except for the next to bottom solid line, 
	we assume the survey is circularly symmetric, with an angular radius of
	$\theta^{\rm max}$. The survey area is $2\pi (1 - {\,\rm cos}\theta^{\rm max.})$,
	or in square degrees $2\pi (1 - {\,\rm cos}\theta^{\rm max.}) \times (180/\pi)^2$.
	The full sky has $\theta^{\rm max} = \pi$ and an area of about $41000$ square degrees.

\bibitem{linearVSnl}
	As explained in the last section, the linear power spectrum is used for computing all velocity terms.
	One might wonder if the conclusion on the subdominance of the Poissonian velocity term 
	would change if we had used the nonlinear power spectrum. We find that using the nonlinear power
	raises $(\sigma_1^{\rm Poiss.,\, vel.})^2$ by less than a factor of $2$, and so the Poissonian velocity
	term remains subdominant. Using the nonlinear power has very little impact on $C_{11}^{\rm vel.}$, as expected,
	since it is dominated by power on large scales.
	Note that using the nonlinear power spectrum is strictly speaking not self-consistent since we make use of the linear
	growth in eq. (\ref{sigPfull}). Doing so, however, might give us a rough idea
	on how large the Poissonian velocity term might be due to nonlinear effects.

\bibitem{fixed}
	Why are we interested in the errorbar on $M$ keeping other parameters fixed, instead of, say, the marginalized error on $M$?
	A low redshift SN survey such as the SNfactory has minimal constraining power on parameters other than $M$.
	The Fisher matrix $F_{\alpha\beta}$ (eq. [\ref{fisher}]) for such a low redshift survey can be approximated by 
	$F_{\alpha\beta} = 0$ for all $\alpha$ and $\beta$ except for $F_{44}$. (Recall that $p_1$ to $p_4$ stand for
	$w_{\rm pivot}$, $w_a$, $\Omega_{\rm de}$ and $M$.) The errorbar on $M$ fixing everything else
	from this low redshift survey is exactly $\sqrt{1/F_{44}}$.
	When it is combined with a high redshift survey, the resulting Fisher matrix 
	can be well-approximated by the Fisher matrix for the high redshift survey, plus the Fisher matrix for the low redshift one.
	In other words, the low redshift contribution to 
	the eventual final errors on $w_{\rm pivot}$, $w_a$, etc for the combined surveys is determined entirely by $F_{44}$.
	This is why, for the low redshift survey, we care about the error on $M$ keeping everything else fixed.
	{\it However}, it is important to stress that the above splitting of the Fisher matrix into
	a low z portion and a high z portion is for pedagogy only
	(i.e. the discussion in this very footnote). 
	We do {\it not} employ this splitting in any of
	our error forecasts; we {\it always} use the full exact Fisher matrix.

\bibitem{martin}
	M. White, submitted to Astroparticle Physics (2004) [astro-ph/0405593].

\bibitem{comments}
	This is strictly unnecessary: as can be seen from eq. [\ref{sigPfull}] and [\ref{Cijfull}],
	$C^{\rm lens.}_{ij}$ as well as $\sigma_i^{\rm Poiss.,\, lens}$ are independent of bin width as long
	as it is small, and only
	depend on the mean redshift $z_i$ or $z_j$.

\bibitem{sys}
	The systematic error is assigned according to the prescription of either
	\cite{kim} ($\sigma^{\rm sys.} = 0.02 \times
	z/1.7)$ or \cite{linderhut05} ($\sigma^{\rm sys.} = 0.02 \times (1+z)/2.7$).
	In both cases, the prescriptions are bin specific: they
	are for each redshift bin of $\Delta z = 0.1$. Our calculations do use $\Delta z = 0.1$
	for the high $z$ surveys, but use $\Delta z = 0.05$ for the low z survey (SNfactory) -- for
	this lowest $z$ bin, we assign systematic error according to the prescriptions as if 
	it has $\Delta z = 0.1$. Note that while the systematic error prescriptions are
	bin specific, all statistical error calculations in this paper (both
	the intrinsic and large scale structure scatter) are insensitive to the precise
	choice of binning (see \cite{bin}).

\bibitem{linderhut05}
	E. V. Linder \& D. Huterer, \prd 72, 043509 (2005).

\bibitem{hamuy}
	M. Hamuy \etal, \aj 106, 2392 (1993).

\bibitem{whelaniben}
	J. Whelan \& I. Iben, \apj 186, 1007 (1973).

\bibitem{dalal}
	N. Dalal, D. E. Holz, X. Chen \& J. A. Frieman, \apjl 585, 11 (2002).

\bibitem{gunnarsson}
	C. Gunnarsson, T. Dahlen, A. Goobar, J. Jonsson, E. Mortsell, submitted to \apj (2005) [astro-ph/0506764].

\bibitem{nealchris}
	N. Dalal \& C. S. Kochanek, \apj 572, 25 (2002).

\bibitem{john}
	D. J. Radburn-Smith, J. R. Lucey \& M. J. Hudson, \mnras 355, 1378 (2004).

\bibitem{williams}
	L. L. R. Williams \& J. Song, \mnras 351, 1387 (2004) [astro-ph/0403680].

\bibitem{wang}
	Y. Wang, JCAP 0503, 005 (2005) [astro-ph/0406635].

\bibitem{menard}
	B. Menard \& N. Dalal, \mnras 358, 101 (2004) [astro-ph/0407023].

\bibitem{hamuy96}
	M. Hamuy, M. M. Phillips, N. B. Suntzeff, R. A. Schommer, J. Maza \& R. Aviles, \aj 112, 2398 (1996).

\bibitem{riess95}
	A. G. Riess, W. H. Press \& R. P. Kirshner, \apjl 445, 91 (1995).

\bibitem{riess97}
	A. G. Riess, M. Davis, J. Baker \& R. P. Kirshner, \apjl 488, 1 (1997).
	
\bibitem{davis}
	M. Davis, A. Nusser \& J. A. Willick, \apj 473, 22 (1996).

\bibitem{willick}
	J. A. Willick, S. Courteau, S. M. Faber, D. Burstein, A. Dekel \& M. A. Strauss, \apj 109, 333 (1997).

%\bibitem{tonry}
%	J. L. Tonry, A. Dressler, J. P. Blakeslee, E. A. Ajhar, A. B. Fletcher, G. A. Luppino, M. R. Metzger, C. B. Moore,
%	\apj 546, 681 (2001).

\bibitem{ed}
	A nice review is: E. Bertschinger, Les Houches Summer School Lectures (1993) [astro-ph/9503125].


\bibitem{wald}
	R. M. Wald, General Relativity, University of Chicago Press (1984).


%\cite{Wood-Vasey:2004pj}
%\bibitem{Wood-Vasey:2004pj}
%  W.~M.~Wood-Vasey {\it et al.},
%  %``The Nearby Supernova Factory,''
%  New Astron.\ Rev.\  {\bf 48}, 637 (2004)
%  [arXiv:astro-ph/0401513].
%  %%CITATION = ASTRO-PH 0401513;%%

%\cite{Riess:1999ka}
%\bibitem{riess99}
%  A.~G.~Riess,
%  %``Peculiar Velocities from Type Ia Supernovae,''
%  arXiv:astro-ph/9908237.
%  %%CITATION = ASTRO-PH 9908237;%%


%\bibitem{pec}
%	Peculiar motion modifies the observed redshift of a SN. Since we are ultimately interested
%	in the relation between magnitude and redshift, one can instead view 
%	peculiar motion as introducing fluctuations in the magnitude for a given observed redshift.
%	This is explained in more detail in \S \ref{flux}. {\bf or is it Appendix \ref{app:dL}??}

%\bibitem{horizon}
%	Fluctuations on scales of the order of the horizon are generally too small to affect error forecasts for SN surveys.

\end{thebibliography}
\end{document}